\documentstyle[12pt,bezier]{article}
\def \sect #1 {\setcounter{equation} 0\section{#1}}
\def \be  {\begin{equation}}
\def \ee  {\end{equation}}
\def \ba  {\begin{eqnarray}}
\def \ea  {\end{eqnarray}}
\def \baa {\begin{eqnarray*}}
\def \eaa {\end{eqnarray*}}
\def \bb  {\begin{thebibliography}}
\def \eb  {\end{thebibliography}}
\newcommand \ci [1] {\cite{#1}}
\newcommand \bi [1] {\bibitem{#1}}
\def \lab #1 {\label{#1}}
\newcommand\re[1]{(\ref{#1})}
\def \qqquad {\qquad\quad}
\def \qqqquad {\qquad\qquad}
\newcommand\lr[1]{{\left({#1}\right)}}

\def \tr {\mbox{tr}}
\newcommand \vev [1] {\langle{#1}\rangle}

\newcommand \ket [1] {|{#1}\rangle}
\newcommand \bra [1] {\langle {#1}|}
\newcommand\bin[2]{\left({#1}\atop{#2}\right)}
\def \e {\mbox{e}}
\def \CO {{\cal O}}

\def \CH {{\cal H}}
\def \CI {{\cal I}}
\def \CQ {q}
\newcommand \widebar [1] {\overline{#1}}
\renewcommand{\Re}{\mathop{\rm Re}\nolimits}
\newcommand\fra[2]{\mbox{\small $\frac{#1}{#2}$}}

\font\cmss=cmss10 \font\cmsss=cmss10 at 7pt
\def\inbar{\,\vrule height1.5ex width.4pt depth0pt}
\def\IC{\relax\hbox{$\inbar\kern-.3em{\rm C}$}}
\def\IZ{\relax\ifmmode\mathchoice
{\hbox{\cmss Z\kern-.4em Z}}{\hbox{\cmss Z\kern-.4em Z}}
{\lower.9pt\hbox{\cmsss Z\kern-.4em Z}}
{\lower1.2pt\hbox{\cmsss Z\kern-.4em Z}}\else{\cmss Z\kern-.4em Z}\fi}
\def\IR{{\hbox{{\rm I}\kern-.2em\hbox{\rm R}}}}

\newcommand \partder [1] {{\partial \over\partial #1}}
\def\Im{\hbox{\rm Im}\,}
\newcommand{\as}{\ifmmode\alpha_{\rm s}\else{$\alpha_{\rm s}$}\fi}

\begin{document}

\def\thefootnote{\fnsymbol{footnote}}
\thispagestyle{empty}
\hfill\parbox{35mm}{{\sc ITP--SB--94--62}\par
                        hep-ph/9501232  \par
                        December, 1994}
\vspace*{28mm}
\begin{center}
{\LARGE Bethe Ansatz for QCD Pomeron
\footnote{The paper is based on talks delivered at CERN (May 94),
Princeton Univ. (June 94) and at the 2nd Workshop on Small$-x$ and
Diffractive Physics at the Tevatron, Fermilab (Sept 94)}
}
\par\vspace*{15mm}\par
{\large G.~P.~Korchemsky}%
\footnote{On leave from the Laboratory of Theoretical Physics,
          JINR, Dubna, Russia}
\par\bigskip\par\medskip

{\em Institute for Theoretical Physics, \par
State University of New York at Stony Brook, \par
Stony Brook, New York 11794 -- 3840, U.S.A.}
\end{center}
\vspace*{15mm}

\begin{abstract}
The equivalence is found between high--energy QCD in the generalized leading
logarithmic approximation and the one-dimensional Heisenberg magnet.
According to Regge theory, the high energy asymptotics of hadronic scattering
amplitudes are related to singularities of partial waves in the complex
angular momentum plane. In QCD, the partial waves are determined by nontrivial
two-dimensional dynamics of the transverse gluonic degrees of freedom. The
``bare'' gluons interact with each other to form a collective excitation,
the Reggeon. The partial waves of the scattering amplitude satisfy the
Bethe-Salpeter equation whose solutions describe the color singlet compound
states of Reggeons -- Pomeron, Odderon and higher Reggeon states. We show that
the QCD Hamiltonian for reggeized gluons coincides in the multi--color limit
with the Hamiltonian of XXX Heisenberg magnet for spin $s=0$ and spin
operators the generators of the conformal $SL(2,\IC)$ group. As a result,
the Schrodinger equation for the compound states of Reggeons has a sufficient
number of conservation laws to be completely integrable. A generalized Bethe
ansatz is developed for the diagonalization of the QCD Hamiltonian and for
the calculation of hadron-hadron scattering. Using the Bethe Ansatz solution
of high--energy QCD we investigate the properties of the Reggeon compound
states which govern the Regge behavior of the total hadron-hadron cross
sections and the small$-x$ behavior of the structure functions of deep
inelastic scattering.
\end{abstract}
\newpage
\def\thefootnote{\arabic{footnote}}
\setcounter{footnote} 0

\newpage

\section{Introduction}

The Regge problem is one of the longstanding problems in high-energy
physics which goes back to sixties \ci{Col}.
In a general sense, it is the problem
of behavior of hadronic scattering amplitudes in
the limit of high-energies $s$ and fixed transferred momentum $t$
\be
s \gg -t \sim M^2
\lab{kin}
\ee
with $M$ being a hadronic mass scale. There are a wealth of
experimental data which still confirm its intricacies, e.g., remarkably
linear Regge trajectories of hadronic families, approximate duality of
hadronic scattering amplitudes \ci{Col},
the growth of total hadronic cross sections \ci{land}
and the small$-x$ behavior of the structure functions of deep
inelastic scattering \ci{dis}.

There were numerous attempts to solve the Regge problems during the pre-QCD
era which have led to the development of the theory of the Regge poles
\ci{Col}.
The theory was inspired by the remarkable observation by Regge that,
in quantum mechanics, the high-energy asymptotics of the scattering
amplitudes in potential scattering is governed by the singularities of
the partial waves in the complex angular momentum plane. It was understood
from general principals of relativistic quantum field theory \ci{Col}
that a similar phenomenon occurs in high-energy hadronic scattering in
Regge kinematics \re{kin}. Namely, after the decomposition of the scattering
amplitude $A(s,t)$ over partial waves as
\be
A(s,t)= i s \int_{\delta-i\infty}^{\delta+i\infty}
\frac{d \omega}{2\pi i}\ \lr{\frac{s}{M^2}}^\omega\ \widetilde A(\omega,t)\,,
\lab{omega}
\ee
the behavior of $A(s,t)$ in the Regge limit \re{kin} depends on the
singularities of the partial waves, $\widetilde A(\omega,t),$
which lie in the
complex $\omega-$plane to
the left from the integration contour.
Possible singularities of $\widetilde A(\omega,t)$
were identified as Regge poles and Regge cuts.
Among all families of Regge poles there is a special one corresponding
to the so called Pomeranchuk poles or Pomerons, which have vacuum quantum
numbers and which provide dominant contributions to the hadronic scattering
amplitude. However,
Regge theory itself does not allow us to calculate the positions of these
singularities and it is still a challenge for QCD as a theory
of strong interaction to give us this input from ``first principals''.
After almost twenty years of a great deal of activity, we have
different phenomenological models to describe the Regge behavior of the
scattering amplitudes \ci{mod0,mod1,mod2,is}, but we do not have a complete
understanding from QCD.

In this paper we concentrate on perturbative QCD approach to the
Pomeron \ci{Lip1}. Namely, we consider hadrons as consisting of partons
whose distributions are described
by the nonperturbative hadronic wave functions in the infinite momentum frame.
Hadrons scatter by exchanging quarks and gluons in the $t-$channel
and we treat their interaction using the $S-$matrix of perturbative QCD.
The hadronic scattering amplitude is a convolution of hadronic wave functions
and partonic scattering amplitude. Although it cannot be calculated
entirely perturbatively, we can find its evolution of as a function of energy
$s$ in perturbative QCD. The corresponding equations are called the
evolution equations.

There are at least two reasons why perturbative approach might make sense.
First, since we do not understand dynamics in the nonperturbative regime
of QCD, the perturbative approach is a first approximation to start
with. Second, there is a simple way that perturbation theory itself could
tell us
about nonperturbative effects. Namely, studying ambiguities of perturbative
series associated with the contribution of the so-called infrared renormalons
one can get some insight into the structure of nonperturbative corrections
\ci{ren1,ren2,ren3,ren4}.

Analyzing the scattering amplitude in the Regge limit \re{kin}, one
might be especially interested to consider the case $t=0$, corresponding
to the elastic forward scattering. Then, the optical theorem \ci{Col}
allows us to relate the imaginary part of the scattering amplitude to the
total hadronic cross section
\be
\sigma^{\rm tot}_{h_{_{\rm A}}h_{_{\rm B}}}(s)
= \frac1{s}\Im A_{h_{_{\rm A}}h_{_{\rm B}}}(s,t=0)\,,
\lab{tot}
\ee
with $A_{h_{_{\rm A}}h_{_{\rm B}}}$
the amplitude of elastic scattering of hadrons $h_{_{\rm A}}$ and
$h_{_{\rm B}}$.
Moreover, replacing one of the incoming hadrons by virtual photon
with $M^2=-Q^2$ we get an analogous relation%
\footnote{The exact relation between electron-proton inclusive cross
section and the proton structure function $F_2(x,Q^2)$ can be found
in \ci{geor}.}
for the structure function of deep inelastic scattering $\gamma^*+h\to X$
\be
F(x,Q^2) = \frac1{s}\Im A_{\gamma^* h}(s,t=0)\,,\qquad s=Q^2/x  \,,
\lab{opt}
\ee
where $-Q^2$ is a photon virtuality and $x$ is the Bjorken
variable. The Regge limit \re{kin} corresponds to $s\gg Q^2$ or,
equivalently, to small$-x$ asymptotics of the structure
function $F(x,Q^2)$. We stress that relations \re{tot} and \re{opt}
have been found
under the additional conditions of unitarity and analyticity of the
scattering amplitude, $A(s,t)$, in its different channels and that
one of the consequences of these conditions is the Froissart bound
\be
\sigma^{\rm tot}_{h_1h_2}(s) \leq {\rm const.\ }\log^2 s \,,\qquad
F(x,Q^2) \leq {\rm const.\ }\log^2 x\,,\qqqquad
s\to \infty\,,\ \ x\to 0\,.
\lab{bound}
\ee
Thus, in order to apply \re{tot} and \re{opt} in perturbative QCD,
we have to calculate hadronic scattering amplitude and
preserve the conditions of unitarity and analyticity.

Expanded in powers of the coupling constant $\as$,
the $2\to 2$ scattering amplitudes have the
following general form \ci{Bar,CW} in Regge kinematics \re{kin},
\be
A(s,t)=i\sum_{m=0}^\infty
\as s \left[(\as\log s)^{m} f_{m,m}(t)
                        + \as (\as\log s)^{m-1} f_{m,m-1}(t)
                        + \ldots + \as^{m} f_{m,0}(t)\right] + \CO(s^0)\,,
\lab{An}
\ee
where the $f_{m,l}(t)$ are some functions of the transferred momentum.
In the case of deep inelastic scattering, the $\CO(s^0)$ term includes
$\log Q^2$ corrections to the scattering amplitude $A_{\gamma^*h}$.
Since we are not able to resum corrections
to all orders in $\as$, we need an effective approximation to the perturbative
series which, first, will correctly describe the high-energy asymptotics
of $A(s,t)$ in perturbative QCD and, second, will be consistent with
the unitarity
and analyticity of the $S-$matrix. Combined together, these two
conditions put severe restrictions on possible approximate schemes
\ci{Lip1,Bar,CW,Cheng}.
One can easily satisfy the first condition by taking the leading logarithmic
approximation
\be
\as \log s/M^2 \sim 1\,,\qqquad \as \ll 1\,,
\lab{LLA}
\ee
in which only the
$f_{m,m}(t)$ term is kept in \re{An} and all nonleading terms
$f_{m,m-1}(t),$ $...$ , $f_{m,0}(t)$ are neglected. However, the
resummation of leading logarithmic terms leads to the result \ci{bfkl}
\be
A_{\rm LLA}(s,t)=i\sum_{m=0}^\infty \as s(\as\log s)^m f_{m,m}(t)\propto
s^{1+\Delta_{\rm BFKL}(t)}
\lab{del}
\ee
with $\Delta_{\rm BFKL}(0)=\frac{\as N}{\pi} 4\log 2>0$,
which violates the unitarity bound \re{bound}.
The $s-$channel
unitarity of the $S-$matrix is explicitly broken in the leading
logarithmic approximation \ci{bfkl}. To restore unitarity one has to take into
account nonleading logarithmic corrections to the scattering amplitude
\re{An} which will involve $f_{m,m-1}(t),$ $f_{m,m-2}(t),$ $...$ terms.
Thus, to understand the Regge behavior in perturbative QCD we need a new
approach beyond the leading logarithmic approximation, which will preserve
the unitarity of the $S-$matrix. It is this requirement which makes the Regge
problem so complicated in perturbative QCD.

One of the
possible ways to cure the unitarity problem, the so-called generalized
leading logarithmic approximation, has been proposed in \ci{Bar,Cheng}
and can be formulated as follows%
\footnote{We should note that the generalized leading logarithmic
approximation restores unitarity of the scattering amplitude in its direct
$s$, $t$ and $u-$channels but not in the subchannels corresponding to the
intermediate states \ci{Bar}.}.
Although the leading logarithmic terms
$f_{m,m}(t)$ violate unitarity of \re{An}, the total sum of both leading and
nonleading terms satisfies the unitarity conditions. Therefore, there should
exist some minimal number of nonleading terms in \re{An} which compensate
nonunitarity of the leading logarithmic result. The generalized leading
logarithmic approximation allows us to identify the ``minimal'' subset of
such nonleading terms $f_{m,m-1}^{\rm min},$ $f_{m,m-2}^{\rm min},$ $...$ .
Of course, this procedure gives us only some pieces of nonleading
terms $f_{m,m-1},$ $f_{m,m-2},$ $...$ and not their exact expressions.
The question remains on how small are nonleading corrections,
$f-f^{\rm min}$, which are left after a minimal set has been found.
We have an answer \ci{Cheng} confirmed by lowest order perturbative
calculations but the general proof is still missing.

In what follows, we use the generalized leading logarithmic approximation
to understand the Regge behavior of the scattering amplitude. We show that
in this approximation QCD turns out to be closely related to two-dimensional
exactly solvable spin chain models \ci{Lip,prog}. Such kind of relation has
been expected for a variety of reasons \ci{bb,PT,VV,LLS,Lip1}. The first
calculations to lowest orders of perturbative QCD \ci{PT}
showed that the contributions of individual Feynman
diagrams to the partonic scattering amplitude \re{An} in the Regge limit
\re{kin} have a form of effective two-dimensional amplitudes in which
$\as\log s$ plays a role of an expansion parameter and the dependence of
$f_{m,m}$ on the transferred momentum $t$ comes from two-dimensional
integrals over transverse components $k_\alpha$ $(\alpha=1,2)$
of the momenta of exchanged gluons.%
\footnote{Here we refer to the Sudakov decomposition of the gluon momenta
over longitudinal and transverse momenta, $K=\alpha P_{_{\rm A}} +
\beta P_{_{\rm B}}+k$
with $P_{_{\rm A}}$ and $P_{_{\rm B}}$ hadron momenta.}
Another argument came from the physical picture of hadron--hadron
scattering \ci{VV}, which suggests that the Regge asymptotics is
determined by a two-dimensional dynamics of transverse partonic
degrees of freedom. In this paper we confirm these observations
and identify the two-dimensional structure behind the high-energy QCD in
the Regge limit as corresponding to the celebrated one-dimensional XXX
Heisenberg magnet for the special value of the spin $s=0$ corresponding to
a unitary representation of the conformal $SL(2,\IC)$ group \ci{prog}. This
correspondence allows us to apply powerful quantum inverse scattering
method \ci{BA1,BA2,BA3,BA4} for the calculation of the nonleading
logarithmic corrections $f_{m,m-1}^{\rm min},$ $f_{m,m-2}^{\rm min},$ $...$
and it opens the
possibility \ci{bb} of finding the Regge asymptotics of scattering amplitude
in QCD by means of the generalized Bethe Ansatz.

The paper is organized as follows. In Sect.~2 we describe the generalized
leading logarithmic approximation and find expressions for large$-s$
behavior of hadronic scattering amplitudes and small$-x$ asymptotics of the
structure functions of deep inelastic scattering in terms of wave
functions and energies of the compound Reggeon states. In Sect.~3 we
review the properties of the Reggeon interaction and consider
simplifications which occur in the limit of multi--color QCD. In Sect.~4
we show that the holomorphic and antiholomorphic Reggeon hamiltonians
coincide with the hamiltonian of XXX Heisenberg magnet for noncompact
spin $s=0$. The Schrodinger equation for the compound Reggeon states
turns out to be completely integrable and the quantum inverse scattering
method is applied to identify hidden conservation laws. In Sect.~5 we
describe the generalized Bethe Ansatz which allows us to find the spectrum
of the Reggeon states using the solutions of the Baxter equation.
In Sect.~6 we discuss solution of the Baxter equation and general
properties of the Reggeon states. Sect.~7 contains concluding remarks.

\section{Generalized leading logarithmic approximation}
\setcounter{equation} 0

We begin the construction of the scattering amplitude $A(s,t)$ in the
generalized leading logarithmic approximation by summarizing the
properties of the leading logarithmic approximation \re{LLA}
in which only the coefficient functions $f_{m,m}(t)$ survive in \re{An}.

In the leading logarithmic approximation \ci{Lip1,bfkl}, the partonic
scattering amplitude is dominated by the contribution of soft gluons
propagating in the $t-$channel between hadrons and interacting with each other.
This interaction leads to the remarkable property of
gluon reggeization \ci{Lip1,bfkl}. The ``bare'' gluons interact with
each other
to form a collective excitation, the Reggeon, which has the
quantum numbers of the gluon. However, in contrast with the
gluon, the properties of the Reggeon depend not only on its momentum but
also on the energy of particles interacting with the Reggeon. Introducing
the Reggeon as a new elementary excitation, it is natural to replace QCD
in the leading logarithmic approximation by an effective field theory
in which the calculation of the scattering amplitudes can be performed
in terms of the propagation of Reggeons in the $t-$channel and their
interaction with each other. Each diagram in the effective theory
is equivalent to an infinite sum of Feynman diagrams involving
bare gluons. Then, the asymptotic behavior of the scattering amplitude is
given in the leading logarithmic approximation by the contribution of
diagrams with only two Reggeons propagating in the $t-$channel,
the famous ladder diagrams \ci{Lip1,bfkl}.

However, two Reggeon diagrams do not satisfy the $s-$channel unitarity
condition and in the generalized leading logarithmic approximation one
has to add additional diagrams to restore unitarity \ci{Bar,CW,Cheng}.
This minimal
set of diagrams is obtained from the two Reggeon diagram by iterating
the number of Reggeons in the $t-$channel.
For example, the first nonleading
correction corresponds to the diagram with three Reggeons propagating
in the $t-$channel. Continuing this procedure, we come to the conclusion
that the scattering amplitude is given in the generalized leading
logarithmic approximation by the sum of the diagrams shown in
fig.~\ref{ladder}.
These diagrams have a form of generalized ladder diagrams
\ci{Bar,Cheng}, in
which summation is performed over all possible numbers of rungs
representing the Reggeon interaction,
and over all possible number of Reggeons
in the $t-$channel, $n=2,$ $3,...$ . It will be shown below that
the scattering of each two Reggeons is
described by the same effective theory as we started with in the leading
logarithmic approximation.

One of peculiar features of the Reggeon
scattering is that it does not change the number of Reggeons in the
$t-$channel. For $n=2$ the diagram of fig.\ref{ladder} represents the leading
logarithmic result for the scattering amplitude and it contributes to the
coefficient functions $f_{m,m}(t)$ to all orders of perturbation theory.
The contribution of the diagram with $n=3$ is suppressed by a power of
$\as$ with respect to that for $n=2$ and it determines the first
nonleading coefficient $f_{m,m-1}^{\rm min}$ in \re{An}. In general,
the contribution of the diagrams of fig.\ref{ladder}
to the partial wave $\widetilde
A(\omega,t)$ can be represented in the following form
\be
\widetilde A(\omega,t)=\widetilde A_2(\omega,t)+\as \widetilde A_3(\omega,t)
+ ... \equiv \sum_{n=2}^\infty \as^{n-2} \widetilde A_n(\omega,t)\,.
\lab{Aon}
\ee
Here, the $n-$th term corresponds to the diagram with $n$ Reggeons in the
$t-$channel, and the functions $A_n(\omega,t)$ have the following expansion
\be
\widetilde A_n(\omega,t)= \sum_{l=0}^\infty \lr{\frac{\as}{\omega}}^{l+1} \ l!\
f_{n-2+l,l}(t)\,.
\lab{lfac}
\ee
Each term in this sum has a singularity at $\omega=0$ which
after integration in \re{omega} is transformed into $\as(\as \log s)^l$
correction. When \re{lfac} and \re{Aon}
are substituted into \re{omega}, they reproduce the perturbative
expansion \re{An}. It does not mean however that the total sum,
$\widetilde A_n(\omega,t)$, has singularities only at $\omega=0$.
If the partial wave $\widetilde A_n(\omega,t)$ sums up to a
singularity at $\omega=\omega_0$, its contribution to the
scattering amplitude \re{omega} would have been
a Regge asymptotic behavior
$\sim s^{\omega_0}$. Indeed, as we shall see below, this turns out to be
the case for $\widetilde A_2(\omega,t)$,
which has a square-root singularity at $\omega=\Delta_{\rm BFKL}
=\frac{\as N}{\pi}4\log 2$,
the so-called BFKL Pomeron \ci{bfkl}. We have to generalize this result
for the higher partial waves $\widetilde A_n(\omega,t)$ with arbitrary
$n$ and find their singularities in the complex $\omega-$plane.

\subsection{The Bethe--Salpeter equation for the partial waves}

The partial wave $\widetilde A_n(\omega,t)$ can be represented in the
following form \ci{Lip1,bfkl} in terms of partonic distributions inside
hadrons,
$\Phi_{_{\rm A}}$ and $\Phi_{_{\rm B}}$,
and partonic scattering amplitude, $T_n$,
\be
\widetilde A_n(\omega,t)=\int [d^2k] \int [d^2k']\
\Phi_{\rm A}
\lr{\{k\};q}\ T_n\lr{\{k\},\{k'\};\omega}\
\Phi_{\rm B}\lr{\{k'\};q}\,,
\lab{fact}
\ee
where $[d^2k] = \prod_{j=1}^{n} d^2 k_j \delta(\sum_jk_j-q)$ and the
definition of $[d^2k']$ is analogous. Here, the hadrons $h_{_{\rm A}}$ and
$h_{_{\rm B}}$ emit
$n$ Reggeons in the $t-$channel with momenta $\{k\}=(k_1,\ldots,k_n)$ and
$\{k'\}=(k_1',\ldots,k_n')$, respectively, and the total transferred
momentum is ${q}=\sum_{j=1}^n{k}_j=\sum_{j=1}^n{k}_j'$, with ${q}^2=-t$.
The functions $\Phi_{\rm A}$ and $\Phi_{\rm B}$
describe the internal structure of the
hadrons, and they can be expressed in terms of hadronic wave functions in
the infinite momentum frame. In the high-energy limit \re{kin},
both functions depend on the transverse momenta of Reggeons \ci{Lip1,bfkl}
and not on the
energy $s$, or equivalently on $\omega$. The function $T_n$ describes
$n$ to $n$ scattering of Reggeons in the $t-$channel, and is the main
object of our consideration. In \re{fact}, the integration is performed
over transverse momenta of $n$ Reggeons, while integration over longitudinal
components is performed inside $T_n(\omega)$. The partonic distributions and
the Reggeon scattering amplitude depend on the color indices of Reggeons
and summation over these indices is implied in \re{fact}.

Considering the partonic distribution functions as states in a
two-dimensional transverse phase space for the $n$ Reggeons, one can
rewrite the scattering amplitude \re{fact} as the following matrix element
\be
\widetilde A_n(\omega,t)=\bra{\Phi_{\rm A}(q)} T_n(\omega)
\ket{\Phi_{\rm B}(q)}
\,,\qqquad t=-q^2\,,
\lab{Tn}
\ee
where the transition operator $T_n(\omega)$ describes the elastic
scattering of $n$ Reggeons. To find the transition operator
$T_n$ we notice that diagrams of fig.\ref{ladder} have a ladder structure,
which suggests a Bethe--Salpeter--like equation
for $T_n(\omega)$ shown in fig.\ref{figBS}. The corresponding
equation has the following form \ci{Bar,KP}
\be
\omega T_n(\omega) = T_n^{(0)} + {\cal H}_n T_n(\omega)\,,
\lab{BS}
\ee
where $T_n^{(0)}$ corresponds to the free propagation of $n$ Reggeons in the
$t-$channel, while the operator ${\cal H}_n$ describes the pair--wise
interactions of $n$ Reggeons. The Reggeon scattering amplitude,
$T_n\lr{\{k\},\{k'\};\omega}$, depends on the transverse momenta and the
color indices of $n$ incoming and $n$ outgoing Reggeons. The hamiltonian
${\cal H}_n$ acts in \re{BS} only on the incoming Reggeons,
and is given by
\be
{\cal H}_n=-\frac{\as}{2\pi} \sum_{n \ge i > j \ge 1} H_{ij} \ t_i^a t_j^a \,,
\lab{ham}
\ee
where the sum goes over all possible pairs $(i,j)$ of Reggeons.
Each term in this sum has a color factor, which is given by the direct
product of the gauge group generators in the adjoint representation of
the $SU(N)$ group, acting in the color space of $i-$th and $j-$th Reggeons,
$$
t^a_i=\underbrace{
I\otimes ... \otimes t^a}_i \otimes ... \otimes I  \,,
\qqquad  (t^a)_{bc}=-if_{abc}\,,
$$
with $f_{abc}$ the structure constants of the $SU(N)$. The
Reggeon interaction \re{ham} is described by the two-particle hamiltonian
$H_{ij}$, which depends only on the transverse momenta of Reggeons.
If $\{k_1,...,k_n\}$ and $\{k_1',...,k_n'\}$ are the transverse momenta of
incoming and outgoing Reggeons, respectively, then the operator $H_{ij}$
acts on Reggeon momenta as follows
\be
\bra{k_1,...,k_n}H_{ij}\ket{k_1',...,k_n'}=
H(k_i,k_j|k_i',k_j')
\prod_{l=1,l\neq i,j}^n\delta^2(k_l-k_l')\,,
\lab{ham1}
\ee
where $H(k_i,k_j|k_i',k_j')$ is given by \ci{bfkl}
\ba
H_{ij}\equiv H(k_i,k_j|k_i',k_j')&=&\frac1{\pi}\Bigg\{
\frac{k_i^2(p-k_i')^2+(p-k_i)^2k_i'^2-(k_i-k_i')^2p^2}
{k_i^2(p-k_i)^2(k_i'-k_i)^2}
\lab{ham2}
\\
&-&
\delta^2(k_i-k_i') \int d^2 k'
\left[\frac{k_i^2}{k'^2+(k_i-k')^2}
      +\frac{(p-k_i)^2}{(p-k')^2+(k_i-k')^2}
\right]
\Bigg\}
\nonumber
\ea
with $p=k_j+k_i=k_j'+k_i'$. The operator $T_n^{(0)}$ is equal to the
product of $n$ Reggeon propagators, and in momentum representation it can be
written as
\be
\bra{k_1,...,k_n}T_n^{(0)}\ket{k_1',...,k_n'}=
\prod_{j=1}^n\delta^2(k_j-k_j')\frac1{k_j^2}\,.
\lab{T0}
\ee
Iteration of the equation \re{BS} reproduces the ladder in fig.~\ref{ladder},
and
gives the perturbative expansion of $T_n(\omega)$ as series in $\as/\omega$
which leads to \re{lfac}.
But this is not what we are interested in because we want to resum the whole
series in \re{lfac}. Using the Bethe--Salpeter
equation \re{BS}, we find the general solution for the transition operator
\be
T_n(\omega)=\frac1{\omega - {\cal H}_n}{T_n^{(0)}}\,.
\lab{sol}
\ee
We recall that the hamiltonian ${\cal H}_n$ corresponds to the pair--wise
interaction \re{ham} of $n$ Reggeons and it describes the evolution of
the $n-$Reggeon state in the $t-$channel.

Suppose for a moment that we know the spectrum of the Reggeon hamiltonian
\be
{\cal H}_n \ket{\chi_{n,\{\alpha\}}} = E_{n,\{\alpha\}}
            \ket{\chi_{n,\{\alpha\}}}\,,
\lab{eig}
\ee
with $\{\alpha\}$ some set of quantum numbers, which parameterize
possible solutions of the equation. Then, the eigenstates of ${\cal H}_n$ can
be identified as compound states of $n$ Reggeons, $\ket{\chi_n}$,
and the corresponding eigenvalues, $E_n$, determine the energies of these
states. A simplest example of such states to be rederived below
is the BFKL Pomeron \ci{bfkl}
which is built from $n=2$ Reggeons. Once we solve \re{eig}, we
can expand the transition amplitude \re{sol} over eigenstates of
Reggeon hamiltonian as%
\footnote{To give a meaning to this expansion one has to specify the
          scalar product on the Hilbert space of the eigenstates and
          find the hermiticity properties of the hamiltonian. This
          will be done in sect.~\ref{scalar}}
\be
T_n(\omega)=\sum_{\{\alpha\}}
\frac1{\omega-E_{n,\{\alpha\}} }
\ket{\chi_{n,\{\alpha\}}}\bra{\chi_{n,\{\alpha\}}} T_n^{(0)}\,,
\lab{exp}
\ee
where the sum over $\alpha$ means the summation over discrete and integration
over continuous $\alpha$. Combined with \re{Tn}, this expression
implies that the singularities of the partial waves
$\widetilde A_n(\omega,t)$
in the $\omega-$plane are determined by the eigenvalues $E_{n,\{\alpha\}}$
of the Reggeon hamiltonian ${\cal H}_n$. Moreover, substituting
\re{exp} into \re{Tn} and \re{omega},
we formally take a residue at $\omega={\cal H}_n$ and
get the following expression for the scattering amplitude
\ba
&&
A(s,t)=\sum_{n=2}^\infty \as^{n-2} A_n(s,t)\,,\qquad
\nonumber
\\
&&
A_n(s,t)=is \bra{\Phi_{\rm A}} s^{{\cal H}_n} T_n^{(0)} \ket{\Phi_{\rm B}}
        =is \sum_{\{\alpha\}} \beta_{\rm A}^{\{\alpha\}}(t)
                              \beta_{\rm B}^{\{\alpha\}}(t)\
                              s^{E_{n,\{\alpha\}}}
\,,
\lab{An-res}
\ea
where the residue functions $\beta_{\rm A}^{\{\alpha\}}$ and
$\beta_{\rm B}^{\{\alpha\}}$ measure the
overlapping between partonic distribution functions and the wave
functions of the compound states of Reggeons
\be
\beta_{\rm A}^{\{\alpha\}}(t) =
\bra{\Phi_{\rm A}(q)} \chi_{n,\{\alpha\}} \rangle\,,\qqquad
\beta_{\rm B}^{\{\alpha\}}(t) =
\bra{\chi_{n,\{\alpha\}}}T_n^{(0)}  \ket{\Phi_{\rm B}(q)}\,,
\qqquad (t=-q^2)\,.
\lab{beta}
\ee
Although the definitions of the residue functions look different, they are
closely related to each other as we will show below in \re{betas}.
We recall that the scalar product of states, used in \re{Tn}, \re{An-res}
and \re{beta}, implies integration over
transverse momenta $k_1,$ $...$ , $k_n$ of $n$ Reggeons as well
as summation over their color indices.

Thus, the scattering amplitude \re{An-res} has
Regge behavior, $A_n(s,t) \sim s^{E_{n,\{\alpha\}}}$, with the intercept
related to the energy of compound states. Moreover,
the scattering amplitude $A_n(s,t)$
gets its dominant contribution from the
compound states with maximum energy $E_n^{\rm max}=
{\rm max}_{\{\alpha\}}E_{n,\{\alpha\}}$.
Once we have expressions for the wave functions of the Reggeon states we
can use either phenomenological or model hadronic distributions
to evaluate $\beta_{\rm A}^{\{\alpha\}}(t)$ and
$\beta_{\rm B}^{\{\alpha\}}(t)$.

The scattering amplitude \re{An-res} is given by a sum of $A_n(s,t)$
over all possible numbers of Reggeons in the $t-$channel.
With each individual $A_n(s,t)$ we associate the family of the compound
Reggeon states that appear as solutions of the eigenstate
equation \re{eig}. Thus, the original problem of calculating the
scattering amplitude $A(s,t)$
in the generalized leading logarithmic approximation
is reduced to the problem \re{eig} of the diagonalization of the Reggeon
hamiltonian
${\cal H}_n$ for an arbitrary number of Reggeons.

\subsection{Evolution equation for the structure function}

Let us apply \re{An-res} in the case of forward
photon-proton scattering, $A_{\gamma^* {\rm p}}(s,t=0)$.
Using \re{opt} and taking the imaginary part
of the scattering amplitude \re{An-res}, we obtain an expansion for the
structure function $F(x,Q^2)$ as a sum of nonleading corrections $F_n$
similar to that in \re{An-res}. The expressions for $F_n$ follow from
expressions \re{An-res} for $A_n(s=Q^2/x,t=0)$.
Finally, we find the small$-x$ asymptotic behavior of the structure function
of deep inelastic scattering in the
generalized leading logarithmic approximation as%
\footnote{Here we assumed that the energies of the compound states,
$E_{n,\{\alpha\}}$, and the product of the residue functions are real
in \re{An-res}. The proof will be given in Sect.~5.3.}
\ba
&&
F(x,Q^2)=\sum_{n=2}^\infty\as^{n-2} F_n(x,Q^2) + \CO(x^0)\,,
\qquad
\nonumber
\\
&&
F_n(x,Q^2)= \sum_{\{\alpha\}} x^{-E_{n,\{\alpha\}}}
\vev{\Phi_{\gamma^*(Q^2)}(0)|\chi_{n,\{\alpha\}}}\
\beta_{\rm p}^{\{\alpha\}}(0)\,,
\lab{Fn-res}
\ea
where the $Q^2-$dependence comes from $\Phi_{\gamma^*(Q^2)}(0)$ and
from $\CO(x^0)$ terms.
To make the correspondence with the parton model, we represent
$F_n(x,Q^2)$ as convolution of a partonic cross section $\sigma_0$ and
a partonic distribution function $f$,
\be
F_n(x,Q^2)=\int [d^2 k]\
\sigma_0^{a_1...a_n}(Q^2;k_1,...,k_n)\
           f^{a_1...a_n}(x;k_1,...,k_n) + \CO(x^0)\,,
\lab{Fn}
\ee
where $\left[d^2 k\right]\equiv d^2 k_1 ... d^2 k_n
\delta^{(2)}(k_1+...+k_n)$, with a delta-function included to ensure
the condition $t=0$. The partonic cross
section, $\sigma_0$, describes the coupling of $n$ Reggeons to the
virtual photon through the quark loop.
For large invariant mass, $-Q^2$,
it can be computed perturbatively \ci{Lip1}
as a scalar product of the photon state
$\Phi_{\gamma^*(Q^2)}(t=0)$ and the state of $n$ Reggeons with transverse
momenta $k_1,... ,k_n$ and color indices $a_1,... ,a_n$,
$$
\sigma_0^{a_1...a_n}(Q^2;k_1,...,k_n)=
\bra{\Phi_{\gamma^*(Q^2)}(0)} k_1,a_1;...;k_n,a_n \rangle\,.
$$
The partonic distribution, $f$, takes into account all effects of Reggeon
interactions and the coupling of Reggeons to the nonperturbative
hadronic state.
In the generalized leading logarithmic approximation it is given by
\be
f^{a_1...a_n}(x;k_1,...,k_n)=\bra{k_1,a_1;...;k_n,a_n} x^{-{\cal H}_n}
T_n^{(0)}
\ket{\Phi_{\rm p}(0)}\,,
\lab{ff}
\ee
where $k_n=-(k_1+...+k_{n-1})$ and the Bjorken variable $x$ is small in
the Regge limit. In terms of the eigenstates and eigenvalues, \re{eig},
the distribution function can be written as
\be
f^{a_1...a_n}(x;k_1,...,k_n)=\sum_{\{\alpha\}} x^{-E_{n,\{\alpha\}}}\
\chi_{n,\{\alpha\}}^{a_1...a_n}(k_1,...,k_n)\ \beta_{\rm p}^{\{\alpha\}}(0)\,,
\lab{f}
\ee
where we have used the following
notation for the wave function of the compound state
of $n$ Reggeons,
$$
\chi_{n,\{\alpha\}}^{a_1...a_n}(k_1,...,k_n)=\bra{k_1,a_1;...;k_n,a_n}
\chi_{n,\{\alpha\}} \rangle\,,
$$
and $\beta_{\rm p}^{\{\alpha\}}(0)$ is nonperturbative residue factor.
Although we cannot compute the distribution function $f$
in perturbative QCD, we may use expression \re{ff} to find the evolution
equation for $f$ as a function of $x$
\be
x\frac{d}{dx} f_n^{\{a\}}(x;\{k\})
= - \int [d^2k']\
{\cal H}_n^{\{a\};\{b\}}\lr{\{k\};\{k'\}}\
f_n^{\{b\}}(x;\{k'\})
\lab{evol}
\ee
where $\{a\}=(a_1,...,a_n)$ and $\{k\}=(k_1,...,k_n)$ and where
the Reggeon hamiltonian was defined in \re{ham}, \re{ham1} and \re{ham2}.
For $n=2$
the evolution equation \re{evol} coincides with the BFKL equation \ci{bfkl}.

Comparing expressions \re{An-res} and \re{Fn-res}, we conclude that the Regge
asymptotics is defined in both cases by the properties of the Reggeon
hamiltonian. The hadronic scattering amplitude and the structure function
of deep inelastic scattering are given by an infinite
sum of terms corresponding to the diagram of fig.~\ref{ladder}
with fixed number $n$ of
Reggeons in the $t-$channel. The first term, $n=2$, describes the leading
logarithmic asymptotics and,
taken alone, it violates the $s-$channel unitarity of the $S-$matrix.
Each next term in the sum over $n$ in \re{An-res} and \re{Fn-res}
defines a nonleading contribution, which is
suppressed by a power of the coupling constant with respect to that of
the $(n-1)-$st term. Examining the high-energy behavior of the $n-$th
term, $A_n(s,t)$ and $F_n(x,Q^2)$,
we recognize that the contribution of compound Reggeon states with positive
energy $E_n >0$ to \re{An-res} and \re{Fn-res} grows as a power of energy,
and thus violates the Froissart bound \re{bound}. We recall however that
these nonleading terms have been defined from the very beginning in such a
way that to restore unitarity of the total scattering amplitude.
This means, that, although each term in the sum \re{An-res} and \re{Fn-res}
violates the unitarity bound, unitarity is restored in their sum.%
\footnote{To get some insight into possible mechanisms of unitarity
          restoration one may consider the following example,
          $s^\lambda - \as s^{2\lambda}/2
          + \as^2 s^{3\lambda}/3+ ... =-\lambda/\as \log s$,
          in which each individual term of the series violates the unitarity
          bound but their sum does not.}

\section{Properties of the Reggeon hamiltonian}
\setcounter{equation} 0

We turn now to the solution of the eigenstate problem \re{eig}
and \re{ham} in order to find
the spectrum of the compound Reggeon states. The Reggeon hamiltonian
\re{ham}, \re{ham1} and \re{ham2} has
been found from the analysis of Feynman diagrams
contributing to the hadronic scattering amplitude in the generalized
leading logarithmic approximation, and it inherits the properties
of high-energy QCD in the Regge limit.

The solution of \re{eig} is known only for $n=2$ and the corresponding
compound state of two Reggeons is called the BFKL Pomeron \ci{bfkl}.
The Reggeon hamiltonian was defined in \re{ham}, \re{ham1} and \re{ham2}
in the two-dimensional
space of transverse momenta of Reggeons. It is more convenient however to
analyse equation \re{eig} in two-dimensional coordinate
space, the so called impact parameter space, rather than in momentum space.
To this end, we perform a
two-dimensional Fourier transformation, and replace in
\re{fact}, \re{ham1} and \re{ham2} the
two-dimensional transverse momenta $k_1,... ,k_n$ and $k_1',... ,k_n'$
by two-dimensional impact vectors $b_1,... ,b_n$ and
$b_1',... ,b_n'$ which describe the transverse coordinates of Reggeons.
Then, for the impact vectors $b_j=(x_j,y_j)$
we define holomorphic and antiholomorphic complex coordinates
$(z_j,\bar z_j)$ as
$$
z_j=x_j+iy_j\,,\qqquad
\bar z_j=x_j-iy_j \,,\qqquad (j=1,...,n)\,,
$$
and analogous coordinates $(z_j',\bar z_j')$ for the impact vectors
$b_j'$. Now one can use \re{ham1} and \re{ham2}
to find the two-particle Reggeon
kernel $H_{ij}$ in the impact parameter space.
It turns out that, expressed in terms of holomorphic and
antiholomorphic coordinates, the kernel $H_{ij}$ becomes
holomorphically separable \ci{Lip2}, i.e.,
\be
H_{ij}=  H(z_i,z_k) + H(\bar z_i,\bar z_k)\,.
\lab{two}
\ee
Here, two operators on the r.h.s. act separately on holomorphic and
antiholomorphic coordinates of Reggeons.  After Fourier transformation
of \re{ham1} and \re{ham2} they are given by
the following {\it equivalent\/} representations
\ba
H(z_i,z_k)&=&-P_i^{-1}\log(z_i-z_k)P_i
           -P_k^{-1}\log(z_i-z_k)P_k
           -\log(P_iP_k)-2\gamma_{_{\rm E}}
\nonumber
\\
     &=&-2\log(z_i-z_k)-(z_i-z_k)\log(P_iP_k)(z_i-z_k)^{-1}-2\gamma_{_{\rm E}}
\,,
\lab{H1}
\ea
where $P_i=i\partder{z_i}$ and $\gamma_{_{\rm E}}$ is the Euler constant.
The same operator can be represented as \ci{Lip2}
\be
H(z_i,z_k)=\sum_{l=0}^\infty \frac{2l+1}{l(l+1)-{\bf L}_{ik}^2}-\frac2{l+1}\,,
\qquad
{\bf L}_{ik}^2=-(z_i-z_k)^2\frac{\partial^2}{\partial z_i\partial z_k}
\,.
\lab{H2}
\ee
Substituting this expression into \re{two} and \re{ham},
we find that the Reggeon hamiltonian $\CH_n$
is invariant under the conformal transformations \ci{Lip1}
\be
z_i \to \frac{az_i+b}{cz_i+d} ,\qqqquad
\bar z_i \to \frac{\bar a \bar z_i+\bar b}{\bar c \bar z_i+\bar d}
\,,
\lab{con}
\ee
with $ad-bc=\bar a\bar d-\bar b\bar c=1$. Indeed, the generators of these
transformations,
\be
S^3= \sum_{k=1}^n z_k\partial_k\,,\qqquad
S^-=-\sum_{k=1}^n \partial_k\,,\qqquad
S^+= \sum_{k=1}^n z_k^2\partial_k
\lab{tra}
\ee
and the analogous antiholomorphic generators $\bar S^3$, $\bar S^-$ and
$\bar S^+$ form the $SL(2,\IC)$ algebra and commute with
${\bf L}_{ik}^2$ and, as a consequence, with ${\cal H}_n$.

Let us consider the properties of the eigenstate $\chi_{n,\{\alpha\}}$ of the
Reggeon hamiltonian \re{eig} in the impact parameter space. These states
are parameterized by quantum numbers $\{\alpha\}$, which should appear as
eigenvalues of some ``hidden'' integrals of motion, and by a two-dimensional
real vector $b_0$,
which represents the center of mass of the compound Reggeon
state. In this notation,
the wave function
$\chi_{n,\{\alpha\}}=\chi_{n,\{\alpha\}}(\{b_i\};b_0)
\equiv\chi_{n,\{\alpha\}}(\{z_i,\bar z_i\};z_0, \bar z_0)$
satisfies the relations \re{eig}, \re{ham} and \re{two}, or equivalently
\be
E_{n,\{\alpha\}}
\chi_{n,\{\alpha\}}(\{z_i,\bar z_i\};z_0, \bar z_0)
= -\frac{\as}{2\pi} \sum_{j,k=1,\, j>k}^n
\left[H(z_j,z_k) + H(\bar z_j,\bar z_k)\right]
\  t^a_j t^a_k \
\chi_{n,\{\alpha\}}(\{z_i,\bar z_i\};z_0, \bar z_0) \,,
\lab{sch}
\ee
which can be interpreted as a two-dimensional Schrodinger equation for a
system of $n$ pair-wise interacting particles with coordinates
$\{z_i,\bar z_i\}$ and internal color degrees of freedom.
One may try to rewrite the total hamiltonian in \re{sch} as a sum of
holomorphic and antiholomorphic parts using the fact that $H(z_j,z_k)$
and $H(\bar z_j,\bar z_k)$ commute. However, the resulting two terms
do not commute with each other due to nontrivial color factor in \re{sch},
and as a consequence holomorphic and antiholomorphic sectors become coupled
to each other via color degrees of freedom. It is this interaction which
makes difficult the solution of the Schrodinger equation \re{sch}.

Each Reggeon carries a color charge $t^a_j$ and the total charge of $n$
Reggeons is equal to their sum $\sum_{j=1}^n t^a_j$.
Since the compound Reggeon states propagate in the $t-$channel
between two hadrons, they carry zero color charge, unchanged by the Reggeon
interaction,
\be
[{\cal H}_n,\sum_{j=1}^n t^a_j]=0\,,\qqqquad
\sum_{j=1}^n t^a_j\ \ket{\chi_{n,\{\alpha\}}}=0
\lab{col}
\ee
An essential simplification occurs in \re{sch} in multi-color limit, $N\to
\infty$ and $\as N={\rm fixed}$, \ci{Lip2,Ven}. In this limit,
only planar diagrams of fig.\ref{ladder} survive,
which have the form of a cylinder attached by both edges to the hadronic
states \ci{Ven}.
Reggeons propagate along the sides of the cylinder and it makes
them possible to interact only with two nearest Reggeons. Using the double
line representation for Reggeon color charge and applying the standard rules
of large $N$ counting, one finds that the color structure in \re{sch} can be
simplified as%
\footnote{There are two special case, $n=2$ and $n=3$, when the
          color factor can be calculated for finite $N$.
          For $n=2$ we use \re{col} in the form $t_1+t_2=0$
          to get $t_1^a t_2^a = -t_1^a t_1^a = -C_A$; for $n=3$ a similar
          condition $t_1+t_2+t_3=0$ leads to
          $t_1^a t_2^a = 1/2(t_1^a+t_2^a)^2-1/2(t_1^at_1^a + t_2^at_2^a)
                       = -C_A/2$ with $C_A=N$ the Casimir operator.}
$$
t_1^a t_2^a \to -N\,,\qquad \mbox{for $n=2$}\,;
\qqqquad
t^a_i t^a_j \to -\frac{N}2 \delta_{i,j+1}\,,\qquad \mbox{for $n\geq 3$}\,,
$$
where $i, j=1,... ,n$ and the Reggeons with $i=1$ and $i=n+1$ are
considering as coinciding. Then, in the multi-color limit we find that,
first, the color factors become trivial in \re{sch} and as a consequence
holomorphic
and antiholomorphic sectors become decoupled and, second, inside each sector
in \re{sch} the pair--wise Reggeon interaction is replaced
by a nearest--neighbour
interaction with periodic boundary conditions. Thus, in the large$-N$ limit,
the two-dimensional Schrodinger equation \re{sch} is replaced by a system of
two one-dimensional Schrodinger equations \ci{Lip2},
\be
H_n \varphi_{n,\{\alpha\}}(\{z_i\};z_0)=
\varepsilon_{n,\{\alpha\}}\varphi_{n,\{\alpha\}}(\{z_i\};z_0)\,,\qquad
{\bar H}_n \bar\varphi_{n,\{\alpha\}}(\{\bar z_i\};\bar z_0)=
\bar\varepsilon_{n,\{\alpha\}}\bar\varphi_{n,\{\alpha\}}(\{\bar z_i\};
\bar z_0)\,.
\lab{reg}
\ee
The hamiltonians $H_n$ and $\bar H_n$ are defined as
\be
H_n = \sum_{k=1}^n H\lr{z_k,z_{k+1}}\,,\qqqquad
{\bar H}_n = \sum_{k=1}^n H\lr{\bar z_k,\bar z_{k+1}}\,,
\lab{near}
\ee
with two-particle hamiltonians given by \re{H1} or \re{H2}
and $z_{n+1}\equiv z_1$. Once we know the solution of \re{reg}, the eigenstates
\re{eig} of the Reggeon hamiltonian in the multi-color limit can be found as
\be
\chi_{n,\{\alpha\}}(\{z_i,\bar z_i\};z_0, \bar z_0)
=\varphi_{n,\{\alpha\}}(\{z_i\};z_0) \
 \widebar{\varphi}_{n,\{\alpha\}}(\{\bar z_i\};\bar z_0) \,,
\lab{chi}
\ee
and the corresponding eigenvalues are given by
\be
E_{n,\{\alpha\}}=\frac{\as N}{4\pi}\lr{
\varepsilon_{n,\{\alpha\}}+\widebar\varepsilon_{n,\{\alpha\}}}\,.
\lab{E}
\ee
We identify \re{reg} as a system of Schrodinger equations for two
one-dimensional lattice models with nearest neighbour interaction \re{near},
and with the number of lattice sites equal to the number of Reggeons.
The hamiltonians of the models are defined by holomorphic and antiholomorphic
parts of the Reggeon hamiltonian \re{two},
and the quantum space for each site of the lattice, ${\tt h}_k$,
$(k=1,...,n)$,
is parameterized by holomorphic and antiholomorphic coordinates,
respectively. Finally,
we notice that $H_n$ and $\bar H_n$ are separately
invariant under conformal
transformations \re{con}.

Thus, using the properties of the Reggeon interaction in the multi-color limit,
we reduced the original problem \re{eig} and \re{sch} of the diagonalization
of the Reggeon hamiltonian to the solution of the one-dimensional lattice
models \re{reg}.
However, this alone does not at all
guarantee that these lattice
models can be exactly solved. To explore this possibility
we recall the famous example
of the exactly solvable one-dimensional lattice model, the XXX Heisenberg
chain of interacting spins $s=1/2$ with the hamiltonian
\be
H_{s=1/2}=-\frac12\sum_{k=1}^n \lr{\sigma_k^a \sigma_{k+1}^a - 1}\,,
\lab{1/2}
\ee
where $\sigma_k^a$ are Pauli matrices acting in the $k-$th site of the
lattice. This model can be generalized \ci{ttf} by means of the
quantum inverse scattering method to give a family of
XXX Heisenberg models for an arbitrary {\it complex\/}
values of spin $s$. The unique feature of these models is that
they are completely integrable, that is, that they contain additional
integrals
of motion whose number is equal to the number of degrees of freedom.
The reason why we might be interested in considering complex spin $s$
is that the one-dimensional lattice models \re{reg}, \re{near} and \re{H2},
which describe the Regge behavior of multi-color QCD, coincide with the XXX
Heisenberg magnet for spin $s=0$.
Let us now discuss how we may make this identification.

\section{Multi-color QCD as XXX Heisenberg magnet}
\setcounter{equation} 0

To identify the lattice models \re{reg} we forget for a moment about
their QCD origin and construct the hamiltonian
of the XXX Heisenberg magnet for spin $s$ \ci{ttf}. Let us consider the
one-dimensional lattice with $n$ sites and assign the spin operators
$S_k^a$ with $a=1,2,3$ to all sites $k=1,... ,n$,
\be
S_k^+=z_k^2\partial_k -2sz_k\,,\qqqquad
S_k^-=-\partial_k\,,\qqqquad
S_k^3=z_k\partial_k -s\,,
\lab{spin}
\ee
where $S^\pm=S^1\pm iS^2$. The spins $\vec S_k$ act as differential
operators on the local quantum space $\tt h_k$ corresponding to the $k-$th
site. The total spin of the lattice $\vec S=\sum_{k=1}^n \vec S_k$ acts
on the full quantum space ${\tt H}_n=\otimes_{k=1}^n {\tt h}_k$, and for
$s=0$ the operators $S^a$ coincide
with the generators of conformal transformations, eq.\re{tra}.
The definition of the integrable XXX spin chain is based on the
existence of a fundamental operator
${\rm R}_{km}(\lambda)$, which acts on the space
${\tt h}_k\otimes {\tt h}_m$,
and which satisfies the Yang--Baxter equation \ci{BA1,BA2,BA3,BA4}
\be
{\rm R}_{k m}(\lambda-\mu)
{\rm R}_{k l}(\lambda-\rho)
{\rm R}_{m l}(\mu-\rho)
=
{\rm R}_{m l}(\mu-\rho)
{\rm R}_{k l}(\lambda-\rho)
{\rm R}_{k m}(\lambda-\mu)\,,
\lab{YB}
\ee
with $\lambda$, $\mu$ and $\rho$ arbitrary complex spectral parameters
and $k,m,l=\IZ_+$.
The solution of this equation for an arbitrary complex $s$ is given
by \ci{ttf,krs}
\be
{\rm R}_{km}(\lambda)
=\frac{\Gamma(i\lambda-2s)\Gamma(i\lambda+2s+1)}
      {\Gamma(i\lambda-J_{km})\Gamma(i\lambda+J_{km}+1)}\,,
\lab{R}
\ee
where the operator $J_{km}$ acts on the space ${\tt h}_k\otimes {\tt h}_m$
and satisfies the equation
$$
J_{km}(1+J_{km})=(\vec S_{k}+\vec S_{m})^2
                        =2 \vec S_{k}\vec S_{m} + 2s(s+1)\,.
$$
The hamiltonian of the XXX Heisenberg magnet for spin $s$ is
defined as \ci{ttf}
\be
H_n=\sum_{k=1}^n H_{k,k+1}\,,\qqqquad
H_{k,k+1}=-i\frac{d}{d\lambda}\log {\rm R}_{k,k+1}(\lambda)\Bigg|_{\lambda=0}
\,.
\lab{two1}
\ee
In the special case, $s=1/2$, one can recover
the hamiltonian \re{1/2} from this expression
 while the limit $s\to\infty$ is related
to the nonlinear Schrodinger equation. Let us consider the
case $s=0$. Using the explicit expressions \re{spin} for the spin operators
for $s=0$ we obtain the two-particle holomorphic
hamiltonian up to an inessential (infinite)
constant%
\footnote{As we will show in Sect.~5.3 infinite constants corresponding to
holomorphic and antiholomorphic hamiltonians, $H_{k,k+1}$ and $\bar H_{k,k+1}$,
are canceled in the sum \re{E}.}
as
\be
H_{k,k+1}=-\psi(-J_{k,k+1})-\psi(1+J_{k,k+1})+2\psi(1)\,,\qquad
J_{k,k+1}(1+J_{k,k+1})=-(z_k-z_{k+1})^2\partial_k\partial_{k+1}
\lab{s=0}
\ee
with $\psi(x)=\frac{d}{dx} \Gamma(x)$. Comparing \re{s=0} and \re{H2}
we find that both expressions for the two-particle hamiltonians
coincide after we identify $J_{k,k+1}(1+J_{k,k+1})={\bf L}^2_{k,k+1}$
and perform
summation over $l$ in \re{H2}. This means \ci{prog} that the holomorphic and
antiholomorphic Reggeon hamiltonians \re{near} are identical to
the hamiltonian \re{s=0} of the XXX Heisenberg magnet for spin $s=0$.
Among other things, this
remarkable property implies that the system of Schrodinger equations
\re{reg},
which describes the Regge asymptotics of multi-color QCD in the
generalized leading logarithmic approximation, is completely integrable,
and that one can apply the quantum inverse scattering method to identify
the ``hidden'' integrals of motion.

To this end, we follow the standard procedure \ci{BA1,BA2,BA3,BA4}
and define the auxiliary
and fundamental Lax operators, $L_{k,a}(\lambda)$ and
${\rm L}_{k,f}(\lambda)$, respectively, for all sites of the lattice,
\be
L_{k,a}(\lambda)=\lambda I_k \otimes I_a + i \vec S_k \otimes \vec\sigma_a
                =\lr{\begin{array}{cc}  \lambda +i S_k^3 & iS_k^-
                                 \\  iS_k^+  & \lambda-iS_k^3
                     \end{array}}\,,
\qqquad
{\rm L}_{k,f}(\lambda)={\rm R}_{k,f}(\lambda)\,.
\lab{Lax}
\ee
These operators act in the space ${\tt h}_k\otimes V$ with the auxiliary
space $V=\IC^2$ for $L_{k,a}$ and $V={\tt h}_f$ for ${\rm L}_{k,f}$
(the auxiliary space ${\tt h}_f$ has the dimension of the local quantum
space ${\tt h}_k$).
The Lax
operators satisfy the Yang--Baxter equations similar to that in \re{YB}.
Now, we multiply Lax operators as matrices in the auxiliary space and
define the auxiliary monodromy matrix
\be
T_a(\lambda)=L_{n,a}(\lambda)L_{n-1,a}(\lambda)\ldots L_{1,a}(\lambda)
            =\lr{\begin{array}{cc}  A(\lambda) & B(\lambda)
                                 \\ C(\lambda) & D(\lambda)
                     \end{array}}
\lab{B}
\ee
and analogously the fundamental monodromy matrix
\be
{\rm T}_f(\lambda)={\rm L}_{n,f}(\lambda){\rm L}_{n-1,f}(\lambda)\ldots
                   {\rm L}_{1,f}(\lambda)\,.
\lab{Tf}
\ee
Finally, we take a trace of the monodromy matrices over auxiliary space
and obtain the auxiliary and fundamental transfer matrices
\be
\Lambda(\lambda) = \tr_a T_a(\lambda)\,,\qqquad
\tau(\lambda)=\tr_f {\rm T}_f(\lambda)\,,
\lab{tau}
\ee
which are operators in the full quantum space ${\tt H}_n$ of the lattice.
Due to the properties of the Lax operators \re{Lax}, the operators
$\tau$ and $\Lambda$ thus defined
commute with each other for different values of
the spectral parameters
\be
[\tau(\lambda), \tau(\mu)] = [\tau(\lambda), \Lambda(\mu)]
=[\Lambda(\lambda), \Lambda(\mu)] = 0 \,.
\lab{lam}
\ee
{}From this relation we find
that the system of Schrodinger equations \re{reg} has
two families of mutually commuting
{\it local\/} integrals of motions
\be
\CI_k = -i\frac{d^k}{d\lambda^k}\log\tau(\lambda)\Bigg|_{\lambda=0}\,,
\qqqquad
\CQ_{n-k} = \frac1{k!}\frac{d^k}{d\lambda^{k}}
              \Lambda(\lambda)\Bigg|_{\lambda=0}\,,
              \qquad
(k=1,2,..)\,.
\lab{int}
\ee
We conclude from \re{int} and \re{lam} that \ci{BA1,BA2,BA3,BA4}
\be
[H_n,\CI_k]=[H_n,\CQ_k]=[\CQ_k,\CQ_j]
=[\CI_k,\CI_j]=[\CQ_k,\CI_j]=0
\,.
\lab{int1}
\ee
Here, the operators $\CI_k$ and $\CQ_k$ describe the interaction between
$(k+1)$ nearest neighbors, and in the special case $k=1$
the operator $\CI_1$ coincides with the holomorphic Reggeon hamiltonian
\re{near}
\be
\CI_1 = H_n =\sum_{k=1}^n H_{k,k+1}
\,.
\lab{ene}
\ee
The expressions for $\CI_k$ with $k\geq 2$ are more complicated, but
one can easily find expressions for the operators $\CQ_k$ by
using the explicit form \ci{Lip,prog}
of the auxiliary Lax operator \re{Lax} for spin $s=0$,
\be
L_{k,a}=\lambda I + i\bin{1}{z_k}\otimes (z_k,-1)\partial_k\,.
\lab{lax}
\ee
Then, using the definition \re{lam} we find the auxiliary transfer
matrix as
\be
\Lambda(\lambda)=2\lambda^n +\CQ_2\lambda^{n-2}
                            +\CQ_3\lambda^{n-3}
                            +\ldots+\CQ_n
\lab{L}
\ee
with the operators $\CQ_k$ given by
\be
\CQ_k=\sum_{n\ge i_1>i_2>\cdots>i_k\geq 1}
i^k z_{i_1i_2} z_{i_2i_3}\ldots z_{i_ki_1}
\partial_{i_1}\partial_{i_2}\ldots\partial_{i_k}
\lab{Qk}
\ee
with $k=2,...,n$ and $z_{jk}\equiv z_j-z_k$.
In particular, for $\CQ_2$ we have expression
\be
\CQ_2=\sum_{n\geq j>k\geq 1} z_{jk}^2\partial_j\partial_k
= - S^3 S^3-\frac12\lr{S^+S^-+S^-S^+} \equiv -h(h-1)
\lab{cas}
\ee
which we identify as the Casimir operator of the conformal group \re{tra}
with $h$ being the conformal weight. Then, the relation $[\CQ_2,H_n]=0$
follows from the invariance of the holomorphic hamiltonian under
conformal transformations \re{con}.

Thus, the holomorphic Schrodinger equations \re{reg} has a sufficient number
of integrals of motion \re{int} and \re{int1} to be exactly solvable.
It is clear that the same consideration can be performed for the
antiholomorphic sector in \re{reg}. The wave functions of the compound
Reggeon state, $\chi_{n\{\alpha\}}$, depend on the quantum numbers
$\{\alpha\}$, which we now identify as the eigenvalues of the
integrals of motion \re{int} and their antiholomorphic partners. Moreover,
the original complicated problem of solving the Schrodinger equations \re{reg}
can be reduced to the diagonalization of the conservation laws \re{int}. We
remember that in the case of the XXX Heisenberg magnet for spin $s=1/2$ this
has been done by means of Algebraic Bethe Ansatz \ci{BA1,BA2,BA3,BA4}.
However, this method
requires the existence of a highest weight for each site of the lattice.
This makes it inapplicable for spin $s=0$, corresponding to
infinite-dimensional representations of the $SL(2,\IC)$ group.
To solve the XXX Heisenberg magnet for spin $s=0$, we have to generalize
the Bethe Ansatz to the case of the noncompact conformal group.

\section{Generalized Bethe Ansatz for Reggeon Hamiltonian}
\setcounter{equation} 0

The Generalized Bethe Ansatz for the XXX Heisenberg chain for spin $s=0$
has been developed in Ref.~\ci{prog}. It is based on the method of the
$Q-$operator \ci{Q,Skl} and on the relation \ci{prog}
between the XXX magnets for spins
$s=0$ and $s=-1$. To understand the latter relation we consider expression
\re{spin} for the spin operators for $s=0$ and $s=-1$ and find the
following identity, $\vec S_k^{(s=-1)}=P_k\ \vec S_k^{(s=0)} P_k^{-1}$
with $P_k=i\partial_k$. It leads to the similar relation between Lax
operators~\re{Lax},
\be
L_{k,a}^{(s=-1)}(\lambda)= P_k\ L_{k,a}^{(s=0)}(\lambda)\ P_k^{-1} \,,
\qqqquad
{\rm L}_{k,f}^{(s=-1)}(\lambda)= \frac{i\lambda+1}{i\lambda-1}
P_k\ {\rm L}_{k,f}^{(s=0)}(\lambda)\ P_k^{-1}\,,
\lab{laxlax}
\ee
where the additional factor in the last relation
follows from the numerator of the
fundamental $R-$operator \re{R}. After its substitution into \re{B} and
\re{Tf} we find the relation between transfer matrices \re{tau} in both models
\be
\Lambda^{(s=-1)}(\lambda)
= P_1...P_n \ \Lambda^{(s=0)}(\lambda)\, \lr{P_1...P_n}^{-1}
= \lr{z_{12}z_{23}...z_{n1}}^{-1}
  \Lambda^{(s=0)}(\lambda)\ z_{12}z_{23}...z_{n1}
\lab{auxaux}
\ee
and
$$
\lr{\frac{i\lambda-1}{i\lambda+1}}^n\tau^{(s=-1)}(\lambda)
= P_1...P_n\ \tau^{(s=0)}(\lambda) \lr{P_1...P_n}^{-1}
= \lr{z_{12}z_{23}...z_{n1}}^{-1}
  \tau^{(s=0)}(\lambda)\ z_{12}z_{23}...z_{n1}\,,
$$
where we took into account that operator
$\CQ_n=z_{12}z_{23}...z_{n1}P_1...P_n$, defined in \re{Qk},
commutes with the transfer matrices \re{tau}. Hence, the eigenstates of the
transfer matrices in the XXX models for spin $s=0$ and $s=-1$
are related as
\be
\varphi_n^{(s=-1)}(\{z_i\};z_0)
= \lr{z_{12}z_{23}...z_{n1}}^{-1}\varphi_n^{(s=0)}(\{z_i\};z_0)
= \CQ_n^{-1} P_1P_2...P_n \ \varphi_n^{(s=0)}(\{z_i\};z_0)\,.
\lab{chi1}
\ee
For corresponding eigenvalues of the hamiltonians and the integrals
of motion we have from \re{int}, \re{ene} and \re{auxaux}
\be
\varepsilon_n^{(s=0)}=\varepsilon_n^{(s=-1)}-2n\,,\qqquad
\CQ_k^{(s=0)}=\CQ_k^{(s=-1)}
\lab{equ}
\ee
and
$$
\CI_{2k}^{(s=0)}=\CI_{2k}^{(s=-1)}\,,\qqquad
\CI_{2k+1}^{(s=0)}=\CI_{2k+1}^{(s=-1)} - 2n (-)^k (2k)!
$$
where $\CQ_k$ and $\CI_k$ stand in \re{chi1} and \re{equ}
for eigenvalues rather than for operators.
These relations establish the equivalence between XXX Heisenberg magnets
for spins $s=0$ and $s=-1$.

We recall that the holomorphic Reggeon hamiltonian
\re{reg} coincides with the hamiltonian
\re{two1} and \re{s=0} of the XXX magnet for
spin $s=0$ and therefore their spectrum are identical
$$
\varphi_n^{(s=0)}(\{z_i\};z_0)=\varphi_{n,\{\alpha\}}(\{z_i\};z_0)\,,
\qqqquad
\varepsilon_n^{(s=0)}=\varepsilon_{n,\{\alpha\}}
$$
It is clear that the same correspondence takes place for the antiholomorphic
Reggeon hamiltonian, $\bar H_n$, in \re{reg}. The spectrum of the
Reggeon hamiltonian depends on the set of quantum numbers with the
conformal weight $h$ being one of them. As it follows from the definition
\re{cas}, parameter $(-h)$ has a meaning of the total spin of the XXX chain.
In the case of the compact spin $s=1/2$ XXX chain it
can take either integer or half integer values \ci{BA1,BA2,BA3,BA4}.
For noncompact spin
$s=0$ chain, possible values of $h$ can be even complex and their explicit
form will be found in Sect.~5.3. However, among all possible $h$ there
is a subset of integer positive $h \ge n$ which plays a special role.

\subsection{Algebraic Bethe Ansatz}

The reason why we prefer to deal with the spin
$s=-1$ in constructing the Bethe Ansatz is that there
exists a pseudovacuum state $\ket{\Omega}$ in the total quantum
space of the lattice which is the highest weight in each site
$$
\ket{\Omega}=\frac1{z_1^2z_2^2...z_n^2}\,,\qqqquad
S_k^+ \ket{\Omega}=0\,,\qqqquad
S_k^3 \ket{\Omega}= - \ket{\Omega}
$$
with the spin operators given by \re{spin} for $s=-1$. This property
allows us to apply the Algebraic Bethe Ansatz \ci{BA1,BA2,BA3,BA4}
and construct the
{\it special subset\/} of the eigenstates corresponding to the integer
positive values of the conformal weight $h\geq n$ in \re{cas}. The
corresponding eigenstates, the so-called Bethe states, are defined as
\ci{prog}
\be
\varphi^{(s=0)}_{n,\{\lambda\}}(\{z_i\};z_0)
=z_{12}z_{23}...z_{n1}\
                B(\lambda_1) B(\lambda_2) ... B(\lambda_l)
                \frac1{z_{10}^2z_{20}^2...z_{n0}^2}\,,
\qquad l=h-n=0,1, ...
\lab{Bet}
\ee
Here, $B(\lambda)$ is the differential operator which acts on holomorphic
coordinates $z_1,... ,z_n$ and which can be found through \re{B} and using
\re{lax} and \re{laxlax} as an element of
the auxiliary monodromy matrix, $T_a^{(s=-1)}$,
for the XXX magnet of spin $s=-1$. As
an example, we give expression for $B(\lambda)$ for $n=2$
\be
B_{n=2}(\lambda)=-\lambda(P_1+P_2) + P_1 P_2\ z_{12}\,.
\lab{B-23}
\ee
The Bethe states \re{Bet} are parameterized by the set of complex numbers
$\{\lambda\}=(\lambda_1,...,\lambda_{h-n})$ which are solutions of the Bethe
equation for spin $s=-1$
\be
\lr{\frac{\lambda_k-i}{\lambda_k+i}}^n
=\prod_{j=1,j\neq k}^{h-n} \frac{\lambda_k-\lambda_j+i}{\lambda_k-\lambda_j-i}
\,.
\lab{Beq}
\ee
The eigenvalues of the holomorphic Reggeon hamiltonian corresponding to the
Bethe states \re{Bet} are given by
\be
\varepsilon_n^{(s=0)}=-2n-2\sum_{k=1}^{h-n} \frac1{\lambda_k^2+1}
\,.
\lab{Ben}
\ee
The Bethe states \re{Bet} diagonalize the transfer matrices
$$
\Lambda(\lambda)=(\lambda-i)^n\prod_{k=1}^{h-n}
\frac{\lambda_k-\lambda+i}{\lambda_k-\lambda}
+(\lambda+i)^n\prod_{k=1}^{h-n}
\frac{\lambda_k-\lambda-i}{\lambda_k-\lambda}
$$
and we can find the eigenvalues of the conserved charges $\{\CQ\}$ from
this expression by expanding it in powers of $\lambda$ and
comparing with \re{L}. The eigenvalues of the operators $\{\CI\}$
can be found from \re{equ} using the relation
$$
\CI_k^{(s=-1)}=i(-)^k (k-1)! \sum_{j=1}^{h-n} \left[
(\lambda_j+i)^{-k}-(\lambda_j-i)^{-k}\right]
\,.
$$
Given a solution of the Bethe equation \re{Beq}, the relations
\re{Bet} and \re{Ben} allow us to find the solution of the
holomorphic Schrodinger equation in \re{reg}, the holomorphic
wave function \re{Bet} and the energy \re{Ben}. We stress that these
results correspond to the special case of positive integer conformal weight
$h\ge n$ while in general we are interesting to choose $h$ to be a
complex.

\subsection{Functional Bethe Ansatz}

The explicit form of the Bethe states \re{Bet} can be used in order to
generalize them for an arbitrary complex $h$.
The generalization is based on the existence \ci{Q,Skl} of the Baxter operator
$Q_n(\lambda)$ which acts on the full quantum space of the XXX spin $s=-1$
model and commutes with itself, $[Q_n(\mu),Q_n(\lambda)]=0$, and
with auxiliary transfer matrix, $[Q_n(\mu),\Lambda(\lambda)]=0$.
Moreover, the eigenvalues of the operator satisfy the Baxter equation
\be
\lr{
2\lambda^n + \CQ_2 \lambda^{n-2}+\CQ_3\lambda^{n-3}+...+\CQ_n
}Q_n(\lambda) = (\lambda+i)^n Q_n(\lambda+i)+(\lambda-i)^n Q_n(\lambda-i)\,,
\lab{Bax}
\ee
where $\CQ_k$ and $Q_n(\lambda)$ denote the eigenvalues of the corresponding
operators in the XXX model for spin $s=-1$. The relation \re{equ}
allows us not to indicate explicitly the value of spin. Solving the Baxter
equation we will get expression for $Q_n(\lambda)$ which depends on the
quantum numbers $h$ and $\CQ_k$ and which is the function of
complex spectral parameter $\lambda$. Then, for given solution
$Q_n(\lambda)$ of the Baxter equation the eigenvalues of the XXX magnet
hamiltonian $H_n^{(s=-1)}$ and of the integral of motions $\CI_k^{(s=-1)}$
are given by
$$
\varepsilon_n^{(s=-1)}=-i\frac{d}{d\lambda}
\log\frac{Q_n(-\lambda-i)}{Q_n(-\lambda+i)}\Bigg|_{\lambda=0}\,,
\qqqquad
\CI_k^{(s=-1)}=-i\frac{d^k}{d\lambda^k}
\log\frac{Q_n(-\lambda-i)}{Q_n(-\lambda+i)}\Bigg|_{\lambda=0}
\,.
$$
Using the relation \re{equ} between XXX models for spins $s=0$ and $s=-1$
we find the spectrum of the holomorphic Reggeon hamiltonian
\be
\varepsilon_n=-i\frac{d}{d\lambda}
\log\frac{\widetilde Q_n(-\lambda-i)}
         {\widetilde Q_n(-\lambda+i)}\Bigg|_{\lambda=0}
\,,\qquad
\widetilde Q_n(\lambda)\equiv \lambda^n Q_n(\lambda)
\lab{Bax-e}
\ee
and the eigenvalues of the integrals of motion
$$
\CI_k=-i\frac{d^k}{d\lambda^k}
\log\frac{\widetilde Q_n(-\lambda-i)}
{\widetilde Q_n(-\lambda+i)}\Bigg|_{\lambda=0}
\,.
$$
To define the corresponding eigenstates we introduce the Sklyanin operators
\ci{Skl} for the XXX magnet of spin $s=-1$. We use the definition \re{Lax} and
\re{spin} of the
auxiliary Lax operator for $s=-1$ and find from \re{B} that the operator
$B(\lambda)$ is a polynomial of order $n-1$ in $\lambda$ which we
represent as
\be
B(\lambda)=i S^- (\lambda - x_1)(\lambda-x_2)...(\lambda-x_{n-1})
\,,
\lab{x}
\ee
where $S^-=-\sum_{k=1}^n\partial_k$ is the total spin of the model
and $x_1,...,x_{n-1}$ are operator zeros of $B(\lambda)$.
The relative order of factors is inessential in \re{x}
since $[S^-,x_k]=[x_k,x_j]=0$. The explicit form of the operator zeros
$x_1,...,x_{n-1}$ can be found for any fixed $n$ from the definitions \re{x}
and \re{B}. In particular, using \re{B-23} we find the explicit expression
for the operator zero at $n=2$
\be
x_1\bigg|_{n=2} = i\frac{\partial_1\partial_2}{\partial_1+\partial_2}z_{12}
\,.
\lab{x1}
\ee
The operators $B(\lambda)$ and $x_k$ have different expressions for
$s=-1$ and $s=0$ models and the relation between them is similar to that
between the auxiliary monodromy matrices, eq.\re{auxaux}.
Then, having expressions for
the operators $x_1,... ,x_{n-1}$ in the XXX model for spin $s=-1$ we use the
solution of the Baxter equation \re{Bax} to find the eigenstates of the
holomorphic Reggeon hamiltonian in the following form
\be
\varphi_{n,\{\CQ\}}(\{z_i\};z_0)=z_{12}z_{23}...z_{n1}\
(iS^-)^{h-n}\ Q_n(x_1) Q_n(x_2) ... Q_n(x_{n-1})
\frac1{z_{10}^2z_{20}^2...z_{n0}^2}
\lab{Bax-s}
\ee
with $h$ a complex conformal weight. Notice that the operators
$Q_n(\lambda)$ and $x_j$ do not commute, and $Q_n(x_j)$ denotes the eigenvalue
of the operator $Q_n(\lambda)$
evaluated for operator value of the spectral parameter
$\lambda=x_j$.

Expressions \re{Bax-e} and \re{Bax-s} give the solution of the
holomorphic Schrodinger equation \re{reg} for an arbitrary complex conformal
weight $h$. Different eigenstates of the Reggeon hamiltonian are parameterized
by different solutions of the Baxter equation \re{Bax}. In particular,
for integer positive conformal weight $h\geq n$ the Baxter equation \re{Bax}
has
a polynomial solution
\be
Q_n(\lambda)={\rm const.}\ \prod_{k=1}^{h-n} (\lambda - \lambda_k)
\lab{Q-pol}
\ee
and taking $\lambda=\lambda_j$ in \re{Bax} we find that the parameters
$\lambda_1$, $...$, $\lambda_{h-n}$ satisfy the Bethe
equation \re{Beq}. After substitution of \re{Q-pol} into \re{Bax-s}
we use the definition \re{x} to identify \re{Bax-e} and \re{Bax-s} with
analogous expressions, \re{Ben} and \re{Bet}, given by the algebraic
Bethe ansatz.

To apply the expressions \re{Bax-e} and \re{Bax-s} one has to solve the
Baxter equation \re{Bax} for an arbitrary values of the quantum numbers
$h$ and $\CQ_k$ and under appropriate boundary conditions. In particular,
the condition for the solution of the Baxter equation \re{Bax} to be
a finite polynomial in $\lambda$ leads to the algebraic Bethe Ansatz
described in the previous section.

\subsection{Scalar product}
\lab{scalar}

The decomposition of the Reggeon hamiltonian over the complete set of the
states has been introduced before in \re{exp}. However this decomposition is
formal unless we specify the scalar product on the Hilbert space of the
Reggeon hamiltonian. Although we have the general expression \re{Bax-s}
for the compound Reggeon states, the choice of a scalar product together
with the condition of finiteness of the norm may rule out some of these
eigenstates and introduce selection rules for the quantum numbers $h$
and $\CQ_k$. In what follow, we use the scalar product proposed in \ci{prog}
and based on the relation between XXX magnets for spin $s=0$ and $s=-1$.

Let us define the following auxiliary hamiltonian
\be
\CH_n^{\rm aux}= H_n^{(s=-1)} + \bar H_n^{(s=0)} =\sum_{k=1}^n
H^{\rm aux}_{k,k+1}\,,
\lab{K}
\ee
which is equal to the sum of holomorphic XXX hamiltonian for spin $s=-1$
and antiholomorphic hamiltonian for spin $s=0$. Using relations \re{equ}
we find the spectrum of this hamiltonian as
\baa
E_n^{\rm aux}&=&\varepsilon_n^{(s=-1)}+\bar\varepsilon_n^{(s=0)}
\,,
\\
\chi^{\rm aux}(\{z_k,\bar z_k\};z_0,\bar z_0)
&=&\varphi^{(s=-1)}(\{z_k\};z_0)\bar\varphi^{(s=0)}(\{\bar z_k\};\bar z_0)
\,.
\eaa
These expressions are related to the eigenvalues and
eigenstates of the Reggeon hamiltonian,
\re{chi} and \re{chi1}, as
\ba
E_n^{\rm aux}&=&\varepsilon_n + \bar\varepsilon_n + 2n
\,,
\nonumber
\\
\chi^{\rm aux}(\{z_k,\bar z_k\};z_0,\bar z_0)&=&
\lr{z_{12}z_{23}...z_{n1}}^{-1} \chi(\{z_k,\bar z_k\};z_0,\bar z_0)
\,.
\lab{one}
\\
&=&\CQ_n^{-1} P_1P_2...P_n\ \chi(\{z_k,\bar z_k\};z_0,\bar z_0)
\nonumber
\ea
Then, in each site of the model \re{K} we have three holomorphic spin $s=-1$
operators and three antiholomorphic spin $s=0$ operators which
form the unitary representation of the principal series, $t^{0,2}$, of the
$SL(2,\IC)$ group \ci{zs}
and we identify $t^{0,2}$ as the local quantum space
${\tt h}_k$ in the $k-$th site. This identification leads to the following
properties. We notice that two-particle hamiltonian $H^{\rm aux}_{k,k+1}$ is
unbounded operator due to singularities of the $\psi-$function on real
negative axis,
\be
H^{\rm aux}_{12} = -\psi(-J_{12})-\psi(1+J_{12})
                   -\psi(-\bar J_{12})-\psi(1+\bar J_{12}) + C
\lab{K12}
\ee
where $J_{12}(J_{12}+1)=-\partial_1\partial_2 z_{12}^2$ and
$\bar J_{12}(\bar J_{12}+1)=-\bar z_{12}^2\bar\partial_1\bar\partial_2$
are two Casimir operators of the $SL(2,\IC)$.
The operators $J_{12}$ and $\bar J_{12}$ act on the tensor product
of two quantum spaces $t^{0,2}$ which can be decomposed \ci{zs}
into the direct sum of the principal series representations
$t^{0,2}\otimes t^{0,2}=\oplus_{\nu,m} t^{\nu,2m}$
with integer $m$ and real $\nu$. For fixed $m$ and $\nu$ the Casimir
operators have the following eigenvalues \ci{zs}
$$
J_{12}=-h\,,\qquad \bar J_{12}=-1+h^*\,,\qquad
h=\frac{1+m}2 -i\nu
\,.
$$
After their substitution into \re{K12} we use the identity
$\psi(x)-\psi(1-x)= - \pi\cot (\pi x)$ and find that singularities
at $\nu=0$ cancel in \re{K12} to give a finite real result
\be
H^{\rm aux}_{12} = -4 \Re \psi\lr{\frac{1+|m|}2+i\nu} + C
\lab{one-1}
\ee
The constant $C$ entering into \re{K12} is defined as%
\footnote{To derive \re{K12} and the expression for $C$
one uses the definition \re{two1} and \re{R} of the
two particle XXX hamiltonian for antiholomorphic spin $s$ and for
holomorphic spin $(-1-s^*)$ with $s=i\delta$ and $\delta\to 0$}
\be
C=\psi(-2i\delta) + \psi(1+2i\delta)
 +\psi(2-2i\delta) + \psi(-1+2i\delta)\bigg|_{\delta\to 0}
 =2+4 \psi(1)\,.
\lab{one-2}
\ee
The total quantum space of the model is the tensor product of $n$ copies
of $t^{0,2}$ and the scalar product on this space is given by \ci{zs}
\be
\vev{\chi_1^{\rm aux}|\chi_2^{\rm aux}} = \int dz d\bar z\
\chi_1^{\rm aux}(\{z_k,\bar z_k\})
\lr{\chi_2^{\rm aux}(\{z_k,\bar z_k\})}^*
\lab{sca}
\ee
where $dz d\bar z=\prod_{k=1}^n dz_k d\bar z_k$. With this choice of the
scalar product the hamiltonian $\CH^{\rm aux}_n$ as well as the
Reggeon hamiltonian, $\CH_n$, are bounded operators on the
quantum space of the model. Using one-to-one correspondence, \re{one},
between the spectra of the hamiltonians $\CH^{\rm aux}_n$ and
$\CH_n$, we obtain two different choices for the scalar product of the
Reggeon states $\chi$ \ci{Lip}
$$
\int dz d\bar z\
\chi_1(z,\bar z) P_1\bar P_1 ... P_n\bar P_n
\lr{\chi_2(z,\bar z)}^* \qquad
{\rm or}\qquad
\int dz d\bar z\
\frac{\chi_1(z,\bar z)\lr{\chi_2(z,\bar z)}^*}
{z_{12}\bar z_{12} ...z_{n,1}\bar z_{n,1}}
\,.
$$
For eigenstates of the Reggeon hamiltonian \re{chi}
both definitions are equivalent.

Another important property of the hamiltonian \re{K} is that it
is a selfadjoint operator,
$$
\lr{{\CH^{\rm aux}_n}}^\dagger=\CH^{\rm aux}_n
$$
since
for $s=-1$ and $s=0$ the spin operators \re{spin} are related as
${S^a}^\dagger|_{s=-1}={\bar S}^a|_{s=0}$. Together with \re{one}
this means that,
first, the eigenvalues of the Reggeon hamiltonian corresponding to the
eigenstates with the finite norm \re{sca} are real
\be
\Im E_n=\frac{\as N}{4\pi}
\Im\lr{\varepsilon_n + \bar\varepsilon_n}=0
\,,
\lab{real}
\ee
where $E_n$ is the energy of the compound state of $n$ Reggeons
defined in \re{eig} and \re{E}.
Second, the eigenstates \re{one} are orthogonal with respect to
the scalar product \re{sca} and, as a consequence, we have the following
orthogonality condition
\be
\delta_{\CQ,\CQ'} \delta(z_0-z_0')\delta(\bar z_0-\bar z_0')=
\int dz d\bar z
\frac{
\lr{\chi_{n,\{\CQ'\}}(\{z_k,\bar z_k\};z_0',\bar z_0')}^*
\chi_{n,\{\CQ\}}(\{z_k,\bar z_k\};z_0,\bar z_0)
}{z_{12}\bar z_{12}z_{23}\bar z_{23}... z_{n1} \bar z_{n1}}
\lab{ort}
\ee
and the completeness condition
\be
\prod_{k=1}^n \delta(z_k-z_k') \delta(\bar z_k-\bar z_k')=
\sum_{\{\CQ\}}\int dz_0 d\bar z_0
\frac{
\chi_{n,\{\CQ\}}(\{z_k,\bar z_k\};z_0,\bar z_0)
\lr{\chi_{n,\{\CQ\}}(\{z_k',\bar z_k'\};z_0,\bar z_0)}^*
}{z_{12}\bar z_{12}z_{23}\bar z_{23}... z_{n1} \bar z_{n1}}
\lab{comp}
\ee
where sum over $\{\CQ\}$ means summation over discrete and
integration over continuum quantum numbers $\CQ_k$.

We recall that relations \re{real} are valid only for the compound Reggeon
states with the finite norm \re{sca} and this condition restricts
possible values of the quantum numbers $\{\CQ_k\}$.
In particular, the energy of the compound state is given by \re{E}
$$
E_n=\frac{\as N}{4\pi}\lr{\varepsilon_n(\CQ) + \varepsilon_n(\bar\CQ)}
$$
with $\varepsilon_n=\varepsilon_n(\CQ)$ defined in \re{Bax-e} and with
analogous expression for $\bar\varepsilon_n=\varepsilon_n(\bar\CQ)$.
To find the energy $\varepsilon_n$ one has to solve the Baxter equation
\re{Bax} and it is natural to assume that for real $\{\CQ\}$ and
$\lambda$ the solution of the Baxter equation, $Q_n(\lambda)$, may be
taken to be real. Then, it follows from \re{Bax-e} and \re{Bax},
that $\varepsilon_n(\CQ)$ is a function of holomorphic quantum numbers
$\{\CQ_k\}$ with real coefficients and, hence, $\lr{\varepsilon_n(\CQ)}^*
=\varepsilon_n(\CQ^*)$.
We notice that in the model \re{K} the operators
$\CQ_k\equiv \CQ_k^{(s=-1)}$ and $\bar\CQ_k\equiv \CQ_k^{(s=0)}$
are related as $\CQ_k^\dagger=\bar\CQ_k$ for arbitrary
$k$. This identity together with \re{equ} leads to a
similar relation between possible values of the quantum numbers
\be
\CQ_k=\bar\CQ_k^*\,, \qqquad (k=2,...,n)
\lab{tri}
\ee
and as a result the energy of the compound Reggeon states is equal to
\be
E_n= \frac{\as N}{4\pi}\lr{\varepsilon_n(\CQ) + \varepsilon_n(\CQ^*)}
   = \frac{\as N}{2\pi}\Re \varepsilon_n(\CQ)
\lab{en-re}
\ee
in accordance with \re{real}. We can find possible values of
$\CQ_2$ and $\bar\CQ_2$ in \re{tri}
using the interpretation \re{cas} of the corresponding operators as the Casimir
operators of the $SL(2,\IC)$ group. Relation \re{equ} implies that the
eigenvalues $\CQ_2$ and $\bar\CQ_2$ are the same for the Reggeon hamiltonian
and for the hamiltonian \re{K}.
In the model \re{K}, the operators
$\CQ_2^{(s=-1)}$ and $\bar\CQ_2^{(s=0)}$ act on the total quantum space
$t^{0,2}\otimes ...\otimes t^{0,2}$ which can be decomposed into the
direct sum of irreducible principal series representations,
$\oplus_{\nu,m} t^{\nu,2m}$, of the $SL(2,\IC)$. Then, for each
$t^{\nu,2m}$ the Casimir operators $\CQ_2$ and $\bar\CQ_2$ have
the following eigenvalues
\be
\CQ_2=-h(h-1)\,,\qquad
\bar\CQ_2=-\bar h(\bar h -1)\,,
\qquad
h=\frac{1+m}{2}-i\nu\,,\qquad
\bar h=\frac{1-m}{2} - i\nu
\lab{m-nu}
\ee
with integer $m$ and real $\nu$. One checks that these expressions
are in accordance with \re{tri}.

Invariance of \re{ort} and \re{comp} under the
conformal transformations \re{con} implies the following
transformation properties of the Reggeon wave function,
$$
\chi_{n,\{\alpha\}}(\{z_i,\bar z_i\};z_0, \bar z_0)
\to
\chi_{n,\{\alpha\}}'(\{z_i',\bar z_i'\};z_0', \bar z_0')
=  (c z_0 + d)^{2h} (\bar c \bar z_0 + \bar d)^{2\bar h}
\chi_{n,\{\alpha\}}(\{z_i,\bar z_i\};z_0, \bar z_0)
$$
with conformal weights $h$ and $\bar h$ defined in \re{m-nu}.
This relation can be rewritten in terms of generators of the conformal group
as follows
$$
\lr{\sum_{k=1}^n \partial_k + \partial_0}\chi_{n,\{\alpha\}}
=\lr{\sum_{k=1}^n z_k\partial_k + z_0\partial_0+h}\chi_{n,\{\alpha\}}
=\lr{\sum_{k=1}^n z_k^2\partial_k + z_0^2\partial_0+2hz_0}\chi_{n,\{\alpha\}}
=0
$$
and analogously for the antiholomorphic generators. Here, the first
equation means that $\chi_{n,\{\alpha\}}(\{z_k,\bar z_k\};z_0, \bar z_0)$
depends on the distance $z_{k0}=z_k-z_0$ to the center of mass, and the
two remaining equations can be represented as
\be
S^3\ \chi_{n,\{\alpha\}}(\{z_i,\bar z_i\}; 0, 0)
 = - h \chi_{n,\{\alpha\}}(\{z_i,\bar z_i\}; 0, 0)\,,\qqquad
S^+\ \chi_{n,\{\alpha\}}(\{z_i,\bar z_i\}; 0, 0) = 0
\lab{const}
\ee
with $S^a$ the total spin for $s=0$. One can show \ci{prog}
that the wave functions of the Reggeon compound states, \re{chi}
and \re{Bax-s}, satisfy \re{const}.

Using relations \re{ort} and \re{comp} we can give a meaning
to the decomposition of the transition amplitude \re{exp} over
the Reggeon states. Comparing \re{comp} with analogous decomposition,
$I=\sum_{\{\alpha\}} \ket{\chi} \bra{\chi}$, used in \re{exp}, we
obtain,
$$
\ket{\chi} = \chi_{n,\{\CQ\}}(\{z_k,\bar z_k\};z_0,\bar z_0)  \,,
\qqqquad
\bra{\chi} = 
\lr{\chi_{n,\{\CQ\}}(\{z_k,\bar z_k\};z_0,\bar z_0)}^*
\lr{z_{12}\bar z_{12} ... z_{n1} \bar z_{n1}}^{-1}
\,.
$$
Let us consider now the state $\bra{\chi}T_n^{(0)}$ which enters into
\re{exp} and \re{beta}. We recall
that operator $T_n^{(0)}$ describes the free propagation of $n$ Reggeons,
eq.\re{T0}, and therefore the state
$\bra{\chi}T_n^{(0)}$ can be represented as
$$
\bra{\chi}T_n^{(0)}= \bra{\chi}\lr{P_1\bar P_1 ... P_n\bar P_n}^{-1}=
\lr{\left(\CQ_n \bar\CQ_n\right)^{-1}\chi_{n,\{\CQ\}}}^*
$$
with operators $\CQ_n$ defined in \re{Qk}. Since
$\chi_{n,\{\CQ\}}$ diagonalize $\CQ_n$ and $\bar\CQ_n$,
the operators can be replaced by their corresponding eigenvalues,
$\CQ_n$ and $\CQ_n^*$, respectively.
After substitution of the last two relations into \re{exp} we find
transition operator as
\be
T_n(\{z,\bar z\},\{z',\bar z'\};\omega)=
\sum_{\{\CQ\}}
\frac1{\CQ_n \CQ_n^*}
\int dz_0 d\bar z_0\
\frac{
\chi_{n,\{\CQ\}}(\{z,\bar z\};z_0,\bar z_0)
\lr{\chi_{n,\{\CQ\}}(\{z',\bar z'\};z_0,\bar z_0)}^*
}
{\omega-E_{n,\{\CQ\}}}
\,.
\lab{TN}
\ee
Here, the sum goes over all possible compound $n-$Reggeon states
propagating in the $t-$channel. Integration is performed
over the position of the center of mass $b_0=(z_0,\bar z_0)$ of the states
and after Fourier transformation $\int d^2 b_0 \exp(i b_0\cdot q_0)$
it can be replaced by integration over the total transverse momentum
$q_0$ carried by the Reggeons in the $t-$channel. Momentum conservation
fixes $q_0$ to be the transferred momentum $q$ in the scalar products
entering \re{Tn} and \re{beta}. As a result, for the residue factors
\re{beta} we obtain the following expressions
\be
\beta_{\rm A}^{\{\CQ\}}(q^2) = \int\!\! dz_0 d\bar z_0\ \e^{ib_0\cdot q}\!\!
\int\!\! dz d\bar z\
\Phi_{\rm A}(\{z,\bar z\}) \chi_{n,\{\CQ\}}(\{z,\bar z\};z_0,\bar z_0)
\,,\quad
\beta_{\rm B}^{\{\CQ\}}(q^2)=
\frac1{\CQ_n \CQ_n^*}\beta_{\rm A}^*(q^2)\big |_{{}_{\rm A\to B}}
\,.
\lab{betas}
\ee
We conclude from \re{real} and \re{betas} that, in accordance with
our expectations, the energy of the Reggeon states and the
product of the residue factors are real in \re{An-res}.

\section{Solving the Baxter equation}
\setcounter{equation} 0

To find the explicit expressions for the spectrum of the Reggeon hamiltonian,
\re{en-re} and \re{Bax-e}, we have to solve the Baxter equation \re{Bax} for
an arbitrary $n$ and for fixed set of the quantum numbers $\{\CQ_k\}$.
The next step will be the identification of the values of $h$ and $\{\CQ_k\}$
corresponding
to maximum eigenvalue $E_n^{\rm max}$, which determines
the dominant contribution to the hadronic scattering amplitudes
\re{An-res} and \re{Fn-res}. To understand better this procedure we first
consider the simplest case $n=2$.

\subsection{Special case: $n=2$}

For $n=2$ the Reggeon hamiltonian \re{near}
is equal to $2(H_{12}+\bar H_{12})$
with the two-particle XXX spin $s=0$ hamiltonians $H_{12}$ and
$\bar H_{12}$ given by \re{s=0}. They depend on the operators
$J_{12}$ and $\bar J_{12}$, which are related to the Casimir operators
of the conformal group as $J_{12}(J_{12}+1)=-\CQ_2=
-z_{12}^2\partial_1\partial_2$ and
$\bar J_{12}(\bar J_{12}+1)=-\bar \CQ_2=-\bar z_{12}^2
\bar\partial_1\bar\partial_2$.
Possible eigenvalues of these operators were found in \re{m-nu}
and the corresponding eigenstates define the wave function of the
compound state of $n=2$ Reggeons, the BFKL Pomeron \ci{bfkl},
\be
\chi_{m,\nu}(\{z,\bar z\};z_0,\bar z_0)
=N_{m,\nu} \lr{\frac{z_{12}}{z_{10}z_{20}}}^{\frac{1+m}2-i\nu}
\lr{\frac{\bar z_{12}}{\bar z_{10}\bar z_{20}}}^{\frac{1-m}2-i\nu}
\,,
\lab{wave}
\ee
where the constant $N_{m,\nu}=\frac1{4\pi^4}
\lr{\nu^2+m^2/4}$ fixes normalization of the state.
To find the energy corresponding to this state one can use either
the definition \re{s=0}, or apply the relations \re{E}, \re{one}, \re{one-1}
and \re{one-2},
\ba
E_2(m,\nu)=\frac{\as N}{\pi}
2 \Re\left[\psi(1)-\psi\left(\frac{1+|m|}{2}+i\nu\right)\right]
\,.
\lab{energy}
\ea
Thus, the energy of $n=2$ Reggeon state is finite and real.
The maximum value of the energy
\be
E_2^{\rm max}=E_2(0,0)=\frac{\as N}{\pi} 4\log 2
\lab{e2max}
\ee
corresponds to $m=\nu=0$, or equivalently to $h=\bar h=\frac12$.
One notice that the expression \re{energy} is
invariant under replacement $h\to h^*$ and $h\to \bar h$,
and the maximum value of $E_2$ corresponds to the values of $h$
and $\bar h$ which are not changed under this transformation.
This simple observation will be generalized in Sect.~6.3
for the spectrum of the Reggeon hamiltonian with $n>2$.

\subsection{Properties of the Baxter equation}

After one introduces the notation \re{Bax-e} for the function
$\widetilde Q_n(\lambda)=\lambda^n Q_n(\lambda)$,
the Baxter equation \re{Bax} takes
a form of the discrete one-dimensional Schrodinger equation
with a singular potential
\be
\widetilde Q_n(\lambda+i)+\widetilde Q_n(\lambda-i) - 2\widetilde Q_n(\lambda)
= \lr{
-\frac{h(h-1)}{\lambda^2}
+ \frac{\CQ_3}{\lambda^3} + ... + \frac{\CQ_n}{\lambda^n}}
\widetilde Q_n(\lambda)
\lab{f-Q}
\ee
where we replaced $\CQ_2$ by the expression \re{m-nu} in terms of the
conformal weight $h$. We notice that \re{f-Q} is invariant under the
replacement $h\to 1-h$ and we may restrict the values of the conformal
weights to be
\be
\Re h \ge 1/2
\lab{h-fun}
\ee
or equivalently $m\ge 0$ in \re{m-nu}.
Taking a naive continuum limit in \re{f-Q}, $\lambda\to \infty$,
we find that for finite $h$ and $\{\CQ\}$ the equation has two
independent solutions corresponding to the different asymptotic behavior
of the solutions
\be
\lambda^nQ_n(\lambda) \stackrel{\lambda\to\infty}{\longrightarrow}
C_1 \lambda^{h} + C_2 \lambda^{1-h}
\lab{as}
\ee
with $C_1$ and $C_2$ arbitrary constants. One of the solutions grows at
infinity while the second one vanishes. To make a choice between them,
we compare \re{as} with the expression \re{Q-pol} for $Q_n(\lambda)$ given
by the algebraic Bethe Ansatz. We find that $C_2=0$ for integer conformal
weights $h \ge n$. This suggests to impose the following asymptotic behavior
on the solutions of the Baxter equation \re{Bax},%
\footnote{This condition fixes the ambiguity of \re{f-Q} those solutions
          can be defined up to factors $\exp(\pm 2\pi \lambda k)$
          for $k=\IZ_+$ which
          lead to the exponential growth of $Q_n(\lambda)$ at infinity.}
\be
Q_n(\lambda)\stackrel{\lambda\to\infty}{\longrightarrow}
 \lambda^{h-n}
\lab{Q-as}
\ee
with the conformal weight $h$ in the fundamental domain \re{h-fun}.

Suppose we know the solution of the Baxter equation,
$Q_n=Q_n(\lambda;h,\{\CQ\})$,
which obeys \re{Q-as}.
Then, replacing $h\to 1-h$ or
$\lambda\to -\lambda$ and $\CQ_k\to (-)^k \CQ_k$ in \re{f-Q} we find that
equation takes the original form, which means that up to inessential
factor
\be
Q_n(\lambda;h,\{\CQ_k\})=Q_n(\lambda;1-h,\{\CQ_k\})=
Q_n(-\lambda;h,\{(-)^k\CQ_k\})
\,.
\lab{Q-sym}
\ee
We substitute this identity into expressions for the energy and for
integrals of motion, \re{Bax-e}, and find that
\be
\varepsilon_n(h,\{\CQ_k\})=\varepsilon_n(1-h,\{\CQ_k\})=
\varepsilon_n(h,\{(-)^k\CQ_k\})
\lab{1-h}
\ee
and
$$
\CI_k(h,\{\CQ_k\})=\CI_k(1-h,\{\CQ_k\})=(-)^k \CI_k(h,\{(-)^k\CQ_k\})
\,.
$$
The last part of these relations can be understood from the properties of
the model under reparameterization of the Reggeon coordinates,
$z_k\to z_{n-k+1}$,
with $k=1...n$. We check that the Reggeon hamiltonian \re{near}
is invariant under this transformation while the operators $\CQ_k$, defined
in \re{Qk}, change sign for odd $k$.

Let us consider the special values of the quantum numbers,
$\CQ_k=\CQ_{k+1}=...=\CQ_n=0$ with $n$ the number of Reggeons and $k\ge 3$.
Then, the solution of the Baxter equation \re{Bax}
for $n$ Reggeons can be expressed as
\be
Q_n(\lambda;h,\CQ_3,...,\CQ_{k-1},0,...,0)=\lambda^{-n+k-1}
Q_{k-1}(\lambda;h,\CQ_3,...,\CQ_{k-1})
\lab{deg}
\ee
where $Q_{k-1}(\lambda;h,\CQ_3,...,\CQ_{k-1})$ is the solution of the
Baxter equation \re{Bax} for $k-1$ Reggeons with the same quantum
numbers $h$ and $\CQ_3,$ $...$ , $\CQ_{k-1}$. To understand
this identity we substitute it into \re{Bax-e} and find the relation
between the eigenvalues of the holomorphic hamiltonians for
different number of Reggeons
\be
\varepsilon_n(h,\CQ_3,...,\CQ_{k-1},0,...,0) =
\varepsilon_{k-1}(h,\CQ_3,...,\CQ_{k-1})
\,.
\lab{en=en}
\ee
As it follows from the definition, \re{Qk}, the operators $\CQ_k$ are given
by a sum of differential operators and each of them acts only on
$k$ holomorphic variables from the total set of $n$ Reggeon coordinates
$z_1,...,z_n$. Therefore, to satisfy  $\CQ_k=\CQ_{k+1}=...=\CQ_n=0$
one can choose the eigenstate to be a function of only $k-1$ holomorphic
Reggeon coordinates, that is to be the holomorphic part of the compound
state of $k-1$ Reggeons. Using \re{TN} we find that these states
give infinite contributions to the transition operator
and therefore should be excluded from the spectrum
of $n-$Reggeon hamiltonian. This means, that solving the Baxter equation,
\re{Bax} and \re{f-Q}, we should satisfy condition
\be
\CQ_n\neq 0\,.
\lab{neq0}
\ee
In particular, for $n=2$ this condition together with \re{m-nu}
and \re{h-fun} implies that $h \neq 1$ and $h\neq 0$.

\subsection{Maximum eigenvalue of the Reggeon hamiltonian}

Let us consider expression \re{en-re} for the energy of the $n-$Reggeon
compound state as a function of possible sets of the quantum numbers
$E_n=E_n(\CQ)$. Then, it follows from \re{en-re} and \re{1-h},
that this function has
the following properties
$$
E_n(\CQ)=E_n(\CQ^*)=E_n(-\CQ)
\,,
$$
where $\CQ=(\CQ_2,\CQ_3,...,\CQ_n)$ and
$-\CQ=(\CQ_2,-\CQ_3,...,(-)^n\CQ_n)$ and where
complex conjugation acts in $\CQ^*$ on all elements of the set.
This means that the spectrum of the Reggeon hamiltonian is
{\it degenerate\/} with respect to quantum numbers
$\{\CQ_k\}$. Hence, if we assume from the very beginning that
the eigenstate of the Reggeon hamiltonian corresponding to maximum
eigenvalue is {\it not degenerate\/}, then the values of the corresponding
quantum numbers can be fixed as%
\footnote{Changing the sign of the Reggeon hamiltonian we can reformulate
this condition as uniqueness of the ground state of the model.}
\be
\CQ_{2k}\bigg|_{\rm max}=\CQ_{2k}^*\bigg|_{\rm max}
=\bar\CQ_{2k}\bigg|_{\rm max}=\bar\CQ_{2k}^*\bigg|_{\rm max}\,,\qquad
\CQ_{2k+1}\bigg|_{\rm max}=\bar\CQ_{2k+1}\bigg|_{\rm max}=0\,.
\lab{maxQn}
\ee
In particular, for quantum numbers $\CQ_2$ and $\bar\CQ_2$
defined in \re{m-nu} this relation leads to
\be
\CQ_2\bigg|_{\rm max}=\bar\CQ_2\bigg|_{\rm max}=\frac14
\,,\qqqquad \mbox{for} \quad h=\bar h=1/2
\,,
\lab{maxQ2}
\ee
which indeed corresponds to the maximal energy \re{e2max} in the
special case $n=2$.

\subsection{Ansatz for the Baxter equation}

To solve the Baxter equation \re{Bax} we choose the following ansatz
\ci{prog}
\be
Q_n(\lambda)=\int_{C_z}
\frac{dz}{2\pi i}\ z^{-i\lambda-1} (z-1)^{i\lambda-1} Q_n(z)
\lab{Cz}
\ee
where the function $Q_n(z)$ is integrated along the closed path $C_z$ in the
complex $z-$plane. Integrating by parts $k$ times we get the
equivalent representation
\be
Q_n(\lambda)=\lambda^{-k}
\int_{C_z} \frac{dz}{2\pi i}\ (z-1)^{i\lambda-1}z^{-i\lambda-1}
\lr{-iz(1-z)\frac{d}{dz}}^k Q_n(z)
\lab{part}
\ee
which allows us to replace the original functional equation \re{Bax}
by the $n-$th order differential equation on the function $Q_n(z)$
\be
\left[ \lr{z(1-z)\frac{d}{dz}}^n
+z(1-z)\sum_{k=0}^{n-2}i^{n-k}\CQ_{n-k}\lr{z(1-z)\frac{d}{dz}}^k
\right]Q_n(z)=0
\lab{diff}
\ee
with $\CQ_2=-h(h-1)$. It is important to notice that the integration by parts
in \re{part} leads to the integrals like $\int_{C_z}
dz\frac{d}{dz} (...)$ where
$(...)$ denotes the product of $Q_n(z)$ and $z^{-i\lambda-1}
(z-1)^{i\lambda-1}$ or their derivatives with respect to $z$. The choice
of the integration contour $C_z$ is fixed by the condition that these
additional terms can be neglected, or equivalently that after encircling the
path $C_z$ the function $(...)$ comes to the initial point with the same
value. To this end we notice that $z^{-i\lambda-1} (z-1)^{i\lambda-1}$
has a nontrivial monodromy at $z=0$ and $z=1$ and the same
points are the singular points for the differential equation \re{diff}.

The asymptotics \re{Q-as} of the solution of the Baxter equation
corresponds to the behavior of the function $Q_n(z)$ at infinity
\be
Q_n(z) \stackrel{z\to \infty}{\longrightarrow} z^{h-n+1}
\,,
\lab{add}
\ee
which may be considered as a possible
additional condition on the solutions of
the differential equation \re{diff}.

\subsection{Solution for $n=2$}

For $n=2$ the differential equation \re{diff} has a form of the
hypergeometric equation
\be
\lr{\frac{d}{dz}z(1-z)\frac{d}{dz}+h(h-1)}Q_2(z)=0
\lab{Leg}
\ee
and its general solution is known as \ci{wat}
\be
Q_2(z)={\rm const.}
\int_{C_w} \frac{dw}{2\pi i} \ w^{h-1} (w-1)^{h-1} (w-z)^{-h}
\,,
\lab{Cw}
\ee
where the integration path $C_w$ is fixed by the condition that the integrand
should be a single valued function on $C_w$, that is after encircling
around $C_w$ it should have the same value.
We substitute \re{Cw} into \re{Cz} and find the general solution of the
Baxter equation for $n=2$ as
\be
Q_2(\lambda)={\rm const.}
             \int_{C_z}\frac{dz}{2\pi i}\ z^{-i\lambda-1}(z-1)^{i\lambda-1}
             \int_{C_w} \frac{dw}{2\pi i} \ w^{h-1} (w-1)^{h-1} (w-z)^{-h}
\,.
\lab{CC}
\ee
This expression depends on the integration paths $C_z$ and $C_w$.
Their choice depends on the values of $\lambda$ and $h$. To show this
let us consider the special case of integer positive conformal weight $h$.
For $h=\IZ_+$, the integrand in \re{CC} has a pole at $w=z$ and
for the integral over $w$ do not vanish the contour $C_w$ in \re{Cw} and
\re{CC} should encircle the point $w=z$.
The result of taking residue at $w=z$ leads to the well-known
Rodrigues's formula for the Legendre polynomials%
\footnote{We should notice that there is a second solution of \re{Leg},
known as the Legendre function of the second kind \ci{wat,hyper}. However,
it does not satisfy the additional condition \re{add} and leads to the
solution of the Baxter equation with asymptotic behavior \re{as} for $C_1=0$.}
$$
Q_2(z)=P_{h-1}(1-2z)=\frac1{(h-1)!}\frac{d^{h-1}}{dz^{h-1}}\lr{z(1-z)}^{h-1}
=F(h,1-h;1;z)\,.
$$
The integral over $z$ in \re{CC} has two branch points
at $z=0$ and $z=1$. For the integrand to be single valued, the contour
$C_z$ should encircle both branch points in the same direction.
Then, we deform the integration path to be close to the segment $[0,1]$
and get the solution of the Baxter equation for integer $h$ as
\be
Q_2(\lambda;h)=\frac{i^{h+1}}{\pi}
\sinh(\pi\lambda)\int_0^1 dz\ z^{-i\lambda-1}(1-z)^{i\lambda-1}
P_{h-1}(1-2z)
\lab{Q-P}
\ee
where the constant has been choosen for the solution to be real for $h=\IZ_+$.
The Legendre polynomials $P_k(x)$ for $k=0,1, ...$ form a system of
orthogonal polynomials on the interval $-1 \leq x \leq 1$ with the weight $1.$
They have the following properties \ci{hyper}
\be
P_{k}(x)=P_{-k-1}(x)\,,\qquad P_k(-x)=(-1)^{k} P_k(x)
\lab{prop0}
\ee
and
\be
z(1-z)\frac{d}{dz}\ P_k(1-2z)=\frac{k(k+1)}{2(2k+1)}
\bigg[{P_{k+1}(1-2z)-P_{k-1}(1-2z)}\bigg]
\,.
\lab{prop}
\ee
Using \re{prop0} we verify that
\be
Q_2(\lambda;h)=i^{1-2h} Q_2(\lambda;1-h)\,,\qquad
Q_2(-\lambda;h)=(-1)^h Q_2(\lambda;h)\,,\qquad
Q_2^*(\lambda;h)=Q_2(\lambda;h)\,.
\lab{sig}
\ee
Applying identity \re{prop} together with \re{part}
we find that $Q_2(\lambda;h)$ satisfies the recurrence relation
\be
\lambda Q_2(\lambda;h)=-
\frac{h(h-1)}{2(2h-1)}\bigg[{Q_2(\lambda;h+1)+Q_2(\lambda;h-1)}\bigg]
\,.
\lab{rec}
\ee
Using expansion of the Legendre polynomial $P_{h-1}(2z-1)$ in powers of $z$
we obtain the representation for the solution of the Baxter equation for
integer conformal weight $h$ as \ci{prog}
\be
Q_2(\lambda;h)=i^h\sum_{k=1}^{\infty}(-)^k
\frac{k}{(k!)^3}
\frac{\Gamma(h+k)}{\Gamma(h-k)}
\frac{\Gamma(k-i\lambda)}{\Gamma(1-i\lambda)}
=i^h h(1-h)\
{_3F_2}\left({1+h,2-h,1-i\lambda \atop 2, 2}; 1
\right)\,.
\lab{3F2}
\ee
For lowest values of the conformal weight $h$ the explicit
expressions for $Q_2(\lambda;h)$ are
\be
Q_2(\lambda;2)= 2 \,,\qquad
Q_2(\lambda;3)=-6 \lambda\,,\qquad
Q_2(\lambda;4)=-2 + 10\lambda^2 \,,\qquad
Q_2(\lambda;5)=\frac{25}{3}\lambda-\frac{35}{3}\lambda^3\,.
\lab{ex}
\ee
As it was expected from the algebraic Bethe Ansatz, \re{Q-pol}, they are
polynomials of degree $h-2$ in $\lambda$ and one checks that the roots of
$Q_2(\lambda;h)$ do satisfy the Bethe equation \re{Beq}.

\subsubsection{Relation to orthogonal polynomials}

The remarkable property of the expression \re{3F2} is that being taken in
the form of $_3F_2$ hypergeometric infinite series for arbitrary complex
$h$ and $\lambda$, the expression for $Q_2(\lambda;h)$ becomes a polynomial
of finite degree either for $h=m$ or for $\lambda=im$ with $m$ integer.
This suggests that there is a deep relation between solutions of the
Baxter equation for $n=2$ (and higher $n$) and some systems of orthogonal
polynomials.
The latter relation was explicitly specified by V.~Spiridonov who has
observed that
solutions of the Baxter equation for $n=2$
are tied to the Hahn class of orthogonal polynomials.

Indeed, if one sets $\lambda=-i(m+1)$ with $m=0,1,...$ and $h=ix+1/2$,
then the Baxter equation \re{Bax} takes the form
of three term recurrence relation which defines the continuous dual Hahn
polynomials of the argument $x^2$ (see, e.g. \ci{aw}),
$$
S_m(x^2; a, b, c) =
(a+b)_m (a+c)_m\; {_3F_2}\left({-m, a+ix, a-ix \atop a+b, a+c}; 1
\right), \qquad a=3/2, \; b=c=1/2\,.
$$
These polynomials are orthogonal on the interval $0\le x < \infty$
with respect to a positive measure and they coincide up to an inessential
factor with the solution $Q_2(-i(m+1);ix+1/2)$.

Due to the well known property of duality of classical orthogonal polynomials
the same solution of the Baxter equation determines
different system of orthogonal polynomials for which $\lambda$ becomes
continuous variable while $h$ plays a role of integer index. Indeed,
putting $h=m+2$ with $m=0,1,...$ and $\lambda=x$ in \re{rec}
we find that \re{rec} takes the form of three term recurrence relation
for continuous symmetric Hahn polynomials \ci{aw},
$$
P_m(x; a,b)=i^m{_3F_2}\left({-m, m+2a+2b-1, a-ix \atop a+b, 2a }; 1
\right), \qquad a=b=1\,,
$$
which are orthogonal on the interval $-\infty < x < \infty$ and which
are proportional to $Q_2(x; m+2)$.

Identification of $Q_2(\lambda; m+2)$ as Hahn polynomials
implies, first, that the solutions,
$\{\lambda\}=(\lambda_1,...\;,\lambda_{k-2})$,
of the Bethe equation \re{Beq} and \re{Q-pol} have the properties of the roots
of orthogonal polynomials \ci{hyper}. Namely, all roots $\{\lambda\}$ are
real and simple, between two consecutive zeros of $Q_2(\lambda;k)$
there is exactly one zero of $Q_2(\lambda;k+1)$ and at least one zero
of $Q_2(\lambda;k+m)$ for each $m>0$.

The relation between orthogonal polynomials and
solutions of the Baxter equation is a general
property of the Baxter equation for XXX magnet of an arbitrary spin,
including the case of compact $SU(2)$ spins. In particular, some solutions
of the Baxter equation \re{Bax} for $n=3$ can be identified with the
Wilson polynomials. The detailed consideration of these and
other analytical solutions of the Baxter equation for generalized
Heisenberg magnet models on the basis of methods developed in \ci{sz}
will be given elsewhere \ci{ksz}.

\subsubsection{Eigenvalues for $n=2$}

Let us use the solution \re{3F2} to obtain the spectrum of the Reggeon
hamiltonian for $n=2$ which has been already defined in sect.~6.1.
We start with the expression for the energy, \re{Bax-e}, and evaluate
$Q_2(\lambda;h)$ and its derivative at $\lambda=-i$ while the same
quantities at $\lambda=i$ can be found from \re{sig}.
We use \re{3F2} to get for $h=\IZ_+$
\be
Q_2(-i;h)=i^h\sum_{k=1}^{h-1} (-1)^k \frac{k}{(k!)^3}
\frac{\Gamma(h+k)}{\Gamma(h-k)}
\frac{\Gamma(k-1)}{\Gamma(0)}
=i^h h(1-h)
\lab{1}
\ee
since only one term with $k=1$ survives in total sum. For the derivative
we obtain
$$
-i\frac{d}{d\lambda} Q_2(-\lambda-i;h)\Bigg|_{\lambda=0}
=i^h\sum_{k=2}^{h-1}(-)^k \frac{\Gamma(h+k)}{(k-1)\Gamma(h-k)\Gamma^2(k+1)}
\,.
$$
To evaluate this sum we rewrite it as a contour integral in the complex
$j-$plane
$$
{-i}\frac{d}{d\lambda} Q_2(-\lambda-i;h)\Bigg|_{\lambda=0}=
i^h\int\frac{dj}{2\pi i}\ \frac{\pi}{\sin(\pi j)}
\frac{\Gamma(h+j)}{(j-1)\Gamma(h-j)\Gamma^2(j+1)}
$$
where integration contour encircles half of the real axis right to $j=2$.
The integral is defined by singularities of the integrand in the complex
$j-$plane. A simple analysis shows that the integrand has poles at
$j=0,1,2,...,h-1$ while singularities at $j=-1,-2,\ldots$
are compensated by $\Gamma^2(j+1)$ in the denumerator. Since the points
$j=2,3,...,h-1$ are inside the integration contour, the integral is
given by the residue at $j=0$ and $j=1$,
\be
{-i}\frac{d}{d\lambda} Q_2(-\lambda-i;h)\Bigg|_{\lambda=0}
=-2i^h h(1-h)\left[\psi(h)-\psi(1)-1\right]
\lab{2}
\ee
with $h$ positive integer conformal weight. Now, we substitute \re{3F2}
into \re{Bax-e}, take into account \re{1}, \re{2} together with
\re{sig} and obtain the following expression for the energy of the
holomorphic Reggeon hamiltonian
\be
\varepsilon_2(h)=-4\left[\psi(h)-\psi(1)\right]\,,\qquad \mbox{for $h=\IZ_+$}
\,.
\lab{fin}
\ee
Expression \re{fin} is valid for positive integer conformal weight $h$ and
in order to find the energy for negative integer $h$ we have to use
the symmetry \re{1-h}. In particular, the conformal weight $\bar h$
corresponding to the antiholomorphic hamiltonian can be found from \re{m-nu}
as $\bar h = 1-h^*$ and for positive $h$ it becomes negative or zero. In
general, for an arbitrary complex $h$ the energy of the $n=2$ Reggeon state
is given by \re{en-re}. We replace $\varepsilon_2(h)$ in \re{en-re}
by its expression \re{fin} for integer $h$ and analytically continue the
result to complex values of the conformal weight, \re{m-nu},
$$
E_2(h) = \frac{\as N}{2\pi} \Re \varepsilon_2(h)
       = 2 \frac{\as N}{\pi}\Re \left[\psi(1)-\psi(h)\right]
\,.
$$
This expression coincides with the energy \re{energy}
of the Reggeon hamiltonian at $n=2$.

\subsubsection{Eigenstates for $n=2$}

For $n=2$ we obtain from \re{Bax-s} the expression for the holomorphic
wave function of $n=2$ Reggeon state as
\be
\varphi_h(z_1,z_2;z_0)=-z_{12}^2\ (iS^-)^{h-2}\ Q_2(x_1;h)\
\frac1{z_{10}^2z_{20}^2}
\,,
\lab{phi2}
\ee
where the operator $x_1$ was defined in \re{x1} and
$S^-=-(\partial_1+\partial_2)$.
Substituting \re{ex} into \re{phi2}, we find that
$$
\varphi_{2}=-2 \lr{\frac{z_{12}}{z_{10}z_{20}}}^2\,,\qquad
\varphi_{3}= 2\cdot 3!\lr{\frac{z_{12}}{z_{10}z_{20}}}^3\,,\quad
\varphi_{4}=-3\cdot 4!\lr{\frac{z_{12}}{z_{10}z_{20}}}^4\,,\quad
\varphi_{5}= 4\cdot 5! \lr{\frac{z_{12}}{z_{10}z_{20}}}^5\,.
$$
To find $\varphi_h$ for an arbitrary $h$ we introduce the notation for the
holomorphic part of the wave function \re{one}
\be
\phi_h = - z_{12}^{-2}\ \varphi_h = (iS^-)^{h-2}\ Q_2(x_1;h)\
\frac1{z_{10}^2z_{20}^2}
\lab{phih}
\ee
and use the property \re{rec} of $Q_2(x_1;h)$ together with \re{phi2}
to transform it into the recurrence relation
$$
\partial_1\partial_2 z_{12} \phi_h =-
\frac{h(h-1)}{2(2h-1)}\lr{\phi_{h+1}- (\partial_1+\partial_2)^2 \phi_{h-1}}
\,.
$$
One can check that the solution of this equation is
$$
\phi_h =(-)^{h}(h-1)N_h \frac{z_{12}^{h-2}}{(z_{10}z_{20})^h}
\,,
$$
where $N_h=hN_{h-1}$. Substituting $\phi_h$ into \re{phih} we obtain the
holomorphic wave function of the $n=2$ Reggeon state in the form \re{wave}.
Thus, for $n=2$ the solution of the Baxter
equation \re{3F2} allows us to reproduce the spectrum of the Reggeon
hamiltonian.

\subsection{Solution for $n\ge 3$}

Analyzing the differential equation \re{diff} for $n\geq 3$, we follow
the same procedure as for $n=2$. Namely, we start with integer positive
conformal weight $h$ and try to find solution of \re{diff} and \re{Cz}
which can be
analytically continued for complex values of $h$ given in \re{m-nu}.
However, for $n\geq 3$ the solution of \re{diff} depends on additional
quantum numbers $\CQ_3,$ $...$, $\CQ_n$ which makes analysis more
complicated.

Let us introduce notation for the derivative
$$
\widetilde Q_n(z)=\lr{-iz(1-z)\frac{d}{dz}}^{n-2}Q_n(z)\,,
\,.
$$
Applying operator $\lr{z(1-z)\frac{d}{dz}}^{n-2}(z(1-z))^{-1}$
to the both sides of \re{diff} we find the differential equation for
$\widetilde Q_n(z)$ as
\be
\left[
\lr{z(1-z)\frac{d}{dz}}^{n-2}\frac{d}{dz}z(1-z)\frac{d}{dz}
+\sum_{k=0}^{n-2} i^{n-k}\CQ_{n-k}\lr{z(1-z)\frac{d}{dz}}^k
\right]
\widetilde Q_n(z)=0
\,.
\lab{fn3}
\ee
Comparison of this equation with \re{Leg} and the property \re{prop}
of the Legendre polynomials suggest us to look for the solution of \re{fn3}
as a linear combination of the Legendre polynomials
\be
\widetilde Q_n(z)=\sum_{k=1}^\infty i^k\ P_{k-1}(1-2z)\ C_k
\,,
\lab{fP}
\ee
where summation is performed over positive integer $k$
and the coefficients $C_k=C_k(h,\{\CQ\})$ have to be found from \re{fn3}.
Once we know these coefficients, we use \re{Cz} and \re{diff}
to find the solution of the Baxter equation as
\be
Q_n(\lambda;h,\{\CQ\}) = \lambda^{2-n} \sum_{k=1}^\infty Q_2(\lambda;k)\
C_k(h,\{\CQ\})
\lab{bax3}
\ee
with $Q_2(\lambda;k)$
the solution of the Baxter equation for $n=2$ defined in \re{3F2}.
Substituting this expansion into \re{Bax-e}, we find expression for
the energy of the holomorphic hamiltonian in terms of the coefficients
$C_k$
\be
\varepsilon_n(h,\{\CQ\})=-4\sum_{k=1}^\infty \left[
\psi(k)-\psi(1)\right]\ \hat C_k(h,\{\CQ\})
\lab{en3}
\ee
with normalized coefficients $\hat C_k$ defined as
\be
\hat C_k= \Re
\frac{i^kk(k-1)C_k}{\sum_{m\geq 1}i^m m(m-1) C_m } \,,\qqqquad
\sum_{k=1}^\infty \hat C_k=1
\,.
\lab{en31}
\ee
For $n=2$ we have $\hat C_k=\delta_{k,h}$ and
only one term with $k=h$ survives in the sum \re{bax3} and \re{en3}.
For $n\geq 3$ we substitute \re{fn3} into \re{diff}
and equate to zero coefficients in front of Legendre polynomials.
To this end, we notice from \re{prop}
that acting on the Legendre polynomial, $P_k(1-2z)$,
the operator $z(1-z)\frac{d}{dz}$ shifts its index, while operator
$\frac{d}{dz}z(1-z)\frac{d}{dz}$ is diagonal. Hence, in the basis
of the Legendre polynomials, $\ket{k}=P_{k-1}(1-2z)$, both operators
can be represented as infinite dimensional matrices
\ba
&&
U_{m,k}
\equiv
i^{k-m-1}\bra{m} z(1-z)\frac{d}{dz} \ket{k} = -\frac{k(k-1)}{2(2k-1)}
\lr{\delta_{m,k+1}+\delta_{m,k-1}}\,, \nonumber
\\
&&
V_{m,k}
\equiv
-\bra{m}\frac{d}{dz} z(1-z)\frac{d}{dz} +h(h-1)\ket{m}
=(k+h-1)(k-h)\delta_{m,k}\,.
\lab{AB}
\ea
In terms of these operators, the resulting equation for the coefficients
$C_k$ has the following matrix form
\be
{W}(\CQ)=
\lr{
U^{n-2}V+\sum_{k=3}^{n}\CQ_{k}\ U^{n-k}
}
\,,\qqqquad
{W}(\CQ)\ket{ C }=0\,,
\lab{Bc}
\ee
where $\ket{C}$ denotes infinite dimensional vector with components $C_k$
and the term with $\CQ_2$ has been included into the definition \re{AB}
of the operator $V$.

The condition \re{add} leads to the asymptotic behavior of the function
$\widetilde Q_n(z)$
\be
\widetilde Q_n(z)\stackrel{z\to\infty}{\rightarrow} z^{h-1}\,,
\qqquad \mbox{for $h=\IZ_+$}\,.
\lab{fnas}
\ee
Let us look for the solutions of \re{fn3} and \re{fP}
which are {\it finite\/} sums
of the Legendre polynomials. Then, from the asymptotics of the Legendre
polynomials at infinity, $P_n(z) \sim z^{n-1}$, and from
comparison of \re{fnas} and \re{fP} we
find that
\be
C_{h+1}=C_{h+2}=...=0
\lab{c=0}
\ee
and nonzero components of $\ket{C}$ form a $h-$dimensional vector.
It is clear from \re{Bc} that the resulting finite-dimensional vector is the
eigenstate of the $h\times h$ minor of the infinite-dimensional
matrix \re{Bc} corresponding to the zero eigenvalue. This implies that
the determinant of the minor should vanish which leads to the condition
on the quantum numbers $\CQ_k$ entering into \re{Bc}. Moreover, as we will
show below, there are $n-3$ additional conditions on $\CQ_k$ which
follow from the relations \re{c=0} and hence the total number of the
independent constraints turns out to be equal to the number of $\CQ_k$'s.
This means, that for the Baxter equation \re{Bax} to have polynomial
solutions, the quantum numbers $\CQ_k$ have to be quantized.

\subsubsection{Special case: $n=3$}

Let us consider the explicit form of \re{Bc} for $n=3$,
\be
f_{k+1}C_{k+1}(\CQ_3) + f_{k-1} C_{k-1}(\CQ_3) = \CQ_3 C_k(\CQ_3)
\,,
\lab{rec3}
\ee
where the coefficients $f_k$ depend on the conformal weight $h$ and
are given by
\be
f_k=\frac{k(k-1)}{2(2k-1)}(k-h)(k+h-1)\,,\qqqquad f_1=f_h=0
\,.
\lab{fk3}
\ee
The last identity ensures the trancation of the recurrence relations \re{rec3}
at $k=h-1$. Introducing the notation for
${W}_h$ to be $h\times h$
minor of the matrix ${W}$ defined in \re{AB} and
for $\ket{C}_h$ to be $h-$dimensional component of the vector
$\ket{C}$,
\be
{W}_h(\CQ_3) =\left(\begin{array}{ccccc}
    \CQ_3        &      -f_2       &      0      &    \ldots   &      0
\\  -f_1         &      \CQ_3      &    -f_3     &             &
\\    0          &      -f_2       &      \CQ_3  &             &
\\  \vdots       &                 &             &             &  -f_h
\\    0          &                 &             &    -f_{h-1}  &  \CQ_3
          \end{array}
          \right) \,,\qqquad
\ket{C}_h=\left(\begin{array}{c} C_1 \\ C_2 \\ C_3 \\ \vdots \\ C_h \end{array}
    \right)\,,
\lab{mat3}
\ee
we rewrite original matrix equation \re{Bc} as
\be
\det {W}_h(\CQ_3) = 0 \,, \qqquad {W}_h (\CQ_3)\ket{C}_h=0
\,.
\lab{mat31}
\ee
Here, the first equation gives us possible values of the quantum numbers
$\CQ_3$ for positive integer conformal weight $h$, while the second one
determines the coefficients entering into
solution of the Baxter equation \re{bax3} and
energy of the Reggeon hamiltonian, \re{en3} and \re{en31}.
It can be easily shown from \re{mat3} that
\be
\det {W}_h(-\CQ_3)=(-1)^h  \det W_h(\CQ_3)\,.
\lab{BC}
\ee
For even $h$ this equation has $h-2$ nontrivial solutions,
$\CQ_3\neq 0$, and a degenerate solution $\CQ_3=0$, which being
substituted into \re{rec3} leads to $C_k=0$. For odd $h$
it has $h-3$ solutions for $\CQ_3\neq 0$ (see fig.~\ref{fig6})
and a solution $\CQ_3=0$ corresponding to the degenerate states
discussed in Sect.~6.2.
Hence, for fixed $h$ we have $2[h/2-1]$ nontrivial
solutions for $\CQ_3$ which satisfy \re{neq0} and which
determine the solutions \re{bax3} of the Baxter equation, or
equivalently $2[h/2-1]$ eigenstates of the holomorphic
Reggeon hamiltonian. Although it is not difficult to solve \re{rec3} or
\re{mat31} for lowest values of $h=\IZ_+$,
the general analytical formula is not available yet.

Since $\det{W}_h(\CQ_3)$ is a polynomial of degree $h$ in $\CQ_3$
with real coefficients and certain parity \re{BC}, its zeros
may be either real or pure imaginary. It turns out that only the
first possibility is realized
\be
\Im \CQ_3 = 0\,, \qquad \mbox{for $h=\IZ_+$}\,.
\lab{im=0}
\ee
To understand this property we notice that the coefficients
$C_k(\CQ_3)$ satisfy the three--term recurrence relations \re{rec3},
which define a system of generalized orthogonal
polynomials in continuous variable $\CQ_3$.%
\footnote{The recurrence relation similar to \re{rec3} and the interpretation
          of its solutions in terms of orthogonal polynomials have been
          also  proposed in \ci{F}.}
Condition $C_{h+1}(\CQ_3)=0$ is equivalent to \re{mat31} and it implies that
the possible values of $\CQ_3$, being roots of orthogonal polynomial, are
real and simple.
We conclude that each term in the sum \re{bax3}
          which determines the solution of the Baxter equation
          for $n=3$ and $h=\IZ$ has a form of
          the product of two orthogonal polynomials, $Q_2(\lambda;k)$
          in continuous variable $\lambda$ and $C_k(\CQ_3)$ in
          continuous variable $\CQ_3$.

One can find $C_k$ from \re{rec3} and \re{mat31} up to some constant
which can be fixed by putting $C_1=1$. Then, it follows from
\re{rec3} and \re{im=0} that $C_k$ are real and
$C_k(-\CQ_3)=(-)^{k+1} C_k(\CQ_3)$.
As a consequence
the solution of the Baxter equation \re{bax3} and the energy of the
holomorphic hamiltonian \re{en3} are also real for integer conformal
weight $h$ and real $\lambda$.

Here, we give few examples of the solutions of the Baxter equation,
$Q_3(\lambda;h,\CQ_3)$,
for lowest values of the conformal weight $h$ and positive $\CQ_3$,
\baa
&&\hspace{-3mm}
Q_3(\lambda;4,2\sqrt 3)=-6-6\sqrt 3\lambda\,,
\\[1.5mm]
&&\hspace{-3mm}
Q_3(\lambda;5,6\sqrt 3)=-\fra{10}3-10\sqrt 3\lambda -\fra{40}3 \lambda^2\,,
\\[1.5mm]
&&\hspace{-3mm}
Q_3(\lambda;6,4\sqrt{30})=-\fra{5\sqrt{30}}2 \lambda - 30 \lambda^2 -
\fra{5\sqrt{30}}2 \lambda^3\,,
\\[1.5mm]
&&\hspace{-3mm}
Q_3(\lambda;6,4\sqrt 3)=-\fra{15}2-\fra{35\sqrt{3}}{4}\lambda + \fra{15}2
\lambda^2 + \fra{25\sqrt 3}4\lambda^3\,,
\\[1.5mm]
&&\hspace{-3mm}
Q_3(\lambda;7,6\sqrt{23}\pm 6\sqrt 3)=
-\fra{21\mp 7\sqrt{69}}{20}-\fra{21\sqrt{23}\mp 49\sqrt 3}{8}\lambda
-\fra{189\pm 7\sqrt{69}}{8}\lambda^2
-\fra{21\sqrt{23}\pm 119\sqrt 3}{8}\lambda^3
-\fra{63\pm 49\sqrt{69}}{40}\lambda^4
\,.
\eaa
For $h < 4$ the Baxter equation has only degenerate solutions \re{deg}
corresponding to $\CQ_3=0$.
As a nontrivial check of these expressions we verify that, similar to the
case $n=2$, the roots of $Q_3(\lambda;h,\CQ_3)$ satisfy the Bethe equation
\re{Beq} for $n=3$. Moreover, we find that the roots $\{\lambda_k\}$
are real and simple. This should be compared with the situation for spin
$s=1/2$ XXX magnet,
where the solutions of the Bethe equation \re{Beq} form strings in
the complex $\lambda-$plane symmetric with respect to real axis
\ci{BA1,BA2,BA3,BA4}.

Unfortunately, we do not have analytical
expressions for $Q_3(\lambda;h,\CQ_3)$
similar to that for $n=2$, eq.\re{3F2}.
For lowest values of $h$ the results of our calculations
of the solution of the Baxter equation and the energy of the
holomorphic Reggeon hamiltonian are shown in fig.~\ref{fig1}--\ref{fig8}.
For fixed $h=\IZ_+$ the energy $\varepsilon_3(h,\CQ_3)$ turns out to be
a decreasing function of $\CQ_3$ and for $\CQ_3^2\to 0$ it approaches
the value $\varepsilon_2(h)=-4[\psi(h)-\psi(1)]$ in accordance
with \re{en=en}.

\subsubsection{Generalizations for $n\geq 4$}

Let us consider the properties of the equations \re{Bc} and \re{c=0} for
$n\geq 4$. Using the explicit form of the matrices \re{AB} one finds that
for an arbitrary $n$ the coefficients $C_k$ satisfy the $(2n-3)-$term
recurrence relations similar to \re{rec3} which
relate the coefficients $C_{k-n+2},...,C_{k+n-2}$ and involve
the quantum numbers $\CQ_3,...,\CQ_n$. If we again define
${W}_h$ as a $h-$dimensional minor of the matrix
\re{Bc}, then the relations \re{mat31} are still valid.
The first equation in \re{mat31} gives the relation
between $\CQ_3,...,\CQ_n$, while
the second one allows us to find the coefficients $C_k$ as functions
of these quantum numbers. We notice that for $n=3$ it is enough to
satisfy $C_{h+1}=0$ in order to get $C_{h+2}=C_{h+3}=...=0$ from
the three-term recurrence relations \re{rec3}. For $n\geq 4$ the
recurrence relations for $C_k$ involve $(2n-3)$ terms and as a result
equation \re{c=0} becomes equivalent to $C_{h+1}=...=C_{h+n-2}=0$.
These relations together with \re{Bc} lead to the following conditions
\be
{W}^{mk} (\CQ) C_k = 0\,,
\lab{addcon}
\ee
with $k=1,...,h$ and $m=h+1,...,h+n-3$.
Once we know expressions for $\ket{C}_h$ from \re{mat31},
we substitute them into \re{addcon} to get $(n-3)-$additional conditions
on $\CQ_3,... ,\CQ_n$
which together with \re{mat31} completely fix allowed values
of these quantum numbers. The problem becomes pure algebraic and we
summarize the properties of the solutions in figs.\ref{fig1}--\ref{fig8}.

It turns out that for $n\geq 4$ and integer positive conformal
weight, $h=\IZ_+$, the Baxter equation \re{Bax}
has nontrivial polynomial solutions
only for $h\ge n$ since for $h < n$ one can not satisfy condition \re{neq0}.
Similar to \re{im=0}, the quantum numbers $\CQ_3,...,\CQ_n$ which one finds
solving \re{mat31} and \re{addcon} are real and simple. Then, the solution
of the Baxter equation, $Q_n(\lambda;h,\{\CQ\})$, becomes a polynomial
of degree $h-n$ in spectral parameter $\lambda$ with real coefficients.
The roots $\lambda_1,...,\lambda_{h-n}$ of $Q_n(\lambda;h,\{\CQ\})$ satisfy
the Bethe equation \re{Beq}. We find that for each solution of the
Baxter equation, the roots are simple and real
\be
\Im \lambda_k = 0
\lab{realroot}
\ee
with $k=1,2,...,h-n$.
Being taken together, the roots corresponding to all solutions of the Baxter
equation for fixed $n$ belong to the interval of real axis in the complex
$\lambda-$plane symmetric with respect to $0$. The length of the interval
increases as $n$ grows.

The property \re{realroot} allows us to find
lower and upper bounds on the energy of the Reggeon hamiltonian,
$\varepsilon_n(h,\CQ)$, corresponding to the polynomial solutions of the
Baxter equation. Using \re{realroot} we find from \re{Ben} the
following relation
\be
-2h \le \varepsilon_n(h,\CQ) \le -2n\,.
\lab{upper}
\ee
One checks, that for $n=2,3$ and $4$ this estimate is in agreement with
our results, \re{fin} and fig.~\ref{fig1}--\ref{fig8}. Moreover, one can find
using \re{Bax-e} and \re{Q-pol} that the maximum
value of energy, $\varepsilon_n=-2n$,
is achieved for the trivial solution of the Baxter
equation \re{Bax},
$$
Q_n(\lambda)={\rm const.}\,,\qquad
h=n\,,\qquad
\CQ_{2k}=\frac{2(-)^k n!}{(2k)! (n-2k)!}\,,\qquad
\CQ_{2k+1}=0\,,\qquad
\varphi_n=\frac{z_{12}z_{23}...z_{n1}}{z_{10}^2z_{20}^2...z_{n0}^2}\,,
$$
which is in agreement with \re{maxQn}. Notice that for odd $n$ this
solution is degenerate since it does not satisfy \re{neq0}.

As we have seen in Sect.~2, the Regge asymptotics of the scattering
amplitude is dominated by the $n-$Reggeon states with maximum value of
the energy \re{en-re}. The relation \re{upper} implies that using the
polynomial solutions of the Baxter equation we cannot exceed the bound
for the energy
$E_n=-\frac{\as N}{\pi} n$. However this value cannot be maximal because
for $n=2$ the maximal value of the energy, \re{e2max}, is positive.
This is in accordance with our expectations, \re{maxQn} and \re{maxQ2},
that the maximal value of energy corresponds to the conformal weight $h=1/2$,
while polynomial solutions exist only for integer $h\ge n$. Nevertheless,
the reason why we might be interested in solving the Baxter equation for
integer conformal weight $h$ is that, similar to the situation for $n=2$
described in Sect.~6.5, we hope to find expressions for the
spectrum of the Reggeon hamiltonian which can be analytically continued
for arbitrary $h$. The analytical expressions for $n=2$ and
the properties of the Baxter equation for $n\ge 3$ (see figs.\ref{fig1}--
\ref{fig8})
indicate that the polynomial solutions of the Baxter equations
should be related to new systems of orthogonal polynomials \ci{ksz}.

\section{Conclusions}
\setcounter{equation} 0

In this paper we studied the asymptotic behavior of hadronic scattering
amplitudes in the Regge limit \re{kin} of high energies and fixed transferred
momentum. The consideration was performed in the
generalized leading logarithmic
approximation which preserves unitarity of the $S-$matrix and which
requires the resummation of infinite series of nonleading logarithmic
corrections, $\as^k \log^m s$, to the scattering amplitude. Although this
approximation is rather crude, e.g., it does not allow us to fix the argument
of the coupling constant in \re{An-res} and \re{Fn-res}, it gives
the first order approximation to the exact result.

After resummation to all orders of perturbation theory, the scattering
amplitude has a Regge behavior which can be associated with the
singularities of the partial waves in the complex angular momentum plane.
In the generalized leading logarithmic approximation,
due to gluon reggeization, longitudinal gluonic
degrees of freedom become semiclassical and nontrivial interaction
occurs only in two-dimensional space of transverse gluon momenta.
As a result of this interaction, reggeized gluons form
color singlet compound
states -- Pomeron, Odderon and higher Reggeon states, which satisfy the
Schrodinger equation with the Reggeon hamiltonian depending on the color
charges of gluons and their transverse momenta. The Reggeon hamiltonian
describes the pair--wise interaction between reggeized gluons and it obeys
the remarkable properties of holomorphic separability and conformal
invariance in the impact parameter space. The properties of the compound
Reggeon states govern the large$-s$ behavior of the total cross section,
\re{tot} and \re{An-res},
and small$-x$ asymptotics of the structure function of deep inelastic
scattering, \re{Fn-res}.

In the impact parameter space, holomorphic and antiholomorphic gluonic
degrees of freedom are coupled to each other only via interaction between
their color charges, $t_k^a t_m^a$.
If one will be able to find the limit in which this
interaction becomes trivial, then  the original two-dimensional Schrodinger
equation \re{sch} can be replaced by the system of holomorphic and
antiholomorphic one-dimensional Schrodinger equations \re{reg}. Moreover,
this correspondence becomes exact in the special case of the compound states
built from $n=2$ and $n=3$ Reggeons -- the BFKL Pomeron and Odderon,
respectively. For $n > 3$ we analyzed the Schrodinger equation \re{sch}
in the multi--color limit, $N\to \infty$, in which holomorphic
and antiholomorphic sectors become decoupled and the
interaction occurs only between nearest Reggeons, \re{near}.%
\footnote{Another possibility would be to consider the $SU(2)$ gauge
          group and take into account \ci{Bar} that for the color singlet
          $n-$Reggeon states $t_k^a t_m^a \ket{\chi}=-2/(n-1) \ket{\chi}$ with
          $(t^a)_{bc}=-i\epsilon_{abc}$. This means that for the $SU(2)$ gauge
          group the Schrodinger equation \re{sch} is equivalent to the system
          \re{reg} but with (anti-)holomorphic hamiltonian
          describing the long--range
          interaction on the one-dimensional lattice,
          $\sum_{k>m} H\lr{z_k,z_m}$.}
Thus, the expression for the energy of the compound $n-$Reggeon
state \re{E} is exact for $n=2,\ 3$, but for $n > 3$ it contains
$\CO(1/N^2)$ corrections which can be fixed from the condition
of the $S-$matrix unitarity in the $1/N-$expansion \ci{Ven}.

We found that holomorphic
and antiholomorphic Reggeon hamiltonians \re{two}, \re{H1} and \re{H2}
coincide with the hamiltonian \re{two1} and \re{s=0}
of generalized XXX Heisenberg magnet for spin $s=0$ corresponding to
the principal series representation of the $SL(2,\IC)$ group. As a result,
the system of Schrodinger equations \re{reg} turns out to be completely
integrable. The holomorphic wave functions and the corresponding energies
of the compound $n-$Reggeon states become identical to the
eigenstates and eigenvalues of the one-dimensional XXX Heisenberg magnet
defined on the one-dimensional lattice with periodic boundary conditions
and with the number of sites equal to the number of Reggeons. In each
site of the lattice the spin operators are realized as generators of
the conformal $SL(2,\IC)$ group acting on the Reggeon holomorphic and
antiholomorphic coordinates. A generalized Bethe ansatz, based on the
method of Baxter
$Q-$operator, was developed for the diagonalization of the Reggeon
hamiltonian. The spectrum of the $n-$Reggeon states, \re{Bax-s} and
\re{Bax-s}, was obtained in terms of the solution of the Baxter equation
\re{Bax}. We studied the general properties of this equation and
found the relation between its solutions and some systems of orthogonal
polynomials. For $n=2$ the solutions of the Baxter equation were identified
as series of Hahn orthogonal polynomials. These solutions were used to
derive the spectrum of the $n=2$ Reggeon state, the BFKL Pomeron. The next
step is the solution of the Baxter for $n\ge 3$. We might hope that the
analytical solution can be found using the relation of Baxter equation
to the orthogonal polynomials.

Another way to solve the Baxter equation \re{Bax} can be to explore the
conformal invariance of the Reggeon interaction and find interpretation
of the differential equation \re{diff} within the framework of
two-dimensional conformal field theories. To this end, we notice the
remarkable similarity between solution of the Baxter equation for $n=2$
and the spectrum of the $SL(2,\IR)/U(1)$ coset conformal field theory
\ci{bl}.

\bigskip\bigskip
\noindent{\Large{\bf Acknowledgements}}
\bigskip

\noindent
This paper arose essentially from a joint work \ci{prog} with L.D. Faddeev
to whom I am deeply indebted. I greatly benefit from stimulating
discussions with J. de Boer, D. Jatkar, V.P. Spiridonov and G. Sterman.
I am grateful for critical remarks to J. Bartels, A.V. Efremov, J.~Ellis,
D. Gross, V.E. Korepin, L.N. Lipatov, B. McCoy, A.V. Radyushkin and
G. Veneziano.
This work was supported in part by the National
Science Foundation under grant PHY 9309888.

\bb{99}
\bi{Col}  S.C. Frautschi, {\it Regge poles and S-matrix theory\/},
          New York, W.A. Benjamin, 1963;
\\        V. de Alfaro and T. Regge, {\it Potential scattering},
          Amsterdam, North-Holland, 1965;
\\        P.D.B. Collins, {\it An introduction to Regge theory
          and high energy physics\/}, Cambridge University Press, 1977.
\bi{land} H1 Collaboration (T. Ahmed et al.), DESY-94-198, Nov 1994,
          Phys. Lett. B338 (1994) 338; \\
          ZEUS Collaboration (M. Derrick, et al.) Z. Phys. C63 (1994) 391.
\bi{dis}  H1 Collaboration. (K. Muller, et al.),  DESY 94-112-D, Jul 1994;
\\        ZEUS Collaboration (M. Derrick, et al.),  DESY-94-192, Oct 1994,
          DESY-94-143, Aug 1994; Phys. Lett. B316 (1993) 412.
\bi{mod0} F. Low, Phys. Rev.  D12 (1975) 163; \\
          S. Nussinov, Phys. Rev. Lett. 34 (1975) 1286; \\
          J. Gunion and D. Soper, Phys. Rev. D15 (1977) 2617.
\bi{mod1} A. Donnachie and P.V. Landshoff, Nucl. Phys. B244 (1984) 322;
          Phys. Lett. B296 (1992) 227.
\bi{mod2} L.V. Gribov, E.M. Levin and M.G. Ryskin, Phys. Rep. 100 (1983) 1;
\\        E.M. Levin and M.G. Ryskin, Phys. Rep. 189 (1990) 267;
\\        E.M. Levin, {\it Low x(B) physics with open eyes\/},
          FERMILAB-CONF-94-068-T, Mar 1994.
\bi{is}   G. Ingelman and P. Schlein, Phys. Lett. 152B (1985) 256;
\\        J. Collins et al., {\it Measuring parton densities in the pomeron\/},
          preprint CTEQ/PUB/02, psu/th/136.
\bi{Lip1} L.N. Lipatov, {\it Pomeron in quantum chromodynamics\/},
          in ``Perturbative QCD'', pp.411--489, ed. A.H. Mueller,
          World Scientific, Singapore, 1989.
\bi{ren1} A.H. Mueller, Nucl. Phys. B250 (1985) 327; in {\it QCD 20 years
          later}, Aachen, 1992, ed. P.M.Zerwas and H.A.Kastrup,
          World Scientific, Singapore, 1993, v.1, p.162.
\bi{ren2} M. Beneke and V.I. Zakharov, Phys. Lett. B312 (1993) 340; \\
          V.I. Zakharov, Nucl. Phys. B385 (1992) 452.
\bi{ren3} G.P. Korchemsky and G. Sterman, {\it Nonperturbative corrections in
          resummed high energy cross sections\/}, Stony Brook preprint
          ITP--SB--94--50, Nov 1994, hep-ph/9411211.
\bi{ren4} E. Levin, {\it Renormalons at low $x$\/},
          Tel--Aviv Univ. preprint TAUP 2221--94, Dec 1994;
          hep-ph/9412345.
\bi{geor} {\it Handbook of Perturbative QCD\/},
          CTEQ Collaboration, ed. G. Sterman, FERMILAB--PUB--93/094,
          Apr 1993.
\bi{Bar}  J. Bartels, Nucl. Phys. B175 (1980) 365.
\bi{CW}   H. Cheng and T.T. Wu, {\it Expanding Protons: Scattering at
          High Energies\/}, MIT Press, Cambridge, Massachusetts, 1987.
\bi{Cheng}H. Cheng, J. Dickinson, C.Y. Lo and K. Olaussen,
          Phys. Rev. D23 (1981) 534.
\bi{bfkl} E.A. Kuraev,  L.N. Lipatov and V.S. Fadin,
          Phys. Lett. B60 (1975) 50;
          Sov. Phys. JETP 44 (1976) 443; 45 (1977) 199;
\\        Ya.Ya. Balitskii and L.N. Lipatov, Sov. J. Nucl. Phys. 28 (1978) 822.
\bi{Lip}  L.N. Lipatov, {\it High-energy asymptotics of multicolor QCD and
          exactly solvable lattice models\/}, Padova preprint DFPD-93-TH-70,
          Oct 1993;  JETP Lett. 59 (1994) 596.
\bi{prog} L.D. Faddeev and G.P. Korchemsky,
          {\it High energy QCD as a completely
          integrable model\/}, Stony Brook preprint ITP--SB--94--14, Apr 1994;
          Phys. Lett. B342 (1994) 311.
\bi{bb}   {\it Feynman's office: the last blackboards\/},
          Physics Today, February 1989, p.88.
\bi{PT}   H.T. Nieh and Y.-P. Yao,  Phys.Rev.Lett. 32 (1974) 1074;
          Phys.Rev.D13 (1976) 1082;
\\        B. McCoy and T.T. Wu, Phys. Rev. D12 (1975) 3257;
\\        C.Y. Lo and H. Cheng, Phys. Rev. D13 (1976) 1131;
\\        L. Tyburski, Phys.Rev. D13 (1976) 1107.
\bi{VV}   H. Verlinde and E. Verlinde, {\it QCD at high energies and
          two-dimensional field theory\/}, Princeton Univ. preprint,
          PUPT-1319, September 93.
\bi{LLS}  E. Laenen, E. Levin and A.G. Shuvaev, Nucl. Phys. B419 (1994) 39.
\bi{BA1}  E.K. Sklyanin, L.A. Takhtajan and L.D.Faddeev,
          Theor. Math. Phys. 40 (1980) 688.
\bi{BA2}  L.A. Takhtajan and L.D. Faddeev, Russ. Math. Survey 34 (1979) 11.
\bi{BA3}  L.D. Faddeev,
          Stony Brook preprint, ITP-SB-94-11, Mar 1994; hep-th/9404013;
          {\it The Bethe ansatz\/}, Andrejewski lectures, Freie Univ.
          preprint, SFB-288-70, Jun 1993;
          in Nankai Lectures on Mathematical Physics,
          Integrable Systems, ed. by X.-C.Song, pp.23-70,
          Singapore: World Scientific, 1990.
\bi{BA4}  V.E. Korepin, N.M.Bogoliubov and A.G. Izergin, {\it Quantum
          inverse scattering method and correlation functions\/},
          Cambridge Univ. Press, 1993.
\bi{KP}   J. Kwiecinski and M. Praszalowicz, Phys. Lett. B94 (1980) 413.
\bi{Lip2} L.N. Lipatov, Phys. Lett. B251 (1990) 284;  B309 (1993) 394.
\bi{Ven}  G. Veneziano, Nucl. Phys. B74 (1974) 365; Phys. Lett. 52B (1974) 220;
\\        A. Schwimmer and G. Veneziano, Nucl. Phys. B81 (1974) 445;
\\        M. Ciafaloni, G. Marchesini and G. Veneziano, Nucl. Phys. B98 (1975)
          472; 493.
\bi{ttf}  V.O. Tarasov, L.A. Takhtajan and L.D. Faddeev, Theor. Math. Phys.
          57 (1983) 163.
\bi{krs}  P.P. Kulish, N.Yu. Reshetikhin and E.K. Sklyanin,
          Lett. Math. Phys. 5 ( 1981) 393.
\bi{Q}    R.J. Baxter, {\it Exactly Solved Models in Statistical
          Mechanics\/}, Academic Press, London, 1982.
\bi{Skl}  E.K. Sklyanin, {\it The quantum Toda chain\/},
          Lecture Notes in Physics (Springer) 226 (1985) 196;
          {\it Quantum Inverse Scattering Method. Selected Topics\/},
          in ``Quantum Group and Quantum Integrable Systems'' (Nankai
          Lectures in Mathematical Physics), ed. Mo-Lin Ge, Singapore:
          World Scientific, 1992, pp.63--97; hep-th/9211111.
\bi{zs}   D.P. Zhelobenko and  A.I. Shtern, {\it Representations of Lie
          groups\/} (in Russian), Nauka, Moscow, 1983, pp.211-220.
\bi{wat}  E.T. Whittaker and G.N. Watson, {\it A course of modern analysis\/},
          Cambridge Univ. Press, 1963.
\bi{hyper}{\it Higher transcendental functions\/}, vols.1 and 2,
          Bateman manuscript project, ed. A. Erdelyi, McGraw-Hill, 1953.
\bi{aw}   R. Askey and J. Wilson, {\it Some basic hypergeometric orthogonal
          polynomials that generalize Jacobi polynomials\/},
          Mem. Am. Math. Soc. {319} (1985) 1-55.
\bi{sz}   V. Spiridonov and A. Zhedanov, {\it Discrete reflectionless
          potentials,
          quantum algebras and q-orthogonal polynomials\/},
          preprint CRM-1928 (1993); Annals of Physics, vol. 237 (1995)
          126; {\it Discrete Darboux transformations, discrete time
          Toda lattice and the Askey-Wilson polynomials\/},
          preprint CRM-1929 (1993).
\bi{ksz}  G.P. Korchemsky, V.P. Spiridonov and A.S. Zhedanov, in preparation.
\bi{F}    L.N. Lipatov, Talk at the 2nd Workshop on Small--$x$ and Diffractive
          Physics at the Tevatron, Fermilab, Sept. 1994.
\bi{bl}   R. Dijkgraaf, E. Verlinde and H. Verlinde,
          Nucl. Phys. B371 (1992) 269;
\\        D. Jatkar, Nucl. Phys. B395 (1993) 167.
\eb

\newpage

\bigskip
\bigskip
\begin{center}
\noindent{\Large{\bf Figures:}}
\end{center}

\newcommand \ladder{
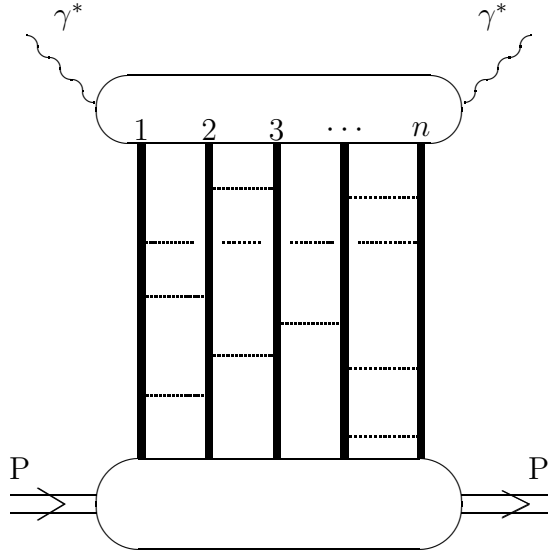
\begin{figure}[h]
\begin{center}
\unitlength=0.6mm
\linethickness{0.4pt}
\begin{picture}(130.00,138.00)(5,0)
\put(70.50,107.50){\oval(81.00,15.00)[]}
\put(70.50,20.00){\oval(81.00,20.00)[]}
\put(111.00,22.00){\line(1,0){19.00}}
\put(130.00,22.00){\line(-1,0){19.00}}
\put(111.00,18.00){\line(1,0){19.00}}
\put(11.00,22.00){\line(1,0){19.00}}
\put(30.00,18.00){\line(-1,0){19.00}}
\put(17.00,23.6){\line(5,-3){6.00}}
\put(23.00,20.00){\line(-5,-3){6.00}}
\put(120.00,17.00){\line(2,1){6.00}}
\put(126.00,20.00){\line(-2,1){6.00}}
\put(13.00,28.00){\makebox(0,0)[cc]{P}}
\put(128.00,28.00){\makebox(0,0)[cc]{P}}
\linethickness{1mm}
\put(40.00,30.00){\line(0,1){70.00}}
\put(55.00,100.00){\line(0,-1){70.00}}
\put(70.00,30.00){\line(0,1){70.00}}
\put(85.00,100.00){\line(0,-1){70.00}}
\put(102.00,30.00){\line(0,1){70.00}}
\linethickness{0.8pt}
\put(40.00,103.00){\makebox(0,0)[cc]{$1$}}
\put(55.00,103.00){\makebox(0,0)[cc]{$2$}}
\put(70.00,103.00){\makebox(0,0)[cc]{$3$}}
\put(102.00,103.00){\makebox(0,0)[cc]{$n$}}
\put(85.00,103.00){\makebox(0,0)[cc]{$\cdots$}}
\bezier{12}(40.00,78.00)(45.50,78.00)(51.00,78.00)
\bezier{7}(58.00,78.00)(62.00,78.00)(66.00,78.00)
\bezier{9}(73.00,78.00)(78.00,78.00)(82.00,78.00)
\bezier{13}(88.00,78.00)(94.00,78.00)(101.50,78.00)
\bezier{16}(55.00,90.00)(63.00,90.00)(70.00,90.00)
\bezier{16}(55.00,66.00)(48.00,66.00)(40.00,66.00)
\bezier{16}(70.00,60.00)(78.00,60.00)(85.00,60.00)
\bezier{16}(55.00,44.00)(48.00,44.00)(40.00,44.00)
\bezier{16}(55.00,53.00)(63.00,53.00)(70.00,53.00)
\bezier{16}(85.00,88.00)(94.00,88.00)(102.00,88.00)
\bezier{16}(85.00,35.00)(94.00,35.00)(102.00,35.00)
\bezier{16}(85.00,50.00)(94.00,50.00)(102.00,50.00)
\put(118.00,128.00){\makebox(0,0)[cc]{$\gamma^*$}}
\put(24.00,128.00){\makebox(0,0)[cc]{$\gamma^*$}}
\linethickness{0.4pt}
\put(111.50,111.50){\oval(5,5)[rb]}
\put(116.50,111.50){\oval(5,5)[lt]}
\put(116.50,116.50){\oval(5,5)[rb]}
\put(121.50,116.50){\oval(5,5)[lt]}
\put(121.50,121.50){\oval(5,5)[rb]}
\put(126.50,121.50){\oval(5,5)[lt]}
\put(29.50,111.50){\oval(5,5)[lb]}
\put(24.50,111.50){\oval(5,5)[rt]}
\put(24.50,116.50){\oval(5,5)[lb]}
\put(19.50,116.50){\oval(5,5)[rt]}
\put(19.50,121.50){\oval(5,5)[lb]}
\put(14.50,121.50){\oval(5,5)[rt]}
\end{picture}
\end{center}
\caption{
Unitary Feynman diagrams contributing to the structure function
of deep inelastic scattering in the generalized leading logarithmic
approximation in the multicolor limit, $N\to\infty$. Solid lines
represent $n$ reggeized
gluons propagating in the $t-$channel. Dotted lines denote interaction
of reggeized gluons with their nearest neighbours. Upper and lower blobs
describe the coupling of gluons to virtual photon $\gamma^*$ and proton
state $\rm p$, respectively. For finite $N$ one has to add the diagrams
with 
pair--wise interactions between $n$ reggeized gluons.}
\label{ladder}
\end{figure}}

\newcommand \BSeq {
\begin{figure}[b]
\begin{center}
\unitlength=0.75mm
\linethickness{0.4pt}
\begin{picture}(175.00,9.00)
\put(30.00,40.00){\oval(40.00,20.00)[]}
\put(155.00,40.00){\oval(40.00,20.00)[]}
\put(60.00,40.00){\makebox(0,0)[cc]{$=$}}
\put(110.00,40.00){\makebox(0,0)[cc]{$+$}}
\put(121.00,38.00){\makebox(0,0)[cc]{$
\begin{array}{c} \sum \\[-2mm] {_{_{1\le j < k \le n}}}
\end{array}
$}}
\put(20.00,57.00){\makebox(0,0)[cc]{$...$}}
\put(30.00,57.00){\makebox(0,0)[cc]{$...$}}
\put(40.00,57.00){\makebox(0,0)[cc]{$...$}}
\put(75.00,57.00){\makebox(0,0)[cc]{$...$}}
\put(85.00,57.00){\makebox(0,0)[cc]{$...$}}
\put(95.00,57.00){\makebox(0,0)[cc]{$...$}}
\put(145.00,57.00){\makebox(0,0)[cc]{$...$}}
\put(155.00,57.00){\makebox(0,0)[cc]{$...$}}
\put(165.00,57.00){\makebox(0,0)[cc]{$...$}}
\put(155.00,40.00){\makebox(0,0)[cc]{$T_n(\omega)$}}
\put(140.00,79.00){\makebox(0,0)[cc]{$1$}}
\put(150.00,79.00){\makebox(0,0)[cc]{$j$}}
\put(160.00,79.00){\makebox(0,0)[cc]{$k$}}
\put(170.00,79.00){\makebox(0,0)[cc]{$n$}}
\put(100.00,79.00){\makebox(0,0)[cc]{$n$}}
\put(90.00,79.00){\makebox(0,0)[cc]{$k$}}
\put(80.00,79.00){\makebox(0,0)[cc]{$j$}}
\put(70.00,79.00){\makebox(0,0)[cc]{$1$}}
\put(15.00,79.00){\makebox(0,0)[cc]{$1$}}
\put(25.00,79.00){\makebox(0,0)[cc]{$j$}}
\put(35.00,79.00){\makebox(0,0)[cc]{$k$}}
\put(45.00,79.00){\makebox(0,0)[cc]{$n$}}
\put(0.00,40.00){\makebox(0,0)[cc]{$\omega$}}
\put(30.00,40.00){\makebox(0,0)[cc]{$T_n(\omega)$}}
\linethickness{1.4pt}
\bezier{8}(150.00,66.00)(154.75,66.00)(159.50,66.00)
\linethickness{1.0mm}
\put(15.00,49.00){\line(0,1){26.00}}
\put(45.00,49.00){\line(0,1){26.00}}
\put(15.00,13.00){\line(0,1){18.00}}
\put(45.00,13.00){\line(0,1){18.00}}
\put(25.00,50.00){\line(0,1){25.00}}
\put(35.00,50.00){\line(0,1){25.00}}
\put(25.00,13.00){\line(0,1){17.00}}
\put(35.00,13.00){\line(0,1){17.00}}
\put(70.00,13.00){\line(0,1){62.00}}
\put(80.00,13.00){\line(0,1){62.00}}
\put(90.00,13.00){\line(0,1){62.00}}
\put(100.00,13.00){\line(0,1){62.00}}
\put(140.00,49.00){\line(0,1){26.00}}
\put(170.00,49.00){\line(0,1){26.00}}
\put(140.00,13.00){\line(0,1){18.00}}
\put(170.00,13.00){\line(0,1){18.00}}
\put(150.00,50.00){\line(0,1){25.00}}
\put(160.00,50.00){\line(0,1){25.00}}
\put(150.00,13.00){\line(0,1){17.00}}
\put(160.00,13.00){\line(0,1){17.00}}
\end{picture}
\end{center}
\vspace*{-6mm}
\caption{The Bethe--Salpeter equation for the transition operator
$T_n(\omega)$, describing $n \to n$ elastic scattering of reggeized
gluons in the $t-$channel. Iterations of this equation reproduce the
ladder diagrams of fig.~\ref{ladder}.}
\label{figBS}
\end{figure}}

\def \Dia {\circle{12.00}}
\newcommand \figa {
\begin{figure}
\begin{center}
\setlength{\unitlength}{0.240900pt}
\ifx\plotpoint\undefined\newsavebox{\plotpoint}\fi
\sbox{\plotpoint}{\rule[-0.200pt]{0.400pt}{0.400pt}}%
\begin{picture}(1500,900)(0,0)
\font\gnuplot=cmr10 at 10pt
\gnuplot
\sbox{\plotpoint}{\rule[-0.200pt]{0.400pt}{0.400pt}}%
\put(220.0,113.0){\rule[-0.200pt]{4.818pt}{0.400pt}}
\put(198,113){\makebox(0,0)[r]{-15}}
\put(1416.0,113.0){\rule[-0.200pt]{4.818pt}{0.400pt}}
\put(220.0,198.0){\rule[-0.200pt]{4.818pt}{0.400pt}}
\put(198,198){\makebox(0,0)[r]{-14}}
\put(1416.0,198.0){\rule[-0.200pt]{4.818pt}{0.400pt}}
\put(220.0,283.0){\rule[-0.200pt]{4.818pt}{0.400pt}}
\put(198,283){\makebox(0,0)[r]{-13}}
\put(1416.0,283.0){\rule[-0.200pt]{4.818pt}{0.400pt}}
\put(220.0,368.0){\rule[-0.200pt]{4.818pt}{0.400pt}}
\put(198,368){\makebox(0,0)[r]{-12}}
\put(1416.0,368.0){\rule[-0.200pt]{4.818pt}{0.400pt}}
\put(220.0,453.0){\rule[-0.200pt]{4.818pt}{0.400pt}}
\put(198,453){\makebox(0,0)[r]{-11}}
\put(1416.0,453.0){\rule[-0.200pt]{4.818pt}{0.400pt}}
\put(220.0,537.0){\rule[-0.200pt]{4.818pt}{0.400pt}}
\put(198,537){\makebox(0,0)[r]{-10}}
\put(1416.0,537.0){\rule[-0.200pt]{4.818pt}{0.400pt}}
\put(220.0,622.0){\rule[-0.200pt]{4.818pt}{0.400pt}}
\put(198,622){\makebox(0,0)[r]{-9}}
\put(1416.0,622.0){\rule[-0.200pt]{4.818pt}{0.400pt}}
\put(220.0,707.0){\rule[-0.200pt]{4.818pt}{0.400pt}}
\put(198,707){\makebox(0,0)[r]{-8}}
\put(1416.0,707.0){\rule[-0.200pt]{4.818pt}{0.400pt}}
\put(220.0,792.0){\rule[-0.200pt]{4.818pt}{0.400pt}}
\put(198,792){\makebox(0,0)[r]{-7}}
\put(1416.0,792.0){\rule[-0.200pt]{4.818pt}{0.400pt}}
\put(220.0,877.0){\rule[-0.200pt]{4.818pt}{0.400pt}}
\put(198,877){\makebox(0,0)[r]{-6}}
\put(1416.0,877.0){\rule[-0.200pt]{4.818pt}{0.400pt}}
\put(220.0,113.0){\rule[-0.200pt]{0.400pt}{4.818pt}}
\put(220,68){\makebox(0,0){3}}
\put(220.0,857.0){\rule[-0.200pt]{0.400pt}{4.818pt}}
\put(372.0,113.0){\rule[-0.200pt]{0.400pt}{4.818pt}}
\put(372,68){\makebox(0,0){4}}
\put(372.0,857.0){\rule[-0.200pt]{0.400pt}{4.818pt}}
\put(524.0,113.0){\rule[-0.200pt]{0.400pt}{4.818pt}}
\put(524,68){\makebox(0,0){5}}
\put(524.0,857.0){\rule[-0.200pt]{0.400pt}{4.818pt}}
\put(676.0,113.0){\rule[-0.200pt]{0.400pt}{4.818pt}}
\put(676,68){\makebox(0,0){6}}
\put(676.0,857.0){\rule[-0.200pt]{0.400pt}{4.818pt}}
\put(828.0,113.0){\rule[-0.200pt]{0.400pt}{4.818pt}}
\put(828,68){\makebox(0,0){7}}
\put(828.0,857.0){\rule[-0.200pt]{0.400pt}{4.818pt}}
\put(980.0,113.0){\rule[-0.200pt]{0.400pt}{4.818pt}}
\put(980,68){\makebox(0,0){8}}
\put(980.0,857.0){\rule[-0.200pt]{0.400pt}{4.818pt}}
\put(1132.0,113.0){\rule[-0.200pt]{0.400pt}{4.818pt}}
\put(1132,68){\makebox(0,0){9}}
\put(1132.0,857.0){\rule[-0.200pt]{0.400pt}{4.818pt}}
\put(1284.0,113.0){\rule[-0.200pt]{0.400pt}{4.818pt}}
\put(1284,68){\makebox(0,0){10}}
\put(1284.0,857.0){\rule[-0.200pt]{0.400pt}{4.818pt}}
\put(1436.0,113.0){\rule[-0.200pt]{0.400pt}{4.818pt}}
\put(1436,68){\makebox(0,0){11}}
\put(1436.0,857.0){\rule[-0.200pt]{0.400pt}{4.818pt}}
\put(220.0,113.0){\rule[-0.200pt]{292.934pt}{0.400pt}}
\put(1436.0,113.0){\rule[-0.200pt]{0.400pt}{184.048pt}}
\put(220.0,877.0){\rule[-0.200pt]{292.934pt}{0.400pt}}
\put(45,495){\makebox(0,0){$\varepsilon_3$}}
\put(828,23){\makebox(0,0){$h$}}
\put(220.0,113.0){\rule[-0.200pt]{0.400pt}{184.048pt}}
\put(220,877){\usebox{\plotpoint}}
\multiput(220.00,875.92)(0.576,-0.497){49}{\rule{0.562pt}{0.120pt}}
\multiput(220.00,876.17)(28.834,-26.000){2}{\rule{0.281pt}{0.400pt}}
\multiput(250.00,849.92)(0.646,-0.496){45}{\rule{0.617pt}{0.120pt}}
\multiput(250.00,850.17)(29.720,-24.000){2}{\rule{0.308pt}{0.400pt}}
\multiput(281.00,825.92)(0.683,-0.496){41}{\rule{0.645pt}{0.120pt}}
\multiput(281.00,826.17)(28.660,-22.000){2}{\rule{0.323pt}{0.400pt}}
\multiput(311.00,803.92)(0.740,-0.496){39}{\rule{0.690pt}{0.119pt}}
\multiput(311.00,804.17)(29.567,-21.000){2}{\rule{0.345pt}{0.400pt}}
\multiput(342.00,782.92)(0.753,-0.496){37}{\rule{0.700pt}{0.119pt}}
\multiput(342.00,783.17)(28.547,-20.000){2}{\rule{0.350pt}{0.400pt}}
\multiput(372.00,762.92)(0.793,-0.495){35}{\rule{0.732pt}{0.119pt}}
\multiput(372.00,763.17)(28.482,-19.000){2}{\rule{0.366pt}{0.400pt}}
\multiput(402.00,743.92)(0.866,-0.495){33}{\rule{0.789pt}{0.119pt}}
\multiput(402.00,744.17)(29.363,-18.000){2}{\rule{0.394pt}{0.400pt}}
\multiput(433.00,725.92)(0.888,-0.495){31}{\rule{0.806pt}{0.119pt}}
\multiput(433.00,726.17)(28.327,-17.000){2}{\rule{0.403pt}{0.400pt}}
\multiput(463.00,708.92)(0.977,-0.494){29}{\rule{0.875pt}{0.119pt}}
\multiput(463.00,709.17)(29.184,-16.000){2}{\rule{0.438pt}{0.400pt}}
\multiput(494.00,692.92)(1.010,-0.494){27}{\rule{0.900pt}{0.119pt}}
\multiput(494.00,693.17)(28.132,-15.000){2}{\rule{0.450pt}{0.400pt}}
\multiput(524.00,677.92)(1.010,-0.494){27}{\rule{0.900pt}{0.119pt}}
\multiput(524.00,678.17)(28.132,-15.000){2}{\rule{0.450pt}{0.400pt}}
\multiput(554.00,662.92)(1.121,-0.494){25}{\rule{0.986pt}{0.119pt}}
\multiput(554.00,663.17)(28.954,-14.000){2}{\rule{0.493pt}{0.400pt}}
\multiput(585.00,648.92)(1.171,-0.493){23}{\rule{1.023pt}{0.119pt}}
\multiput(585.00,649.17)(27.877,-13.000){2}{\rule{0.512pt}{0.400pt}}
\multiput(615.00,635.92)(1.210,-0.493){23}{\rule{1.054pt}{0.119pt}}
\multiput(615.00,636.17)(28.813,-13.000){2}{\rule{0.527pt}{0.400pt}}
\multiput(646.00,622.92)(1.171,-0.493){23}{\rule{1.023pt}{0.119pt}}
\multiput(646.00,623.17)(27.877,-13.000){2}{\rule{0.512pt}{0.400pt}}
\multiput(676.00,609.92)(1.272,-0.492){21}{\rule{1.100pt}{0.119pt}}
\multiput(676.00,610.17)(27.717,-12.000){2}{\rule{0.550pt}{0.400pt}}
\multiput(706.00,597.92)(1.315,-0.492){21}{\rule{1.133pt}{0.119pt}}
\multiput(706.00,598.17)(28.648,-12.000){2}{\rule{0.567pt}{0.400pt}}
\multiput(737.00,585.92)(1.392,-0.492){19}{\rule{1.191pt}{0.118pt}}
\multiput(737.00,586.17)(27.528,-11.000){2}{\rule{0.595pt}{0.400pt}}
\multiput(767.00,574.92)(1.439,-0.492){19}{\rule{1.227pt}{0.118pt}}
\multiput(767.00,575.17)(28.453,-11.000){2}{\rule{0.614pt}{0.400pt}}
\multiput(798.00,563.92)(1.392,-0.492){19}{\rule{1.191pt}{0.118pt}}
\multiput(798.00,564.17)(27.528,-11.000){2}{\rule{0.595pt}{0.400pt}}
\multiput(828.00,552.92)(1.538,-0.491){17}{\rule{1.300pt}{0.118pt}}
\multiput(828.00,553.17)(27.302,-10.000){2}{\rule{0.650pt}{0.400pt}}
\multiput(858.00,542.92)(1.590,-0.491){17}{\rule{1.340pt}{0.118pt}}
\multiput(858.00,543.17)(28.219,-10.000){2}{\rule{0.670pt}{0.400pt}}
\multiput(889.00,532.92)(1.538,-0.491){17}{\rule{1.300pt}{0.118pt}}
\multiput(889.00,533.17)(27.302,-10.000){2}{\rule{0.650pt}{0.400pt}}
\multiput(919.00,522.93)(1.776,-0.489){15}{\rule{1.478pt}{0.118pt}}
\multiput(919.00,523.17)(27.933,-9.000){2}{\rule{0.739pt}{0.400pt}}
\multiput(950.00,513.93)(1.718,-0.489){15}{\rule{1.433pt}{0.118pt}}
\multiput(950.00,514.17)(27.025,-9.000){2}{\rule{0.717pt}{0.400pt}}
\multiput(980.00,504.93)(1.718,-0.489){15}{\rule{1.433pt}{0.118pt}}
\multiput(980.00,505.17)(27.025,-9.000){2}{\rule{0.717pt}{0.400pt}}
\multiput(1010.00,495.93)(1.776,-0.489){15}{\rule{1.478pt}{0.118pt}}
\multiput(1010.00,496.17)(27.933,-9.000){2}{\rule{0.739pt}{0.400pt}}
\multiput(1041.00,486.93)(1.947,-0.488){13}{\rule{1.600pt}{0.117pt}}
\multiput(1041.00,487.17)(26.679,-8.000){2}{\rule{0.800pt}{0.400pt}}
\multiput(1071.00,478.93)(2.013,-0.488){13}{\rule{1.650pt}{0.117pt}}
\multiput(1071.00,479.17)(27.575,-8.000){2}{\rule{0.825pt}{0.400pt}}
\multiput(1102.00,470.93)(1.718,-0.489){15}{\rule{1.433pt}{0.118pt}}
\multiput(1102.00,471.17)(27.025,-9.000){2}{\rule{0.717pt}{0.400pt}}
\multiput(1132.00,461.93)(2.247,-0.485){11}{\rule{1.814pt}{0.117pt}}
\multiput(1132.00,462.17)(26.234,-7.000){2}{\rule{0.907pt}{0.400pt}}
\multiput(1162.00,454.93)(2.013,-0.488){13}{\rule{1.650pt}{0.117pt}}
\multiput(1162.00,455.17)(27.575,-8.000){2}{\rule{0.825pt}{0.400pt}}
\multiput(1193.00,446.93)(1.947,-0.488){13}{\rule{1.600pt}{0.117pt}}
\multiput(1193.00,447.17)(26.679,-8.000){2}{\rule{0.800pt}{0.400pt}}
\multiput(1223.00,438.93)(2.323,-0.485){11}{\rule{1.871pt}{0.117pt}}
\multiput(1223.00,439.17)(27.116,-7.000){2}{\rule{0.936pt}{0.400pt}}
\multiput(1254.00,431.93)(2.247,-0.485){11}{\rule{1.814pt}{0.117pt}}
\multiput(1254.00,432.17)(26.234,-7.000){2}{\rule{0.907pt}{0.400pt}}
\multiput(1284.00,424.93)(2.247,-0.485){11}{\rule{1.814pt}{0.117pt}}
\multiput(1284.00,425.17)(26.234,-7.000){2}{\rule{0.907pt}{0.400pt}}
\multiput(1314.00,417.93)(2.323,-0.485){11}{\rule{1.871pt}{0.117pt}}
\multiput(1314.00,418.17)(27.116,-7.000){2}{\rule{0.936pt}{0.400pt}}
\multiput(1345.00,410.93)(2.247,-0.485){11}{\rule{1.814pt}{0.117pt}}
\multiput(1345.00,411.17)(26.234,-7.000){2}{\rule{0.907pt}{0.400pt}}
\multiput(1375.00,403.93)(2.323,-0.485){11}{\rule{1.871pt}{0.117pt}}
\multiput(1375.00,404.17)(27.116,-7.000){2}{\rule{0.936pt}{0.400pt}}
\multiput(1406.00,396.93)(2.660,-0.482){9}{\rule{2.100pt}{0.116pt}}
\multiput(1406.00,397.17)(25.641,-6.000){2}{\rule{1.050pt}{0.400pt}}
\put(220,877){\Dia}
\put(372,750){\Dia}
\put(372,750){\Dia}
\put(524,679){\Dia}
\put(524,636){\Dia}
\put(524,636){\Dia}
\put(676,601){\Dia}
\put(676,601){\Dia}
\put(676,537){\Dia}
\put(676,537){\Dia}
\put(828,554){\Dia}
\put(828,451){\Dia}
\put(828,522){\Dia}
\put(828,522){\Dia}
\put(828,451){\Dia}
\put(980,445){\Dia}
\put(980,445){\Dia}
\put(980,376){\Dia}
\put(980,376){\Dia}
\put(980,498){\Dia}
\put(980,498){\Dia}
\put(1132,463){\Dia}
\put(1132,310){\Dia}
\put(1132,310){\Dia}
\put(1132,375){\Dia}
\put(1132,375){\Dia}
\put(1132,436){\Dia}
\put(1132,436){\Dia}
\put(1284,373){\Dia}
\put(1284,373){\Dia}
\put(1284,251){\Dia}
\put(1284,251){\Dia}
\put(1284,419){\Dia}
\put(1284,419){\Dia}
\put(1284,310){\Dia}
\put(1284,310){\Dia}
\put(1436,392){\Dia}
\put(1436,252){\Dia}
\put(1436,252){\Dia}
\put(1436,198){\Dia}
\put(1436,198){\Dia}
\put(1436,368){\Dia}
\put(1436,368){\Dia}
\put(1436,313){\Dia}
\put(1436,313){\Dia}
\end{picture}
\end{center}
\caption{The energy of the holomorphic Reggeon hamiltonian for $n=3$
corresponding to the polynomial solutions of the Baxter equation. The solid
line represents the energy for $n=2$ defined in \re{fin}. Points on this
line correspond to the degenerate solutions, $\CQ_3=0$, of the Baxter
equation for $n=3$.}
\label{fig1}
\end{figure}
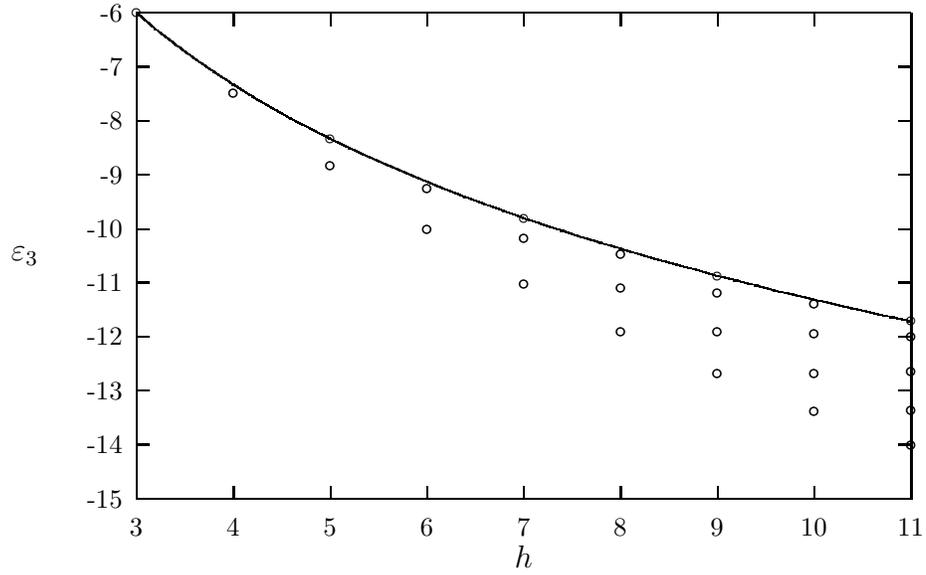}

\newcommand \figb {
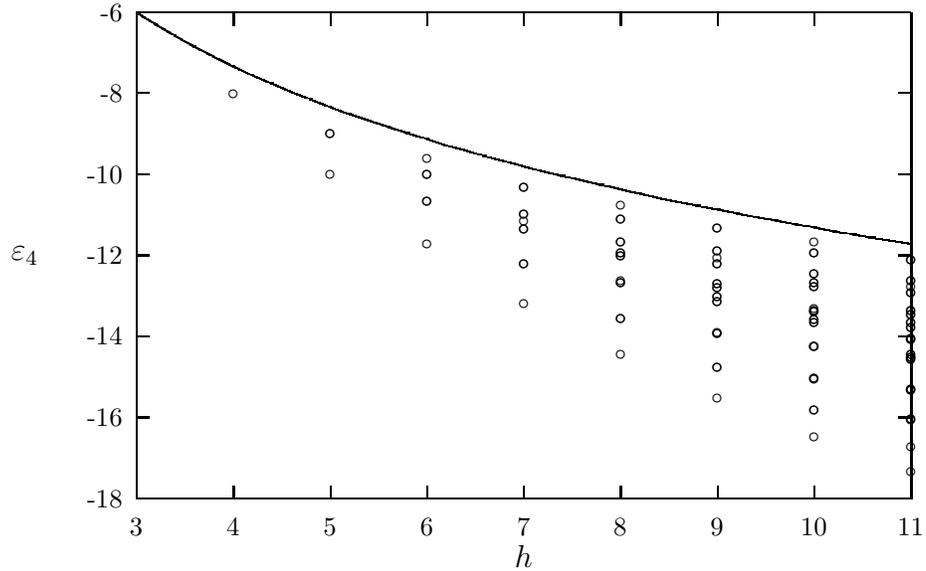
\begin{figure}
\begin{center}
\setlength{\unitlength}{0.240900pt}
\ifx\plotpoint\undefined\newsavebox{\plotpoint}\fi
\sbox{\plotpoint}{\rule[-0.200pt]{0.400pt}{0.400pt}}%
\begin{picture}(1500,900)(0,0)
\font\gnuplot=cmr10 at 10pt
\gnuplot
\sbox{\plotpoint}{\rule[-0.200pt]{0.400pt}{0.400pt}}%
\put(220.0,113.0){\rule[-0.200pt]{4.818pt}{0.400pt}}
\put(198,113){\makebox(0,0)[r]{-18}}
\put(1416.0,113.0){\rule[-0.200pt]{4.818pt}{0.400pt}}
\put(220.0,240.0){\rule[-0.200pt]{4.818pt}{0.400pt}}
\put(198,240){\makebox(0,0)[r]{-16}}
\put(1416.0,240.0){\rule[-0.200pt]{4.818pt}{0.400pt}}
\put(220.0,368.0){\rule[-0.200pt]{4.818pt}{0.400pt}}
\put(198,368){\makebox(0,0)[r]{-14}}
\put(1416.0,368.0){\rule[-0.200pt]{4.818pt}{0.400pt}}
\put(220.0,495.0){\rule[-0.200pt]{4.818pt}{0.400pt}}
\put(198,495){\makebox(0,0)[r]{-12}}
\put(1416.0,495.0){\rule[-0.200pt]{4.818pt}{0.400pt}}
\put(220.0,622.0){\rule[-0.200pt]{4.818pt}{0.400pt}}
\put(198,622){\makebox(0,0)[r]{-10}}
\put(1416.0,622.0){\rule[-0.200pt]{4.818pt}{0.400pt}}
\put(220.0,750.0){\rule[-0.200pt]{4.818pt}{0.400pt}}
\put(198,750){\makebox(0,0)[r]{-8}}
\put(1416.0,750.0){\rule[-0.200pt]{4.818pt}{0.400pt}}
\put(220.0,877.0){\rule[-0.200pt]{4.818pt}{0.400pt}}
\put(198,877){\makebox(0,0)[r]{-6}}
\put(1416.0,877.0){\rule[-0.200pt]{4.818pt}{0.400pt}}
\put(220.0,113.0){\rule[-0.200pt]{0.400pt}{4.818pt}}
\put(220,68){\makebox(0,0){3}}
\put(220.0,857.0){\rule[-0.200pt]{0.400pt}{4.818pt}}
\put(372.0,113.0){\rule[-0.200pt]{0.400pt}{4.818pt}}
\put(372,68){\makebox(0,0){4}}
\put(372.0,857.0){\rule[-0.200pt]{0.400pt}{4.818pt}}
\put(524.0,113.0){\rule[-0.200pt]{0.400pt}{4.818pt}}
\put(524,68){\makebox(0,0){5}}
\put(524.0,857.0){\rule[-0.200pt]{0.400pt}{4.818pt}}
\put(676.0,113.0){\rule[-0.200pt]{0.400pt}{4.818pt}}
\put(676,68){\makebox(0,0){6}}
\put(676.0,857.0){\rule[-0.200pt]{0.400pt}{4.818pt}}
\put(828.0,113.0){\rule[-0.200pt]{0.400pt}{4.818pt}}
\put(828,68){\makebox(0,0){7}}
\put(828.0,857.0){\rule[-0.200pt]{0.400pt}{4.818pt}}
\put(980.0,113.0){\rule[-0.200pt]{0.400pt}{4.818pt}}
\put(980,68){\makebox(0,0){8}}
\put(980.0,857.0){\rule[-0.200pt]{0.400pt}{4.818pt}}
\put(1132.0,113.0){\rule[-0.200pt]{0.400pt}{4.818pt}}
\put(1132,68){\makebox(0,0){9}}
\put(1132.0,857.0){\rule[-0.200pt]{0.400pt}{4.818pt}}
\put(1284.0,113.0){\rule[-0.200pt]{0.400pt}{4.818pt}}
\put(1284,68){\makebox(0,0){10}}
\put(1284.0,857.0){\rule[-0.200pt]{0.400pt}{4.818pt}}
\put(1436.0,113.0){\rule[-0.200pt]{0.400pt}{4.818pt}}
\put(1436,68){\makebox(0,0){11}}
\put(1436.0,857.0){\rule[-0.200pt]{0.400pt}{4.818pt}}
\put(220.0,113.0){\rule[-0.200pt]{292.934pt}{0.400pt}}
\put(1436.0,113.0){\rule[-0.200pt]{0.400pt}{184.048pt}}
\put(220.0,877.0){\rule[-0.200pt]{292.934pt}{0.400pt}}
\put(45,495){\makebox(0,0){$\varepsilon_4$}}
\put(828,23){\makebox(0,0){$h$}}
\put(220.0,113.0){\rule[-0.200pt]{0.400pt}{184.048pt}}
\put(220,877){\usebox{\plotpoint}}
\multiput(220.00,875.92)(0.793,-0.495){35}{\rule{0.732pt}{0.119pt}}
\multiput(220.00,876.17)(28.482,-19.000){2}{\rule{0.366pt}{0.400pt}}
\multiput(250.00,856.92)(0.866,-0.495){33}{\rule{0.789pt}{0.119pt}}
\multiput(250.00,857.17)(29.363,-18.000){2}{\rule{0.394pt}{0.400pt}}
\multiput(281.00,838.92)(0.888,-0.495){31}{\rule{0.806pt}{0.119pt}}
\multiput(281.00,839.17)(28.327,-17.000){2}{\rule{0.403pt}{0.400pt}}
\multiput(311.00,821.92)(0.977,-0.494){29}{\rule{0.875pt}{0.119pt}}
\multiput(311.00,822.17)(29.184,-16.000){2}{\rule{0.438pt}{0.400pt}}
\multiput(342.00,805.92)(1.010,-0.494){27}{\rule{0.900pt}{0.119pt}}
\multiput(342.00,806.17)(28.132,-15.000){2}{\rule{0.450pt}{0.400pt}}
\multiput(372.00,790.92)(1.084,-0.494){25}{\rule{0.957pt}{0.119pt}}
\multiput(372.00,791.17)(28.013,-14.000){2}{\rule{0.479pt}{0.400pt}}
\multiput(402.00,776.92)(1.210,-0.493){23}{\rule{1.054pt}{0.119pt}}
\multiput(402.00,777.17)(28.813,-13.000){2}{\rule{0.527pt}{0.400pt}}
\multiput(433.00,763.92)(1.171,-0.493){23}{\rule{1.023pt}{0.119pt}}
\multiput(433.00,764.17)(27.877,-13.000){2}{\rule{0.512pt}{0.400pt}}
\multiput(463.00,750.92)(1.315,-0.492){21}{\rule{1.133pt}{0.119pt}}
\multiput(463.00,751.17)(28.648,-12.000){2}{\rule{0.567pt}{0.400pt}}
\multiput(494.00,738.92)(1.272,-0.492){21}{\rule{1.100pt}{0.119pt}}
\multiput(494.00,739.17)(27.717,-12.000){2}{\rule{0.550pt}{0.400pt}}
\multiput(524.00,726.92)(1.392,-0.492){19}{\rule{1.191pt}{0.118pt}}
\multiput(524.00,727.17)(27.528,-11.000){2}{\rule{0.595pt}{0.400pt}}
\multiput(554.00,715.92)(1.590,-0.491){17}{\rule{1.340pt}{0.118pt}}
\multiput(554.00,716.17)(28.219,-10.000){2}{\rule{0.670pt}{0.400pt}}
\multiput(585.00,705.92)(1.538,-0.491){17}{\rule{1.300pt}{0.118pt}}
\multiput(585.00,706.17)(27.302,-10.000){2}{\rule{0.650pt}{0.400pt}}
\multiput(615.00,695.92)(1.590,-0.491){17}{\rule{1.340pt}{0.118pt}}
\multiput(615.00,696.17)(28.219,-10.000){2}{\rule{0.670pt}{0.400pt}}
\multiput(646.00,685.93)(1.718,-0.489){15}{\rule{1.433pt}{0.118pt}}
\multiput(646.00,686.17)(27.025,-9.000){2}{\rule{0.717pt}{0.400pt}}
\multiput(676.00,676.92)(1.538,-0.491){17}{\rule{1.300pt}{0.118pt}}
\multiput(676.00,677.17)(27.302,-10.000){2}{\rule{0.650pt}{0.400pt}}
\multiput(706.00,666.93)(2.013,-0.488){13}{\rule{1.650pt}{0.117pt}}
\multiput(706.00,667.17)(27.575,-8.000){2}{\rule{0.825pt}{0.400pt}}
\multiput(737.00,658.93)(1.718,-0.489){15}{\rule{1.433pt}{0.118pt}}
\multiput(737.00,659.17)(27.025,-9.000){2}{\rule{0.717pt}{0.400pt}}
\multiput(767.00,649.93)(2.013,-0.488){13}{\rule{1.650pt}{0.117pt}}
\multiput(767.00,650.17)(27.575,-8.000){2}{\rule{0.825pt}{0.400pt}}
\multiput(798.00,641.93)(1.947,-0.488){13}{\rule{1.600pt}{0.117pt}}
\multiput(798.00,642.17)(26.679,-8.000){2}{\rule{0.800pt}{0.400pt}}
\multiput(828.00,633.93)(1.947,-0.488){13}{\rule{1.600pt}{0.117pt}}
\multiput(828.00,634.17)(26.679,-8.000){2}{\rule{0.800pt}{0.400pt}}
\multiput(858.00,625.93)(2.323,-0.485){11}{\rule{1.871pt}{0.117pt}}
\multiput(858.00,626.17)(27.116,-7.000){2}{\rule{0.936pt}{0.400pt}}
\multiput(889.00,618.93)(2.247,-0.485){11}{\rule{1.814pt}{0.117pt}}
\multiput(889.00,619.17)(26.234,-7.000){2}{\rule{0.907pt}{0.400pt}}
\multiput(919.00,611.93)(2.323,-0.485){11}{\rule{1.871pt}{0.117pt}}
\multiput(919.00,612.17)(27.116,-7.000){2}{\rule{0.936pt}{0.400pt}}
\multiput(950.00,604.93)(2.247,-0.485){11}{\rule{1.814pt}{0.117pt}}
\multiput(950.00,605.17)(26.234,-7.000){2}{\rule{0.907pt}{0.400pt}}
\multiput(980.00,597.93)(2.247,-0.485){11}{\rule{1.814pt}{0.117pt}}
\multiput(980.00,598.17)(26.234,-7.000){2}{\rule{0.907pt}{0.400pt}}
\multiput(1010.00,590.93)(2.323,-0.485){11}{\rule{1.871pt}{0.117pt}}
\multiput(1010.00,591.17)(27.116,-7.000){2}{\rule{0.936pt}{0.400pt}}
\multiput(1041.00,583.93)(2.660,-0.482){9}{\rule{2.100pt}{0.116pt}}
\multiput(1041.00,584.17)(25.641,-6.000){2}{\rule{1.050pt}{0.400pt}}
\multiput(1071.00,577.93)(2.751,-0.482){9}{\rule{2.167pt}{0.116pt}}
\multiput(1071.00,578.17)(26.503,-6.000){2}{\rule{1.083pt}{0.400pt}}
\multiput(1102.00,571.93)(2.660,-0.482){9}{\rule{2.100pt}{0.116pt}}
\multiput(1102.00,572.17)(25.641,-6.000){2}{\rule{1.050pt}{0.400pt}}
\multiput(1132.00,565.93)(2.660,-0.482){9}{\rule{2.100pt}{0.116pt}}
\multiput(1132.00,566.17)(25.641,-6.000){2}{\rule{1.050pt}{0.400pt}}
\multiput(1162.00,559.93)(2.751,-0.482){9}{\rule{2.167pt}{0.116pt}}
\multiput(1162.00,560.17)(26.503,-6.000){2}{\rule{1.083pt}{0.400pt}}
\multiput(1193.00,553.93)(3.270,-0.477){7}{\rule{2.500pt}{0.115pt}}
\multiput(1193.00,554.17)(24.811,-5.000){2}{\rule{1.250pt}{0.400pt}}
\multiput(1223.00,548.93)(2.751,-0.482){9}{\rule{2.167pt}{0.116pt}}
\multiput(1223.00,549.17)(26.503,-6.000){2}{\rule{1.083pt}{0.400pt}}
\multiput(1254.00,542.93)(3.270,-0.477){7}{\rule{2.500pt}{0.115pt}}
\multiput(1254.00,543.17)(24.811,-5.000){2}{\rule{1.250pt}{0.400pt}}
\multiput(1284.00,537.93)(2.660,-0.482){9}{\rule{2.100pt}{0.116pt}}
\multiput(1284.00,538.17)(25.641,-6.000){2}{\rule{1.050pt}{0.400pt}}
\multiput(1314.00,531.93)(3.382,-0.477){7}{\rule{2.580pt}{0.115pt}}
\multiput(1314.00,532.17)(25.645,-5.000){2}{\rule{1.290pt}{0.400pt}}
\multiput(1345.00,526.93)(3.270,-0.477){7}{\rule{2.500pt}{0.115pt}}
\multiput(1345.00,527.17)(24.811,-5.000){2}{\rule{1.250pt}{0.400pt}}
\multiput(1375.00,521.93)(3.382,-0.477){7}{\rule{2.580pt}{0.115pt}}
\multiput(1375.00,522.17)(25.645,-5.000){2}{\rule{1.290pt}{0.400pt}}
\multiput(1406.00,516.93)(3.270,-0.477){7}{\rule{2.500pt}{0.115pt}}
\multiput(1406.00,517.17)(24.811,-5.000){2}{\rule{1.250pt}{0.400pt}}
\put(372,750){\Dia}
\put(524,622){\Dia}
\put(524,686){\Dia}
\put(524,686){\Dia}
\put(676,513){\Dia}
\put(676,647){\Dia}
\put(676,580){\Dia}
\put(676,580){\Dia}
\put(676,622){\Dia}
\put(676,622){\Dia}
\put(828,420){\Dia}
\put(828,549){\Dia}
\put(828,537){\Dia}
\put(828,537){\Dia}
\put(828,482){\Dia}
\put(828,560){\Dia}
\put(828,602){\Dia}
\put(828,482){\Dia}
\put(828,560){\Dia}
\put(828,602){\Dia}
\put(980,340){\Dia}
\put(980,574){\Dia}
\put(980,456){\Dia}
\put(980,553){\Dia}
\put(980,553){\Dia}
\put(980,453){\Dia}
\put(980,453){\Dia}
\put(980,500){\Dia}
\put(980,500){\Dia}
\put(980,396){\Dia}
\put(980,495){\Dia}
\put(980,516){\Dia}
\put(980,396){\Dia}
\put(980,495){\Dia}
\put(980,516){\Dia}
\put(1132,271){\Dia}
\put(1132,492){\Dia}
\put(1132,374){\Dia}
\put(1132,451){\Dia}
\put(1132,451){\Dia}
\put(1132,373){\Dia}
\put(1132,482){\Dia}
\put(1132,373){\Dia}
\put(1132,482){\Dia}
\put(1132,444){\Dia}
\put(1132,503){\Dia}
\put(1132,539){\Dia}
\put(1132,423){\Dia}
\put(1132,431){\Dia}
\put(1132,320){\Dia}
\put(1132,444){\Dia}
\put(1132,503){\Dia}
\put(1132,539){\Dia}
\put(1132,423){\Dia}
\put(1132,431){\Dia}
\put(1132,320){\Dia}
\put(1284,210){\Dia}
\put(1284,301){\Dia}
\put(1284,411){\Dia}
\put(1284,517){\Dia}
\put(1284,499){\Dia}
\put(1284,499){\Dia}
\put(1284,394){\Dia}
\put(1284,452){\Dia}
\put(1284,394){\Dia}
\put(1284,452){\Dia}
\put(1284,302){\Dia}
\put(1284,390){\Dia}
\put(1284,407){\Dia}
\put(1284,302){\Dia}
\put(1284,390){\Dia}
\put(1284,407){\Dia}
\put(1284,408){\Dia}
\put(1284,446){\Dia}
\put(1284,466){\Dia}
\put(1284,352){\Dia}
\put(1284,353){\Dia}
\put(1284,253){\Dia}
\put(1284,408){\Dia}
\put(1284,446){\Dia}
\put(1284,466){\Dia}
\put(1284,352){\Dia}
\put(1284,353){\Dia}
\put(1284,253){\Dia}
\put(1436,155){\Dia}
\put(1436,236){\Dia}
\put(1436,335){\Dia}
\put(1436,446){\Dia}
\put(1436,408){\Dia}
\put(1436,408){\Dia}
\put(1436,332){\Dia}
\put(1436,332){\Dia}
\put(1436,238){\Dia}
\put(1436,335){\Dia}
\put(1436,437){\Dia}
\put(1436,365){\Dia}
\put(1436,238){\Dia}
\put(1436,335){\Dia}
\put(1436,437){\Dia}
\put(1436,365){\Dia}
\put(1436,340){\Dia}
\put(1436,402){\Dia}
\put(1436,456){\Dia}
\put(1436,489){\Dia}
\put(1436,363){\Dia}
\put(1436,382){\Dia}
\put(1436,390){\Dia}
\put(1436,285){\Dia}
\put(1436,284){\Dia}
\put(1436,340){\Dia}
\put(1436,402){\Dia}
\put(1436,456){\Dia}
\put(1436,489){\Dia}
\put(1436,363){\Dia}
\put(1436,382){\Dia}
\put(1436,390){\Dia}
\put(1436,285){\Dia}
\put(1436,284){\Dia}
\put(1436,194){\Dia}
\end{picture}
\end{center}
\caption{The holomophic energy for $n=4$ corresponding to the
polynomial solutions of the Baxter equation. The solid line
represents $\varepsilon_2(h)$.}
\label{fig2}
\end{figure}}

\def \figc {
\begin{figure}
\begin{center}
\setlength{\unitlength}{0.240900pt}
\ifx\plotpoint\undefined\newsavebox{\plotpoint}\fi
\sbox{\plotpoint}{\rule[-0.200pt]{0.400pt}{0.400pt}}%
\begin{picture}(1500,900)(0,0)
\font\gnuplot=cmr10 at 10pt
\gnuplot
\sbox{\plotpoint}{\rule[-0.200pt]{0.400pt}{0.400pt}}%
\put(828.0,113.0){\rule[-0.200pt]{0.400pt}{184.048pt}}
\put(220.0,113.0){\rule[-0.200pt]{4.818pt}{0.400pt}}
\put(198,113){\makebox(0,0)[r]{-15}}
\put(1416.0,113.0){\rule[-0.200pt]{4.818pt}{0.400pt}}
\put(220.0,198.0){\rule[-0.200pt]{4.818pt}{0.400pt}}
\put(198,198){\makebox(0,0)[r]{-14}}
\put(1416.0,198.0){\rule[-0.200pt]{4.818pt}{0.400pt}}
\put(220.0,283.0){\rule[-0.200pt]{4.818pt}{0.400pt}}
\put(198,283){\makebox(0,0)[r]{-13}}
\put(1416.0,283.0){\rule[-0.200pt]{4.818pt}{0.400pt}}
\put(220.0,368.0){\rule[-0.200pt]{4.818pt}{0.400pt}}
\put(198,368){\makebox(0,0)[r]{-12}}
\put(1416.0,368.0){\rule[-0.200pt]{4.818pt}{0.400pt}}
\put(220.0,453.0){\rule[-0.200pt]{4.818pt}{0.400pt}}
\put(198,453){\makebox(0,0)[r]{-11}}
\put(1416.0,453.0){\rule[-0.200pt]{4.818pt}{0.400pt}}
\put(220.0,537.0){\rule[-0.200pt]{4.818pt}{0.400pt}}
\put(198,537){\makebox(0,0)[r]{-10}}
\put(1416.0,537.0){\rule[-0.200pt]{4.818pt}{0.400pt}}
\put(220.0,622.0){\rule[-0.200pt]{4.818pt}{0.400pt}}
\put(198,622){\makebox(0,0)[r]{-9}}
\put(1416.0,622.0){\rule[-0.200pt]{4.818pt}{0.400pt}}
\put(220.0,707.0){\rule[-0.200pt]{4.818pt}{0.400pt}}
\put(198,707){\makebox(0,0)[r]{-8}}
\put(1416.0,707.0){\rule[-0.200pt]{4.818pt}{0.400pt}}
\put(220.0,792.0){\rule[-0.200pt]{4.818pt}{0.400pt}}
\put(198,792){\makebox(0,0)[r]{-7}}
\put(1416.0,792.0){\rule[-0.200pt]{4.818pt}{0.400pt}}
\put(220.0,877.0){\rule[-0.200pt]{4.818pt}{0.400pt}}
\put(198,877){\makebox(0,0)[r]{-6}}
\put(1416.0,877.0){\rule[-0.200pt]{4.818pt}{0.400pt}}
\put(220.0,113.0){\rule[-0.200pt]{0.400pt}{4.818pt}}
\put(220,68){\makebox(0,0){-200}}
\put(220.0,857.0){\rule[-0.200pt]{0.400pt}{4.818pt}}
\put(372.0,113.0){\rule[-0.200pt]{0.400pt}{4.818pt}}
\put(372,68){\makebox(0,0){-150}}
\put(372.0,857.0){\rule[-0.200pt]{0.400pt}{4.818pt}}
\put(524.0,113.0){\rule[-0.200pt]{0.400pt}{4.818pt}}
\put(524,68){\makebox(0,0){-100}}
\put(524.0,857.0){\rule[-0.200pt]{0.400pt}{4.818pt}}
\put(676.0,113.0){\rule[-0.200pt]{0.400pt}{4.818pt}}
\put(676,68){\makebox(0,0){-50}}
\put(676.0,857.0){\rule[-0.200pt]{0.400pt}{4.818pt}}
\put(828.0,113.0){\rule[-0.200pt]{0.400pt}{4.818pt}}
\put(828,68){\makebox(0,0){0}}
\put(828.0,857.0){\rule[-0.200pt]{0.400pt}{4.818pt}}
\put(980.0,113.0){\rule[-0.200pt]{0.400pt}{4.818pt}}
\put(980,68){\makebox(0,0){50}}
\put(980.0,857.0){\rule[-0.200pt]{0.400pt}{4.818pt}}
\put(1132.0,113.0){\rule[-0.200pt]{0.400pt}{4.818pt}}
\put(1132,68){\makebox(0,0){100}}
\put(1132.0,857.0){\rule[-0.200pt]{0.400pt}{4.818pt}}
\put(1284.0,113.0){\rule[-0.200pt]{0.400pt}{4.818pt}}
\put(1284,68){\makebox(0,0){150}}
\put(1284.0,857.0){\rule[-0.200pt]{0.400pt}{4.818pt}}
\put(1436.0,113.0){\rule[-0.200pt]{0.400pt}{4.818pt}}
\put(1436,68){\makebox(0,0){200}}
\put(1436.0,857.0){\rule[-0.200pt]{0.400pt}{4.818pt}}
\put(220.0,113.0){\rule[-0.200pt]{292.934pt}{0.400pt}}
\put(1436.0,113.0){\rule[-0.200pt]{0.400pt}{184.048pt}}
\put(220.0,877.0){\rule[-0.200pt]{292.934pt}{0.400pt}}
\put(45,495){\makebox(0,0){$\varepsilon_3$}}
\put(828,23){\makebox(0,0){$\CQ_3$}}
\put(220.0,113.0){\rule[-0.200pt]{0.400pt}{184.048pt}}
\put(828,877){\Dia}
\put(839,750){\Dia}
\put(817,750){\Dia}
\put(828,679){\Dia}
\put(860,636){\Dia}
\put(796,636){\Dia}
\put(849,601){\Dia}
\put(807,601){\Dia}
\put(895,537){\Dia}
\put(761,537){\Dia}
\put(828,554){\Dia}
\put(947,451){\Dia}
\put(884,522){\Dia}
\put(772,522){\Dia}
\put(709,451){\Dia}
\put(936,445){\Dia}
\put(720,445){\Dia}
\put(1021,376){\Dia}
\put(635,376){\Dia}
\put(863,498){\Dia}
\put(793,498){\Dia}
\put(828,463){\Dia}
\put(1119,310){\Dia}
\put(537,310){\Dia}
\put(1009,375){\Dia}
\put(647,375){\Dia}
\put(914,436){\Dia}
\put(742,436){\Dia}
\put(986,373){\Dia}
\put(670,373){\Dia}
\put(1245,251){\Dia}
\put(411,251){\Dia}
\put(879,419){\Dia}
\put(777,419){\Dia}
\put(1106,310){\Dia}
\put(550,310){\Dia}
\put(828,392){\Dia}
\put(1231,252){\Dia}
\put(425,252){\Dia}
\put(1402,198){\Dia}
\put(254,198){\Dia}
\put(949,368){\Dia}
\put(707,368){\Dia}
\put(1081,313){\Dia}
\put(575,313){\Dia}
\end{picture}
\end{center}
\caption{The holomorphic energy $\varepsilon_3$ for $n=3$ as a
function of quantized $\CQ_3$. The maximum value of the energy
$\varepsilon_3=-6$
corresponds to the degenerate solution $\CQ_3=0$.}
\label{fig3}
\end{figure}
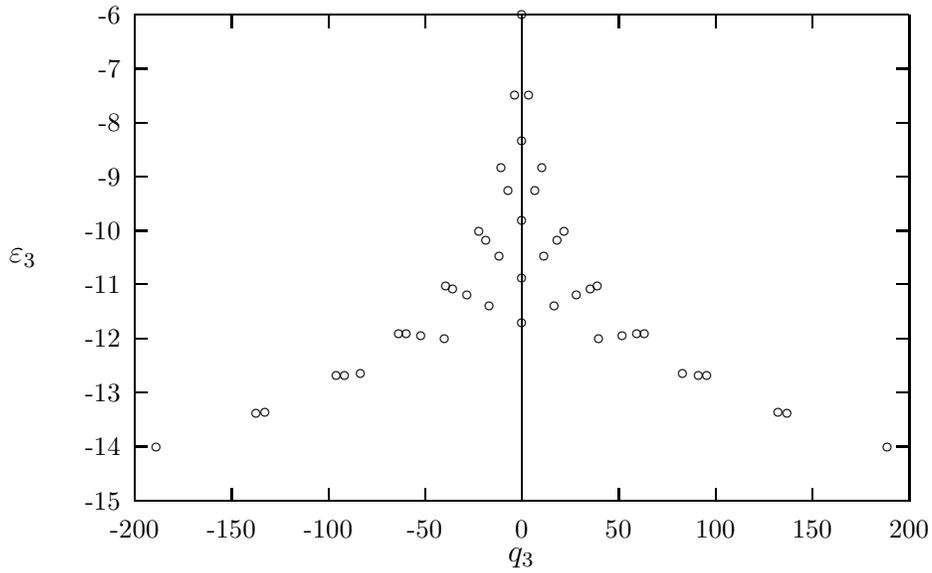}

\def \figd {
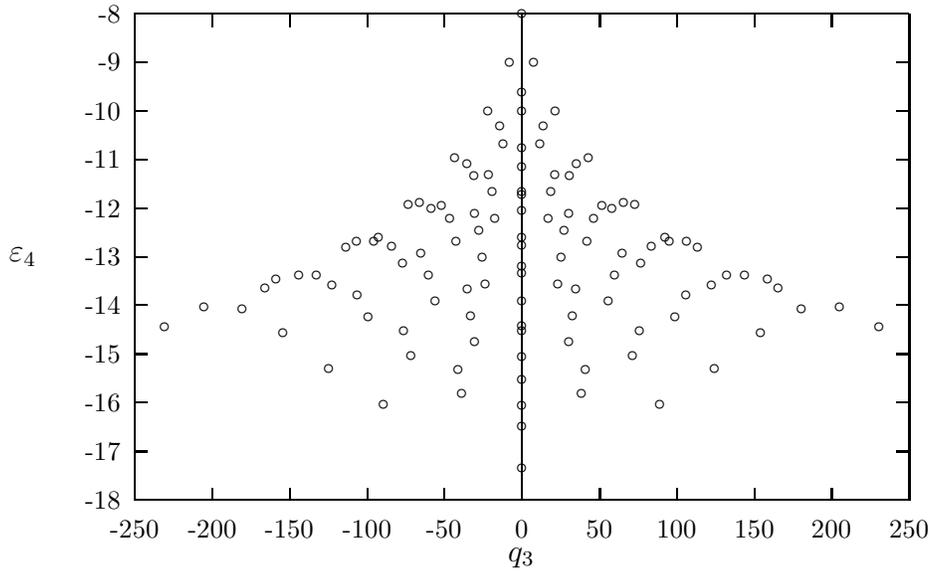
\begin{figure}
\begin{center}
\setlength{\unitlength}{0.240900pt}
\ifx\plotpoint\undefined\newsavebox{\plotpoint}\fi
\begin{picture}(1500,900)(0,0)
\font\gnuplot=cmr10 at 10pt
\gnuplot
\sbox{\plotpoint}{\rule[-0.200pt]{0.400pt}{0.400pt}}%
\put(828.0,113.0){\rule[-0.200pt]{0.400pt}{184.048pt}}
\put(220.0,113.0){\rule[-0.200pt]{4.818pt}{0.400pt}}
\put(198,113){\makebox(0,0)[r]{-18}}
\put(1416.0,113.0){\rule[-0.200pt]{4.818pt}{0.400pt}}
\put(220.0,189.0){\rule[-0.200pt]{4.818pt}{0.400pt}}
\put(198,189){\makebox(0,0)[r]{-17}}
\put(1416.0,189.0){\rule[-0.200pt]{4.818pt}{0.400pt}}
\put(220.0,266.0){\rule[-0.200pt]{4.818pt}{0.400pt}}
\put(198,266){\makebox(0,0)[r]{-16}}
\put(1416.0,266.0){\rule[-0.200pt]{4.818pt}{0.400pt}}
\put(220.0,342.0){\rule[-0.200pt]{4.818pt}{0.400pt}}
\put(198,342){\makebox(0,0)[r]{-15}}
\put(1416.0,342.0){\rule[-0.200pt]{4.818pt}{0.400pt}}
\put(220.0,419.0){\rule[-0.200pt]{4.818pt}{0.400pt}}
\put(198,419){\makebox(0,0)[r]{-14}}
\put(1416.0,419.0){\rule[-0.200pt]{4.818pt}{0.400pt}}
\put(220.0,495.0){\rule[-0.200pt]{4.818pt}{0.400pt}}
\put(198,495){\makebox(0,0)[r]{-13}}
\put(1416.0,495.0){\rule[-0.200pt]{4.818pt}{0.400pt}}
\put(220.0,571.0){\rule[-0.200pt]{4.818pt}{0.400pt}}
\put(198,571){\makebox(0,0)[r]{-12}}
\put(1416.0,571.0){\rule[-0.200pt]{4.818pt}{0.400pt}}
\put(220.0,648.0){\rule[-0.200pt]{4.818pt}{0.400pt}}
\put(198,648){\makebox(0,0)[r]{-11}}
\put(1416.0,648.0){\rule[-0.200pt]{4.818pt}{0.400pt}}
\put(220.0,724.0){\rule[-0.200pt]{4.818pt}{0.400pt}}
\put(198,724){\makebox(0,0)[r]{-10}}
\put(1416.0,724.0){\rule[-0.200pt]{4.818pt}{0.400pt}}
\put(220.0,801.0){\rule[-0.200pt]{4.818pt}{0.400pt}}
\put(198,801){\makebox(0,0)[r]{-9}}
\put(1416.0,801.0){\rule[-0.200pt]{4.818pt}{0.400pt}}
\put(220.0,877.0){\rule[-0.200pt]{4.818pt}{0.400pt}}
\put(198,877){\makebox(0,0)[r]{-8}}
\put(1416.0,877.0){\rule[-0.200pt]{4.818pt}{0.400pt}}
\put(220.0,113.0){\rule[-0.200pt]{0.400pt}{4.818pt}}
\put(220,68){\makebox(0,0){-250}}
\put(220.0,857.0){\rule[-0.200pt]{0.400pt}{4.818pt}}
\put(342.0,113.0){\rule[-0.200pt]{0.400pt}{4.818pt}}
\put(342,68){\makebox(0,0){-200}}
\put(342.0,857.0){\rule[-0.200pt]{0.400pt}{4.818pt}}
\put(463.0,113.0){\rule[-0.200pt]{0.400pt}{4.818pt}}
\put(463,68){\makebox(0,0){-150}}
\put(463.0,857.0){\rule[-0.200pt]{0.400pt}{4.818pt}}
\put(585.0,113.0){\rule[-0.200pt]{0.400pt}{4.818pt}}
\put(585,68){\makebox(0,0){-100}}
\put(585.0,857.0){\rule[-0.200pt]{0.400pt}{4.818pt}}
\put(706.0,113.0){\rule[-0.200pt]{0.400pt}{4.818pt}}
\put(706,68){\makebox(0,0){-50}}
\put(706.0,857.0){\rule[-0.200pt]{0.400pt}{4.818pt}}
\put(828.0,113.0){\rule[-0.200pt]{0.400pt}{4.818pt}}
\put(828,68){\makebox(0,0){0}}
\put(828.0,857.0){\rule[-0.200pt]{0.400pt}{4.818pt}}
\put(950.0,113.0){\rule[-0.200pt]{0.400pt}{4.818pt}}
\put(950,68){\makebox(0,0){50}}
\put(950.0,857.0){\rule[-0.200pt]{0.400pt}{4.818pt}}
\put(1071.0,113.0){\rule[-0.200pt]{0.400pt}{4.818pt}}
\put(1071,68){\makebox(0,0){100}}
\put(1071.0,857.0){\rule[-0.200pt]{0.400pt}{4.818pt}}
\put(1193.0,113.0){\rule[-0.200pt]{0.400pt}{4.818pt}}
\put(1193,68){\makebox(0,0){150}}
\put(1193.0,857.0){\rule[-0.200pt]{0.400pt}{4.818pt}}
\put(1314.0,113.0){\rule[-0.200pt]{0.400pt}{4.818pt}}
\put(1314,68){\makebox(0,0){200}}
\put(1314.0,857.0){\rule[-0.200pt]{0.400pt}{4.818pt}}
\put(1436.0,113.0){\rule[-0.200pt]{0.400pt}{4.818pt}}
\put(1436,68){\makebox(0,0){250}}
\put(1436.0,857.0){\rule[-0.200pt]{0.400pt}{4.818pt}}
\put(220.0,113.0){\rule[-0.200pt]{292.934pt}{0.400pt}}
\put(1436.0,113.0){\rule[-0.200pt]{0.400pt}{184.048pt}}
\put(220.0,877.0){\rule[-0.200pt]{292.934pt}{0.400pt}}
\put(45,495){\makebox(0,0){$\varepsilon_4$}}
\put(828,23){\makebox(0,0){$\CQ_3$}}
\put(220.0,113.0){\rule[-0.200pt]{0.400pt}{184.048pt}}
\put(828,877){\Dia}
\put(828,724){\Dia}
\put(847,801){\Dia}
\put(809,801){\Dia}
\put(828,593){\Dia}
\put(828,754){\Dia}
\put(857,673){\Dia}
\put(799,673){\Dia}
\put(881,724){\Dia}
\put(775,724){\Dia}
\put(828,481){\Dia}
\put(828,636){\Dia}
\put(903,622){\Dia}
\put(753,622){\Dia}
\put(870,556){\Dia}
\put(933,650){\Dia}
\put(862,700){\Dia}
\put(786,556){\Dia}
\put(723,650){\Dia}
\put(794,700){\Dia}
\put(828,386){\Dia}
\put(828,666){\Dia}
\put(828,525){\Dia}
\put(914,641){\Dia}
\put(742,641){\Dia}
\put(931,520){\Dia}
\put(725,520){\Dia}
\put(1006,578){\Dia}
\put(650,578){\Dia}
\put(885,452){\Dia}
\put(970,571){\Dia}
\put(874,597){\Dia}
\put(771,452){\Dia}
\put(686,571){\Dia}
\put(782,597){\Dia}
\put(828,303){\Dia}
\put(828,568){\Dia}
\put(828,426){\Dia}
\put(1060,519){\Dia}
\put(596,519){\Dia}
\put(964,425){\Dia}
\put(941,556){\Dia}
\put(692,425){\Dia}
\put(715,556){\Dia}
\put(1104,510){\Dia}
\put(988,581){\Dia}
\put(880,624){\Dia}
\put(1015,485){\Dia}
\put(890,495){\Dia}
\put(902,361){\Dia}
\put(552,510){\Dia}
\put(668,581){\Dia}
\put(776,624){\Dia}
\put(641,485){\Dia}
\put(766,495){\Dia}
\put(754,361){\Dia}
\put(828,229){\Dia}
\put(828,338){\Dia}
\put(828,470){\Dia}
\put(828,598){\Dia}
\put(954,576){\Dia}
\put(702,576){\Dia}
\put(1126,450){\Dia}
\put(1087,520){\Dia}
\put(530,450){\Dia}
\put(569,520){\Dia}
\put(1002,340){\Dia}
\put(1231,446){\Dia}
\put(974,466){\Dia}
\put(654,340){\Dia}
\put(425,446){\Dia}
\put(682,466){\Dia}
\put(1178,467){\Dia}
\put(1032,512){\Dia}
\put(895,536){\Dia}
\put(1069,400){\Dia}
\put(908,402){\Dia}
\put(922,281){\Dia}
\put(478,467){\Dia}
\put(624,512){\Dia}
\put(761,536){\Dia}
\put(587,400){\Dia}
\put(748,402){\Dia}
\put(734,281){\Dia}
\put(828,164){\Dia}
\put(828,261){\Dia}
\put(828,379){\Dia}
\put(828,513){\Dia}
\put(1150,467){\Dia}
\put(506,467){\Dia}
\put(1203,375){\Dia}
\put(453,375){\Dia}
\put(1045,263){\Dia}
\put(1013,379){\Dia}
\put(986,501){\Dia}
\put(1327,416){\Dia}
\put(611,263){\Dia}
\put(643,379){\Dia}
\put(670,501){\Dia}
\put(329,416){\Dia}
\put(1389,385){\Dia}
\put(1214,460){\Dia}
\put(1053,525){\Dia}
\put(902,564){\Dia}
\put(1267,413){\Dia}
\put(1086,435){\Dia}
\put(913,445){\Dia}
\put(1131,320){\Dia}
\put(928,318){\Dia}
\put(267,385){\Dia}
\put(442,460){\Dia}
\put(603,525){\Dia}
\put(754,564){\Dia}
\put(389,413){\Dia}
\put(570,435){\Dia}
\put(743,445){\Dia}
\put(525,320){\Dia}
\put(728,318){\Dia}
\end{picture}
\end{center}
\caption{The holomorphic energy $\varepsilon_4$ for $n=4$ as a
function of quantized $\CQ_3$. The maximum value of energy
$\varepsilon_4=-8$
corresponds to
$\CQ_3=0$ but in contrast with fig.~\ref{fig3}
this solution is not degenerate
because it corresponds to $\CQ_4=2$.}
\label{fig4}
\end{figure}}

\def \fige {
\begin{figure}
\begin{center}
\setlength{\unitlength}{0.240900pt}
\ifx\plotpoint\undefined\newsavebox{\plotpoint}\fi
\begin{picture}(1500,900)(0,0)
\font\gnuplot=cmr10 at 10pt
\gnuplot
\sbox{\plotpoint}{\rule[-0.200pt]{0.400pt}{0.400pt}}%
\put(524.0,113.0){\rule[-0.200pt]{0.400pt}{184.048pt}}
\put(220.0,113.0){\rule[-0.200pt]{4.818pt}{0.400pt}}
\put(198,113){\makebox(0,0)[r]{-18}}
\put(1416.0,113.0){\rule[-0.200pt]{4.818pt}{0.400pt}}
\put(220.0,189.0){\rule[-0.200pt]{4.818pt}{0.400pt}}
\put(198,189){\makebox(0,0)[r]{-17}}
\put(1416.0,189.0){\rule[-0.200pt]{4.818pt}{0.400pt}}
\put(220.0,266.0){\rule[-0.200pt]{4.818pt}{0.400pt}}
\put(198,266){\makebox(0,0)[r]{-16}}
\put(1416.0,266.0){\rule[-0.200pt]{4.818pt}{0.400pt}}
\put(220.0,342.0){\rule[-0.200pt]{4.818pt}{0.400pt}}
\put(198,342){\makebox(0,0)[r]{-15}}
\put(1416.0,342.0){\rule[-0.200pt]{4.818pt}{0.400pt}}
\put(220.0,419.0){\rule[-0.200pt]{4.818pt}{0.400pt}}
\put(198,419){\makebox(0,0)[r]{-14}}
\put(1416.0,419.0){\rule[-0.200pt]{4.818pt}{0.400pt}}
\put(220.0,495.0){\rule[-0.200pt]{4.818pt}{0.400pt}}
\put(198,495){\makebox(0,0)[r]{-13}}
\put(1416.0,495.0){\rule[-0.200pt]{4.818pt}{0.400pt}}
\put(220.0,571.0){\rule[-0.200pt]{4.818pt}{0.400pt}}
\put(198,571){\makebox(0,0)[r]{-12}}
\put(1416.0,571.0){\rule[-0.200pt]{4.818pt}{0.400pt}}
\put(220.0,648.0){\rule[-0.200pt]{4.818pt}{0.400pt}}
\put(198,648){\makebox(0,0)[r]{-11}}
\put(1416.0,648.0){\rule[-0.200pt]{4.818pt}{0.400pt}}
\put(220.0,724.0){\rule[-0.200pt]{4.818pt}{0.400pt}}
\put(198,724){\makebox(0,0)[r]{-10}}
\put(1416.0,724.0){\rule[-0.200pt]{4.818pt}{0.400pt}}
\put(220.0,801.0){\rule[-0.200pt]{4.818pt}{0.400pt}}
\put(198,801){\makebox(0,0)[r]{-9}}
\put(1416.0,801.0){\rule[-0.200pt]{4.818pt}{0.400pt}}
\put(220.0,877.0){\rule[-0.200pt]{4.818pt}{0.400pt}}
\put(198,877){\makebox(0,0)[r]{-8}}
\put(1416.0,877.0){\rule[-0.200pt]{4.818pt}{0.400pt}}
\put(220.0,113.0){\rule[-0.200pt]{0.400pt}{4.818pt}}
\put(220,68){\makebox(0,0){-200}}
\put(220.0,857.0){\rule[-0.200pt]{0.400pt}{4.818pt}}
\put(372.0,113.0){\rule[-0.200pt]{0.400pt}{4.818pt}}
\put(372,68){\makebox(0,0){-100}}
\put(372.0,857.0){\rule[-0.200pt]{0.400pt}{4.818pt}}
\put(524.0,113.0){\rule[-0.200pt]{0.400pt}{4.818pt}}
\put(524,68){\makebox(0,0){0}}
\put(524.0,857.0){\rule[-0.200pt]{0.400pt}{4.818pt}}
\put(676.0,113.0){\rule[-0.200pt]{0.400pt}{4.818pt}}
\put(676,68){\makebox(0,0){100}}
\put(676.0,857.0){\rule[-0.200pt]{0.400pt}{4.818pt}}
\put(828.0,113.0){\rule[-0.200pt]{0.400pt}{4.818pt}}
\put(828,68){\makebox(0,0){200}}
\put(828.0,857.0){\rule[-0.200pt]{0.400pt}{4.818pt}}
\put(980.0,113.0){\rule[-0.200pt]{0.400pt}{4.818pt}}
\put(980,68){\makebox(0,0){300}}
\put(980.0,857.0){\rule[-0.200pt]{0.400pt}{4.818pt}}
\put(1132.0,113.0){\rule[-0.200pt]{0.400pt}{4.818pt}}
\put(1132,68){\makebox(0,0){400}}
\put(1132.0,857.0){\rule[-0.200pt]{0.400pt}{4.818pt}}
\put(1284.0,113.0){\rule[-0.200pt]{0.400pt}{4.818pt}}
\put(1284,68){\makebox(0,0){500}}
\put(1284.0,857.0){\rule[-0.200pt]{0.400pt}{4.818pt}}
\put(1436.0,113.0){\rule[-0.200pt]{0.400pt}{4.818pt}}
\put(1436,68){\makebox(0,0){600}}
\put(1436.0,857.0){\rule[-0.200pt]{0.400pt}{4.818pt}}
\put(220.0,113.0){\rule[-0.200pt]{292.934pt}{0.400pt}}
\put(1436.0,113.0){\rule[-0.200pt]{0.400pt}{184.048pt}}
\put(220.0,877.0){\rule[-0.200pt]{292.934pt}{0.400pt}}
\put(45,495){\makebox(0,0){$\varepsilon_4$}}
\put(828,23){\makebox(0,0){$\CQ_4$}}
\put(220.0,113.0){\rule[-0.200pt]{0.400pt}{184.048pt}}
\put(527,877){\Dia}
\put(539,724){\Dia}
\put(527,801){\Dia}
\put(527,801){\Dia}
\put(568,593){\Dia}
\put(529,754){\Dia}
\put(542,673){\Dia}
\put(542,673){\Dia}
\put(524,724){\Dia}
\put(524,724){\Dia}
\put(622,481){\Dia}
\put(548,636){\Dia}
\put(542,622){\Dia}
\put(542,622){\Dia}
\put(576,556){\Dia}
\put(516,650){\Dia}
\put(529,700){\Dia}
\put(576,556){\Dia}
\put(516,650){\Dia}
\put(529,700){\Dia}
\put(714,386){\Dia}
\put(531,666){\Dia}
\put(588,525){\Dia}
\put(524,641){\Dia}
\put(524,641){\Dia}
\put(582,520){\Dia}
\put(582,520){\Dia}
\put(500,578){\Dia}
\put(500,578){\Dia}
\put(639,452){\Dia}
\put(537,571){\Dia}
\put(551,597){\Dia}
\put(639,452){\Dia}
\put(537,571){\Dia}
\put(551,597){\Dia}
\put(858,303){\Dia}
\put(558,568){\Dia}
\put(660,426){\Dia}
\put(524,519){\Dia}
\put(524,519){\Dia}
\put(653,425){\Dia}
\put(550,556){\Dia}
\put(653,425){\Dia}
\put(550,556){\Dia}
\put(472,510){\Dia}
\put(513,581){\Dia}
\put(531,624){\Dia}
\put(582,485){\Dia}
\put(597,495){\Dia}
\put(744,361){\Dia}
\put(472,510){\Dia}
\put(513,581){\Dia}
\put(531,624){\Dia}
\put(582,485){\Dia}
\put(597,495){\Dia}
\put(744,361){\Dia}
\put(1072,229){\Dia}
\put(779,338){\Dia}
\put(611,470){\Dia}
\put(534,598){\Dia}
\put(524,576){\Dia}
\put(524,576){\Dia}
\put(575,450){\Dia}
\put(491,520){\Dia}
\put(575,450){\Dia}
\put(491,520){\Dia}
\put(770,340){\Dia}
\put(430,446){\Dia}
\put(602,466){\Dia}
\put(770,340){\Dia}
\put(430,446){\Dia}
\put(602,466){\Dia}
\put(500,467){\Dia}
\put(542,512){\Dia}
\put(561,536){\Dia}
\put(662,400){\Dia}
\put(679,402){\Dia}
\put(905,281){\Dia}
\put(500,467){\Dia}
\put(542,512){\Dia}
\put(561,536){\Dia}
\put(662,400){\Dia}
\put(679,402){\Dia}
\put(905,281){\Dia}
\put(1373,164){\Dia}
\put(958,261){\Dia}
\put(704,379){\Dia}
\put(569,513){\Dia}
\put(524,467){\Dia}
\put(524,467){\Dia}
\put(664,375){\Dia}
\put(664,375){\Dia}
\put(948,263){\Dia}
\put(694,379){\Dia}
\put(558,501){\Dia}
\put(461,416){\Dia}
\put(948,263){\Dia}
\put(694,379){\Dia}
\put(558,501){\Dia}
\put(461,416){\Dia}
\put(369,385){\Dia}
\put(457,460){\Dia}
\put(509,525){\Dia}
\put(533,564){\Dia}
\put(557,413){\Dia}
\put(601,435){\Dia}
\put(621,445){\Dia}
\put(792,320){\Dia}
\put(811,318){\Dia}
\put(369,385){\Dia}
\put(457,460){\Dia}
\put(509,525){\Dia}
\put(533,564){\Dia}
\put(557,413){\Dia}
\put(601,435){\Dia}
\put(621,445){\Dia}
\put(792,320){\Dia}
\put(811,318){\Dia}
\end{picture}
\end{center}
\caption{The holomorphic energy $\varepsilon_4$ for $n=4$ as a
function of quantized $\CQ_4$. The maximum value of the energy
$\varepsilon_4=-8$
corresponds to $\CQ_4=2$.}
\label{fig5}
\end{figure}
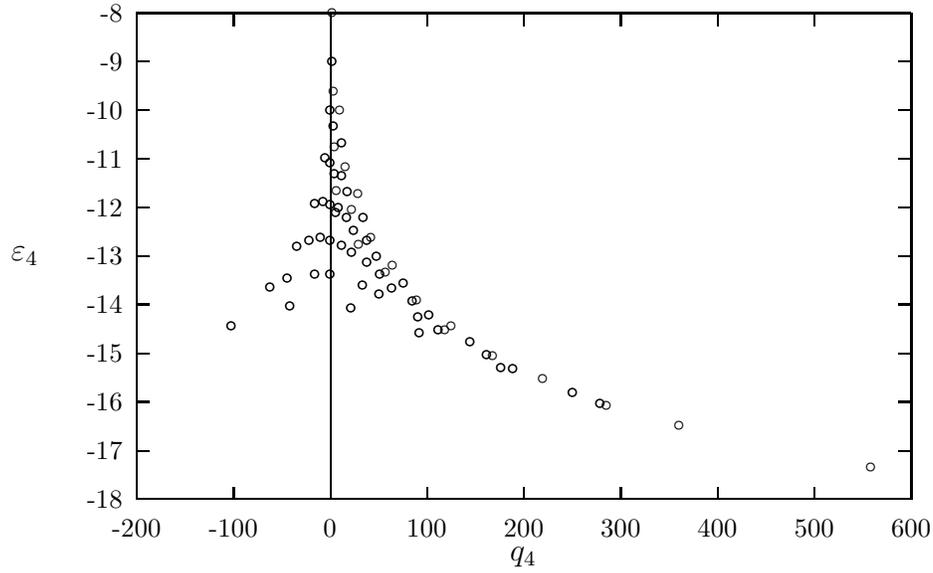}

\def \figf {
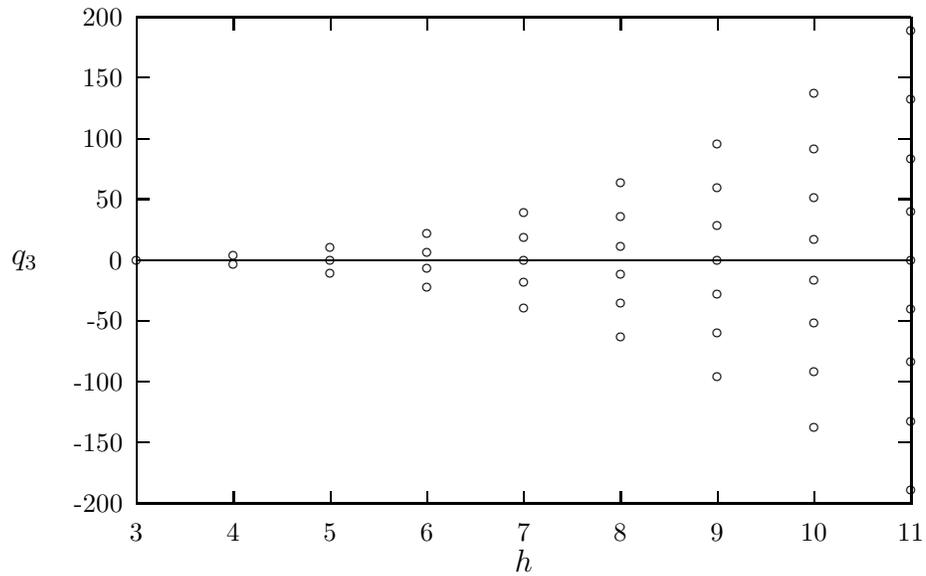
\begin{figure}
\begin{center}
\setlength{\unitlength}{0.240900pt}
\ifx\plotpoint\undefined\newsavebox{\plotpoint}\fi
\sbox{\plotpoint}{\rule[-0.200pt]{0.400pt}{0.400pt}}%
\begin{picture}(1500,900)(0,0)
\font\gnuplot=cmr10 at 10pt
\gnuplot
\sbox{\plotpoint}{\rule[-0.200pt]{0.400pt}{0.400pt}}%
\put(220.0,495.0){\rule[-0.200pt]{292.934pt}{0.400pt}}
\put(220.0,113.0){\rule[-0.200pt]{4.818pt}{0.400pt}}
\put(198,113){\makebox(0,0)[r]{-200}}
\put(1416.0,113.0){\rule[-0.200pt]{4.818pt}{0.400pt}}
\put(220.0,209.0){\rule[-0.200pt]{4.818pt}{0.400pt}}
\put(198,209){\makebox(0,0)[r]{-150}}
\put(1416.0,209.0){\rule[-0.200pt]{4.818pt}{0.400pt}}
\put(220.0,304.0){\rule[-0.200pt]{4.818pt}{0.400pt}}
\put(198,304){\makebox(0,0)[r]{-100}}
\put(1416.0,304.0){\rule[-0.200pt]{4.818pt}{0.400pt}}
\put(220.0,400.0){\rule[-0.200pt]{4.818pt}{0.400pt}}
\put(198,400){\makebox(0,0)[r]{-50}}
\put(1416.0,400.0){\rule[-0.200pt]{4.818pt}{0.400pt}}
\put(220.0,495.0){\rule[-0.200pt]{4.818pt}{0.400pt}}
\put(198,495){\makebox(0,0)[r]{0}}
\put(1416.0,495.0){\rule[-0.200pt]{4.818pt}{0.400pt}}
\put(220.0,591.0){\rule[-0.200pt]{4.818pt}{0.400pt}}
\put(198,591){\makebox(0,0)[r]{50}}
\put(1416.0,591.0){\rule[-0.200pt]{4.818pt}{0.400pt}}
\put(220.0,686.0){\rule[-0.200pt]{4.818pt}{0.400pt}}
\put(198,686){\makebox(0,0)[r]{100}}
\put(1416.0,686.0){\rule[-0.200pt]{4.818pt}{0.400pt}}
\put(220.0,782.0){\rule[-0.200pt]{4.818pt}{0.400pt}}
\put(198,782){\makebox(0,0)[r]{150}}
\put(1416.0,782.0){\rule[-0.200pt]{4.818pt}{0.400pt}}
\put(220.0,877.0){\rule[-0.200pt]{4.818pt}{0.400pt}}
\put(198,877){\makebox(0,0)[r]{200}}
\put(1416.0,877.0){\rule[-0.200pt]{4.818pt}{0.400pt}}
\put(220.0,113.0){\rule[-0.200pt]{0.400pt}{4.818pt}}
\put(220,68){\makebox(0,0){3}}
\put(220.0,857.0){\rule[-0.200pt]{0.400pt}{4.818pt}}
\put(372.0,113.0){\rule[-0.200pt]{0.400pt}{4.818pt}}
\put(372,68){\makebox(0,0){4}}
\put(372.0,857.0){\rule[-0.200pt]{0.400pt}{4.818pt}}
\put(524.0,113.0){\rule[-0.200pt]{0.400pt}{4.818pt}}
\put(524,68){\makebox(0,0){5}}
\put(524.0,857.0){\rule[-0.200pt]{0.400pt}{4.818pt}}
\put(676.0,113.0){\rule[-0.200pt]{0.400pt}{4.818pt}}
\put(676,68){\makebox(0,0){6}}
\put(676.0,857.0){\rule[-0.200pt]{0.400pt}{4.818pt}}
\put(828.0,113.0){\rule[-0.200pt]{0.400pt}{4.818pt}}
\put(828,68){\makebox(0,0){7}}
\put(828.0,857.0){\rule[-0.200pt]{0.400pt}{4.818pt}}
\put(980.0,113.0){\rule[-0.200pt]{0.400pt}{4.818pt}}
\put(980,68){\makebox(0,0){8}}
\put(980.0,857.0){\rule[-0.200pt]{0.400pt}{4.818pt}}
\put(1132.0,113.0){\rule[-0.200pt]{0.400pt}{4.818pt}}
\put(1132,68){\makebox(0,0){9}}
\put(1132.0,857.0){\rule[-0.200pt]{0.400pt}{4.818pt}}
\put(1284.0,113.0){\rule[-0.200pt]{0.400pt}{4.818pt}}
\put(1284,68){\makebox(0,0){10}}
\put(1284.0,857.0){\rule[-0.200pt]{0.400pt}{4.818pt}}
\put(1436.0,113.0){\rule[-0.200pt]{0.400pt}{4.818pt}}
\put(1436,68){\makebox(0,0){11}}
\put(1436.0,857.0){\rule[-0.200pt]{0.400pt}{4.818pt}}
\put(220.0,113.0){\rule[-0.200pt]{292.934pt}{0.400pt}}
\put(1436.0,113.0){\rule[-0.200pt]{0.400pt}{184.048pt}}
\put(220.0,877.0){\rule[-0.200pt]{292.934pt}{0.400pt}}
\put(45,495){\makebox(0,0){$\CQ_3$}}
\put(828,23){\makebox(0,0){$h$}}
\put(220.0,113.0){\rule[-0.200pt]{0.400pt}{184.048pt}}
\put(220,495){\Dia}
\put(372,502){\Dia}
\put(372,488){\Dia}
\put(524,495){\Dia}
\put(524,515){\Dia}
\put(524,475){\Dia}
\put(676,508){\Dia}
\put(676,482){\Dia}
\put(676,537){\Dia}
\put(676,453){\Dia}
\put(828,495){\Dia}
\put(828,570){\Dia}
\put(828,530){\Dia}
\put(828,460){\Dia}
\put(828,420){\Dia}
\put(980,563){\Dia}
\put(980,427){\Dia}
\put(980,616){\Dia}
\put(980,374){\Dia}
\put(980,517){\Dia}
\put(980,473){\Dia}
\put(1132,495){\Dia}
\put(1132,678){\Dia}
\put(1132,312){\Dia}
\put(1132,609){\Dia}
\put(1132,381){\Dia}
\put(1132,549){\Dia}
\put(1132,441){\Dia}
\put(1284,594){\Dia}
\put(1284,396){\Dia}
\put(1284,757){\Dia}
\put(1284,233){\Dia}
\put(1284,527){\Dia}
\put(1284,463){\Dia}
\put(1284,670){\Dia}
\put(1284,320){\Dia}
\put(1436,495){\Dia}
\put(1436,748){\Dia}
\put(1436,242){\Dia}
\put(1436,856){\Dia}
\put(1436,134){\Dia}
\put(1436,571){\Dia}
\put(1436,419){\Dia}
\put(1436,654){\Dia}
\put(1436,336){\Dia}
\end{picture}
\end{center}
\caption{Quantized values of $\CQ_3$ as a function of conformal
weight $h$ for $n=3$ Baxter equation. The symmetry $\CQ_3\to -\CQ_3$
is in agreement with \re{Q-sym} and \re{1-h}. The points $\CQ_3=0$
correspond to the degenerate solutions. The distribution of $\CQ_3$
for $n=4$ Baxter equation follows the same pattern.}
\label{fig6}
\end{figure}}

\def \figg {
\begin{figure}
\begin{center}
\setlength{\unitlength}{0.240900pt}
\ifx\plotpoint\undefined\newsavebox{\plotpoint}\fi
\begin{picture}(1500,900)(0,0)
\font\gnuplot=cmr10 at 10pt
\gnuplot
\sbox{\plotpoint}{\rule[-0.200pt]{0.400pt}{0.400pt}}%
\put(220.0,304.0){\rule[-0.200pt]{292.934pt}{0.400pt}}
\put(220.0,113.0){\rule[-0.200pt]{4.818pt}{0.400pt}}
\put(198,113){\makebox(0,0)[r]{-200}}
\put(1416.0,113.0){\rule[-0.200pt]{4.818pt}{0.400pt}}
\put(220.0,209.0){\rule[-0.200pt]{4.818pt}{0.400pt}}
\put(198,209){\makebox(0,0)[r]{-100}}
\put(1416.0,209.0){\rule[-0.200pt]{4.818pt}{0.400pt}}
\put(220.0,304.0){\rule[-0.200pt]{4.818pt}{0.400pt}}
\put(198,304){\makebox(0,0)[r]{0}}
\put(1416.0,304.0){\rule[-0.200pt]{4.818pt}{0.400pt}}
\put(220.0,400.0){\rule[-0.200pt]{4.818pt}{0.400pt}}
\put(198,400){\makebox(0,0)[r]{100}}
\put(1416.0,400.0){\rule[-0.200pt]{4.818pt}{0.400pt}}
\put(220.0,495.0){\rule[-0.200pt]{4.818pt}{0.400pt}}
\put(198,495){\makebox(0,0)[r]{200}}
\put(1416.0,495.0){\rule[-0.200pt]{4.818pt}{0.400pt}}
\put(220.0,591.0){\rule[-0.200pt]{4.818pt}{0.400pt}}
\put(198,591){\makebox(0,0)[r]{300}}
\put(1416.0,591.0){\rule[-0.200pt]{4.818pt}{0.400pt}}
\put(220.0,686.0){\rule[-0.200pt]{4.818pt}{0.400pt}}
\put(198,686){\makebox(0,0)[r]{400}}
\put(1416.0,686.0){\rule[-0.200pt]{4.818pt}{0.400pt}}
\put(220.0,782.0){\rule[-0.200pt]{4.818pt}{0.400pt}}
\put(198,782){\makebox(0,0)[r]{500}}
\put(1416.0,782.0){\rule[-0.200pt]{4.818pt}{0.400pt}}
\put(220.0,877.0){\rule[-0.200pt]{4.818pt}{0.400pt}}
\put(198,877){\makebox(0,0)[r]{600}}
\put(1416.0,877.0){\rule[-0.200pt]{4.818pt}{0.400pt}}
\put(220.0,113.0){\rule[-0.200pt]{0.400pt}{4.818pt}}
\put(220,68){\makebox(0,0){4}}
\put(220.0,857.0){\rule[-0.200pt]{0.400pt}{4.818pt}}
\put(394.0,113.0){\rule[-0.200pt]{0.400pt}{4.818pt}}
\put(394,68){\makebox(0,0){5}}
\put(394.0,857.0){\rule[-0.200pt]{0.400pt}{4.818pt}}
\put(567.0,113.0){\rule[-0.200pt]{0.400pt}{4.818pt}}
\put(567,68){\makebox(0,0){6}}
\put(567.0,857.0){\rule[-0.200pt]{0.400pt}{4.818pt}}
\put(741.0,113.0){\rule[-0.200pt]{0.400pt}{4.818pt}}
\put(741,68){\makebox(0,0){7}}
\put(741.0,857.0){\rule[-0.200pt]{0.400pt}{4.818pt}}
\put(915.0,113.0){\rule[-0.200pt]{0.400pt}{4.818pt}}
\put(915,68){\makebox(0,0){8}}
\put(915.0,857.0){\rule[-0.200pt]{0.400pt}{4.818pt}}
\put(1089.0,113.0){\rule[-0.200pt]{0.400pt}{4.818pt}}
\put(1089,68){\makebox(0,0){9}}
\put(1089.0,857.0){\rule[-0.200pt]{0.400pt}{4.818pt}}
\put(1262.0,113.0){\rule[-0.200pt]{0.400pt}{4.818pt}}
\put(1262,68){\makebox(0,0){10}}
\put(1262.0,857.0){\rule[-0.200pt]{0.400pt}{4.818pt}}
\put(1436.0,113.0){\rule[-0.200pt]{0.400pt}{4.818pt}}
\put(1436,68){\makebox(0,0){11}}
\put(1436.0,857.0){\rule[-0.200pt]{0.400pt}{4.818pt}}
\put(220.0,113.0){\rule[-0.200pt]{292.934pt}{0.400pt}}
\put(1436.0,113.0){\rule[-0.200pt]{0.400pt}{184.048pt}}
\put(220.0,877.0){\rule[-0.200pt]{292.934pt}{0.400pt}}
\put(45,495){\makebox(0,0){$\CQ_4$}}
\put(828,23){\makebox(0,0){$h$}}
\put(220.0,113.0){\rule[-0.200pt]{0.400pt}{184.048pt}}
\put(220,306){\Dia}
\put(394,314){\Dia}
\put(394,306){\Dia}
\put(394,306){\Dia}
\put(567,331){\Dia}
\put(567,307){\Dia}
\put(567,315){\Dia}
\put(567,315){\Dia}
\put(567,304){\Dia}
\put(567,304){\Dia}
\put(741,365){\Dia}
\put(741,319){\Dia}
\put(741,315){\Dia}
\put(741,315){\Dia}
\put(741,337){\Dia}
\put(741,299){\Dia}
\put(741,307){\Dia}
\put(741,337){\Dia}
\put(741,299){\Dia}
\put(741,307){\Dia}
\put(915,423){\Dia}
\put(915,309){\Dia}
\put(915,344){\Dia}
\put(915,304){\Dia}
\put(915,304){\Dia}
\put(915,340){\Dia}
\put(915,340){\Dia}
\put(915,289){\Dia}
\put(915,289){\Dia}
\put(915,376){\Dia}
\put(915,312){\Dia}
\put(915,321){\Dia}
\put(915,376){\Dia}
\put(915,312){\Dia}
\put(915,321){\Dia}
\put(1089,514){\Dia}
\put(1089,325){\Dia}
\put(1089,390){\Dia}
\put(1089,304){\Dia}
\put(1089,304){\Dia}
\put(1089,385){\Dia}
\put(1089,320){\Dia}
\put(1089,385){\Dia}
\put(1089,320){\Dia}
\put(1089,272){\Dia}
\put(1089,297){\Dia}
\put(1089,308){\Dia}
\put(1089,341){\Dia}
\put(1089,350){\Dia}
\put(1089,442){\Dia}
\put(1089,272){\Dia}
\put(1089,297){\Dia}
\put(1089,308){\Dia}
\put(1089,341){\Dia}
\put(1089,350){\Dia}
\put(1089,442){\Dia}
\put(1262,648){\Dia}
\put(1262,464){\Dia}
\put(1262,359){\Dia}
\put(1262,310){\Dia}
\put(1262,304){\Dia}
\put(1262,304){\Dia}
\put(1262,336){\Dia}
\put(1262,283){\Dia}
\put(1262,336){\Dia}
\put(1262,283){\Dia}
\put(1262,459){\Dia}
\put(1262,245){\Dia}
\put(1262,353){\Dia}
\put(1262,459){\Dia}
\put(1262,245){\Dia}
\put(1262,353){\Dia}
\put(1262,289){\Dia}
\put(1262,315){\Dia}
\put(1262,327){\Dia}
\put(1262,391){\Dia}
\put(1262,402){\Dia}
\put(1262,543){\Dia}
\put(1262,289){\Dia}
\put(1262,315){\Dia}
\put(1262,327){\Dia}
\put(1262,391){\Dia}
\put(1262,402){\Dia}
\put(1262,543){\Dia}
\put(1436,838){\Dia}
\put(1436,577){\Dia}
\put(1436,417){\Dia}
\put(1436,332){\Dia}
\put(1436,304){\Dia}
\put(1436,304){\Dia}
\put(1436,392){\Dia}
\put(1436,392){\Dia}
\put(1436,571){\Dia}
\put(1436,411){\Dia}
\put(1436,325){\Dia}
\put(1436,264){\Dia}
\put(1436,571){\Dia}
\put(1436,411){\Dia}
\put(1436,325){\Dia}
\put(1436,264){\Dia}
\put(1436,207){\Dia}
\put(1436,262){\Dia}
\put(1436,295){\Dia}
\put(1436,310){\Dia}
\put(1436,325){\Dia}
\put(1436,352){\Dia}
\put(1436,365){\Dia}
\put(1436,472){\Dia}
\put(1436,484){\Dia}
\put(1436,691){\Dia}
\put(1436,207){\Dia}
\put(1436,262){\Dia}
\put(1436,295){\Dia}
\put(1436,310){\Dia}
\put(1436,325){\Dia}
\put(1436,352){\Dia}
\put(1436,365){\Dia}
\put(1436,472){\Dia}
\put(1436,484){\Dia}
\put(1436,691){\Dia}
\end{picture}
\end{center}
\caption{Quantized values of $\CQ_4$ as a function of
conformal weight $h$ for $n=4$ Baxter equation.}
\label{fig7}
\end{figure}
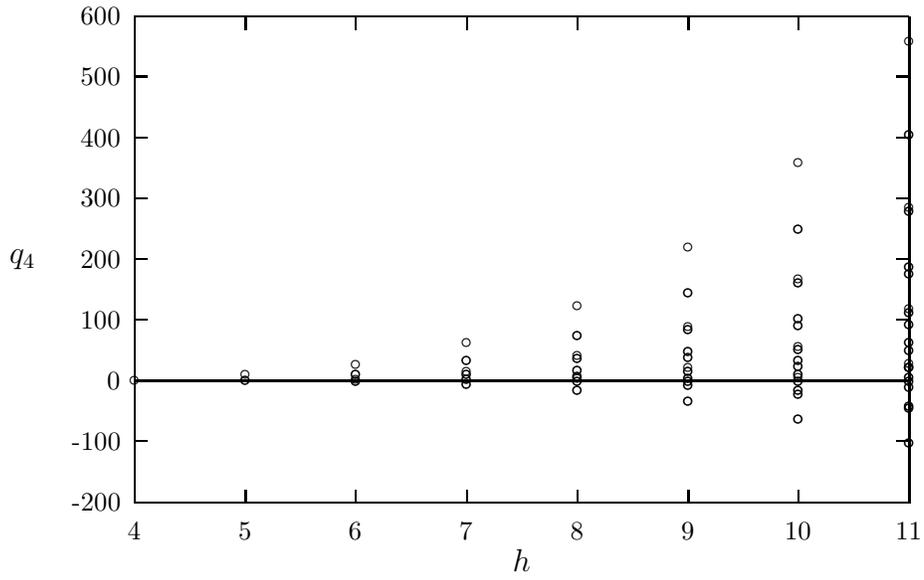}

\def \figh {
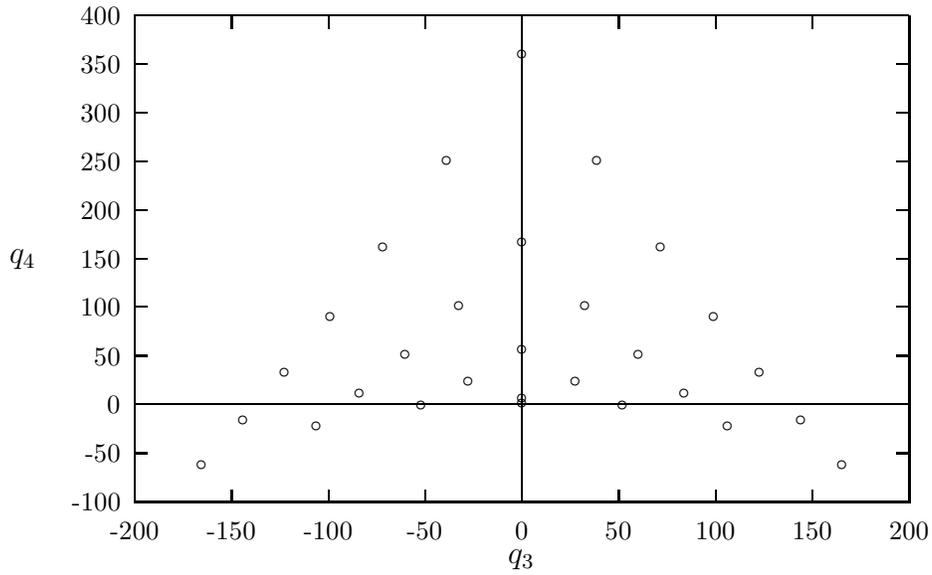
\begin{figure}
\begin{center}
\setlength{\unitlength}{0.240900pt}
\ifx\plotpoint\undefined\newsavebox{\plotpoint}\fi
\begin{picture}(1500,900)(0,0)
\font\gnuplot=cmr10 at 10pt
\gnuplot
\sbox{\plotpoint}{\rule[-0.200pt]{0.400pt}{0.400pt}}%
\put(220.0,266.0){\rule[-0.200pt]{292.934pt}{0.400pt}}
\put(828.0,113.0){\rule[-0.200pt]{0.400pt}{184.048pt}}
\put(220.0,113.0){\rule[-0.200pt]{4.818pt}{0.400pt}}
\put(198,113){\makebox(0,0)[r]{-100}}
\put(1416.0,113.0){\rule[-0.200pt]{4.818pt}{0.400pt}}
\put(220.0,189.0){\rule[-0.200pt]{4.818pt}{0.400pt}}
\put(198,189){\makebox(0,0)[r]{-50}}
\put(1416.0,189.0){\rule[-0.200pt]{4.818pt}{0.400pt}}
\put(220.0,266.0){\rule[-0.200pt]{4.818pt}{0.400pt}}
\put(198,266){\makebox(0,0)[r]{0}}
\put(1416.0,266.0){\rule[-0.200pt]{4.818pt}{0.400pt}}
\put(220.0,342.0){\rule[-0.200pt]{4.818pt}{0.400pt}}
\put(198,342){\makebox(0,0)[r]{50}}
\put(1416.0,342.0){\rule[-0.200pt]{4.818pt}{0.400pt}}
\put(220.0,419.0){\rule[-0.200pt]{4.818pt}{0.400pt}}
\put(198,419){\makebox(0,0)[r]{100}}
\put(1416.0,419.0){\rule[-0.200pt]{4.818pt}{0.400pt}}
\put(220.0,495.0){\rule[-0.200pt]{4.818pt}{0.400pt}}
\put(198,495){\makebox(0,0)[r]{150}}
\put(1416.0,495.0){\rule[-0.200pt]{4.818pt}{0.400pt}}
\put(220.0,571.0){\rule[-0.200pt]{4.818pt}{0.400pt}}
\put(198,571){\makebox(0,0)[r]{200}}
\put(1416.0,571.0){\rule[-0.200pt]{4.818pt}{0.400pt}}
\put(220.0,648.0){\rule[-0.200pt]{4.818pt}{0.400pt}}
\put(198,648){\makebox(0,0)[r]{250}}
\put(1416.0,648.0){\rule[-0.200pt]{4.818pt}{0.400pt}}
\put(220.0,724.0){\rule[-0.200pt]{4.818pt}{0.400pt}}
\put(198,724){\makebox(0,0)[r]{300}}
\put(1416.0,724.0){\rule[-0.200pt]{4.818pt}{0.400pt}}
\put(220.0,801.0){\rule[-0.200pt]{4.818pt}{0.400pt}}
\put(198,801){\makebox(0,0)[r]{350}}
\put(1416.0,801.0){\rule[-0.200pt]{4.818pt}{0.400pt}}
\put(220.0,877.0){\rule[-0.200pt]{4.818pt}{0.400pt}}
\put(198,877){\makebox(0,0)[r]{400}}
\put(1416.0,877.0){\rule[-0.200pt]{4.818pt}{0.400pt}}
\put(220.0,113.0){\rule[-0.200pt]{0.400pt}{4.818pt}}
\put(220,68){\makebox(0,0){-200}}
\put(220.0,857.0){\rule[-0.200pt]{0.400pt}{4.818pt}}
\put(372.0,113.0){\rule[-0.200pt]{0.400pt}{4.818pt}}
\put(372,68){\makebox(0,0){-150}}
\put(372.0,857.0){\rule[-0.200pt]{0.400pt}{4.818pt}}
\put(524.0,113.0){\rule[-0.200pt]{0.400pt}{4.818pt}}
\put(524,68){\makebox(0,0){-100}}
\put(524.0,857.0){\rule[-0.200pt]{0.400pt}{4.818pt}}
\put(676.0,113.0){\rule[-0.200pt]{0.400pt}{4.818pt}}
\put(676,68){\makebox(0,0){-50}}
\put(676.0,857.0){\rule[-0.200pt]{0.400pt}{4.818pt}}
\put(828.0,113.0){\rule[-0.200pt]{0.400pt}{4.818pt}}
\put(828,68){\makebox(0,0){0}}
\put(828.0,857.0){\rule[-0.200pt]{0.400pt}{4.818pt}}
\put(980.0,113.0){\rule[-0.200pt]{0.400pt}{4.818pt}}
\put(980,68){\makebox(0,0){50}}
\put(980.0,857.0){\rule[-0.200pt]{0.400pt}{4.818pt}}
\put(1132.0,113.0){\rule[-0.200pt]{0.400pt}{4.818pt}}
\put(1132,68){\makebox(0,0){100}}
\put(1132.0,857.0){\rule[-0.200pt]{0.400pt}{4.818pt}}
\put(1284.0,113.0){\rule[-0.200pt]{0.400pt}{4.818pt}}
\put(1284,68){\makebox(0,0){150}}
\put(1284.0,857.0){\rule[-0.200pt]{0.400pt}{4.818pt}}
\put(1436.0,113.0){\rule[-0.200pt]{0.400pt}{4.818pt}}
\put(1436,68){\makebox(0,0){200}}
\put(1436.0,857.0){\rule[-0.200pt]{0.400pt}{4.818pt}}
\put(220.0,113.0){\rule[-0.200pt]{292.934pt}{0.400pt}}
\put(1436.0,113.0){\rule[-0.200pt]{0.400pt}{184.048pt}}
\put(220.0,877.0){\rule[-0.200pt]{292.934pt}{0.400pt}}
\put(45,495){\makebox(0,0){$\CQ_4$}}
\put(828,23){\makebox(0,0){$\CQ_3$}}
\put(220.0,113.0){\rule[-0.200pt]{0.400pt}{184.048pt}}
\put(828,269){\Dia}
\put(828,817){\Dia}
\put(828,522){\Dia}
\put(828,353){\Dia}
\put(828,276){\Dia}
\put(986,266){\Dia}
\put(670,266){\Dia}
\put(1201,317){\Dia}
\put(1151,233){\Dia}
\put(455,317){\Dia}
\put(505,233){\Dia}
\put(1046,513){\Dia}
\put(1331,172){\Dia}
\put(1011,345){\Dia}
\put(610,513){\Dia}
\put(325,172){\Dia}
\put(645,345){\Dia}
\put(1266,242){\Dia}
\put(1083,284){\Dia}
\put(912,303){\Dia}
\put(1129,405){\Dia}
\put(927,422){\Dia}
\put(946,649){\Dia}
\put(390,242){\Dia}
\put(573,284){\Dia}
\put(744,303){\Dia}
\put(527,405){\Dia}
\put(729,422){\Dia}
\put(710,649){\Dia}
\end{picture}
\end{center}
\caption{Distribution of the quantized values of $\CQ_4$ versus
$\CQ_3$ for $n=4$ and $h=10$.}
\label{fig8}
\end{figure}}

\ladder

\bigskip

\BSeq

\figa

\figb

\figc

\figd

\fige

\figf

\figg

\figh

\end{document}